\pdfoutput=1
\documentclass[11pt,reqno,preprint]{article}
\usepackage{jheppub,epsfig,amssymb,amsmath,mathrsfs,hyperref,graphicx}
\usepackage{multirow,bbm}
\usepackage[makeroom]{cancel}
\usepackage{caption}
\usepackage{subcaption}

\def\be{\begin{equation}}
\def\ee{\end{equation}}
\def\ba{\begin{eqnarray}}
\def\ea{\end{eqnarray}}
\newcommand{\bea}{\begin{eqnarray}}
\newcommand{\eea}{\end{eqnarray}}
\newcommand{\nn}{\nonumber}

\def\z#1{\zeta_{#1}}

\def\ah{\hat{a}}
\def\bh{\hat{b}}
\def\ch{\hat{c}}
\def\dh{\hat{d}}
\def\eh{\hat{e}}
\def\fh{\hat{f}}
\def\uh{\hat{u}}
\def\vh{\hat{v}}
\def\wh{\hat{w}}

\newcommand\zb{\bar{z}}
\newcommand{\del}{\partial}
\newcommand{\de}{\delta}
\newcommand{\lnden}{\ln|\de|}

\newcommand{\lndene}[1]{\ln^{#1}|\de|}

\def\sect#1{section~{\ref{#1}}}
\def\eqn#1{eq.~(\ref{#1})}

\def\eqns#1#2{eqs.~(\ref{#1}) and~(\ref{#2})}

\def\eqn#1{eq.~(\ref{#1})}
\def\eqns#1#2{eqs.~(\ref{#1}) and~(\ref{#2})}

\newcommand{\fwboxL}[2]{\text{\makebox[#1][l]{$#2$}}}

\def\lr{\leftrightarrow}
\def\RR{\mathcal{R}}
\def\EE{\mathcal{E}}

\def\Hhex{{\cal H}^{\rm hex}}
\def\Et{\tilde{E}}

\def\Gcusp{\Gamma_{\rm cusp}}
\def\Gvirt{\Gamma_{\rm virt}}

\def\uh{\hat{u}}
\def\vh{\hat{v}}
\def\wh{\hat{w}}

\def\blue#1{{\color{blue}#1}}
\def\green#1{{\color{green}#1}}

\title{An Eight Loop Amplitude via Antipodal Duality}

\author{Lance~J.~Dixon$^{1}$}
\author{and Yu-Ting Liu$^{1}$}

\affiliation{$^1$ SLAC National Accelerator Laboratory,
Stanford University, Stanford, CA 94309, USA}

\abstract{We compute the six-particle maximally-helicity-violating
(MHV) amplitude in planar ${\cal N}=4$ super-Yang-Mills theory at
eight loops, using antipodal duality and the recently computed
eight-loop three-point form factor for the chiral stress energy tensor
multiplet. Antipodal duality maps the form factor symbol to the amplitude
symbol on a two-dimensional parity-preserving surface in the three-dimensional
amplitude kinematics. There are remarkably few ambiguities in lifting
from two to three dimensions, nor in promoting the symbol to a function.
The amplitude passes many tests, including near-collinear, multi-Regge,
factorization, self-crossing and origin limits. These checks also
constitute a validation of antipodal duality at eight loops.}

\emailAdd{lance@slac.stanford.edu}\emailAdd{aytliu@stanford.edu}

\preprint{ \begin{flushright} SLAC--PUB--17693 \end{flushright}}

\begin{document}
\hypersetup{pageanchor=false}
\maketitle
\hypersetup{pageanchor=true}

\section{Introduction}
\label{sec:intro}

Perturbative scattering amplitudes in gauge theory lead to notoriously
complicated functions of the kinematical variables, especially beyond one loop.
On the other hand, there are a few examples where the functions can be
expressed in terms of multiple
polylogarithms (MPLs)~\cite{Chen,G91b,Goncharov:1998kja,Remiddi:1999ew,%
Borwein:1999js,Moch:2001zr},
with a fixed symbol alphabet~\cite{Goncharov:2010jf},
giving hope that the all-orders behavior can be elucidated for
generic kinematics.  In particular, in planar ${\cal N}=4$ super-Yang-Mills
theory (SYM), thanks to dual conformal
symmetry~\cite{Drummond:2006rz,Bern:2006ew,Bern:2007ct,Alday:2007hr,
  Drummond:2008vq},
the first nontrivial scattering amplitudes are for six particles, and
depend on three dual conformal cross ratios~\cite{Bern:2008ap,Drummond:2008aq,%
  DelDuca:2009au,DelDuca:2010zg}.
The $L$-loop amplitudes are weight $2L$ polylogarithmic functions, which
have symbols with words of length $2L$, whose letters in this case
are drawn from a nine-letter alphabet~\cite{Goncharov:2010jf}.
This symbol alphabet is associated with a cluster algebra for the
Grassmannian ${\rm Gr}(4,6)$ with Dynkin label
$A_3$~\cite{Golden:2013xva,Golden:2014pua}.
The hexagon alphabet describes both maximally helicity violating (MHV)
and next-to-MHV (NMHV) six-particle scattering amplitudes, which can be
bootstrapped to seven loops~\cite{Dixon:2011pw,Dixon:2011nj,Dixon:2013eka,%
  Dixon:2014iba,Dixon:2014voa,Dixon:2015iva,Caron-Huot:2016owq,%
  Caron-Huot:2019vjl,Caron-Huot:2020bkp,DDToAppear}. 
One also uses branch-cut information in the guise of
first-entry conditions~\cite{Gaiotto:2011dt}
and extended Steinmann relations~\cite{Caron-Huot:2018dsv,%
Caron-Huot:2019vjl,Caron-Huot:2019bsq,He:2021mme},
or equivalently cluster adjacency~\cite{Drummond:2017ssj,Drummond:2018caf}.
A basis of hexagon functions obeying these constraints has been
constructed through weight 11~\cite{Caron-Huot:2019vjl,Caron-Huot:2019bsq}.
Boundary conditions, as well as cross-checks, for the hexagon
amplitude bootstrap come from the near-collinear limit, which
corresponds to an operator product expansion (OPE)~\cite{Alday:2010ku,%
  Gaiotto:2011dt,Basso:2013vsa,Basso:2013aha,Basso:2014koa} of the
dual light-like closed polygonal Wilson loop~\cite{Alday:2007hr,Drummond:2007aua,Brandhuber:2007yx,Alday:2007he,Drummond:2007au,Alday:2008yw,Adamo:2011pv,Bern:2008ap,Drummond:2008aq,Ben-Israel:2018ckc}.

More recently, an even simpler quantity has been bootstrapped one loop
further, to eight loops~\cite{Dixon:2020bbt,Dixon:2022rse}.
That quantity
is the three-particle (MHV) form factor for the chiral stress tensor
operator multiplet, computed earlier through two
loops~\cite{Brandhuber:2010ad,Brandhuber:2012vm}.
It depends on only two dimensionless variables and its symbol alphabet
has only six letters~\cite{Brandhuber:2012vm,Dixon:2020bbt}.
The boundary conditions for this bootstrap are provided by
the recently-developed Form Factor OPE
(FFOPE)~\cite{Sever:2020jjx,Sever:2021nsq,Sever:2021xga}.

Quite remarkably, the MHV six-particle amplitude and the three-particle form
factor are antipodally dual to each other~\cite{Dixon:2021tdw},
which means that their symbols are precisely written backwards
from each other, loop order by loop order.  The duality has been
verified through seven loops, because the MHV six-particle amplitude
was only known through that order.
The duality also involves a kinematic map,
from the two-dimensional form factor kinematics into a two-dimensional
parity-preserving slice of the three-dimensional amplitude kinematics.

More recently, antipodal duality has been understood to follow
from a particular limit of an antipodal \emph{self}-duality
discovered for the \emph{four}-particle
MHV form factor at two loops~\cite{Dixon:2022xqh}.  The
double collinear limit of the four-particle form factor (remainder)
goes smoothly into the three-particle form factor remainder,
while the triple collinear limit becomes the MHV six-particle amplitude
remainder function.
(The triple-collinear behavior follows from the universality of factorization
limits and dual conformal invariance~\cite{Bern:2008ap}, and it
can also be understood from the FFOPE~\cite{Sever:2020jjx,Dixon:2022xqh}.)
The four-particle form factor antipodal self-dual kinematic map neatly maps
the double and triple collinear limits into each other, which means
that its antipodal self-duality {\it subsumes} the earlier ``6--3''
antipodal duality.
Of course, antipodal self-duality is only directly established
to two loops (with three loops to appear~\cite{MoreASDToAppear}),
while the 6--3 duality has been verified to seven loops.

The purpose of this paper is to exploit the 6--3 duality, and the knowledge
of the eight-loop form factor~\cite{Dixon:2022rse}, in order to compute
the eight-loop MHV six-particle amplitude.  The general idea is that
applying antipodal duality to the eight-loop form factor determines
the symbol of the eight-loop amplitude on the two-dimensional
parity-preserving slice, and this symbol constitutes a very strong
boundary condition on the full three-variable answer;
there are almost no ambiguities in lifting from two to three dimensions. 
The power of this method suggests that it may be possible, more broadly,
to bootstrap amplitudes and form factors on simpler subspaces
as intermediate steps, before lifting them up to their full
dimensionality.

Here we briefly outline the different steps in our computation;
in the remaining sections of the paper
we provide more detail and characterize the result.
The first step in the computation
is to apply the 6--3 kinematic map, which is a simple substitution,
and then reverse the ordering of the entries in the symbol, which is
computationally nontrivial, because the fully expanded symbol
has over 1.67 billion terms.  Because of its large size, we generally
encode the symbol information in a nested format, which is organized by
declaring how the derivatives of functions are related to lower-weight
functions.  Reversing the symbol requires undoing the nesting, but
it does not need to be all undone at once; one can take advantage of
initial and final-entry relations to make the computation easier.

\begin{figure}
\begin{center}
\includegraphics[width=3.2in]{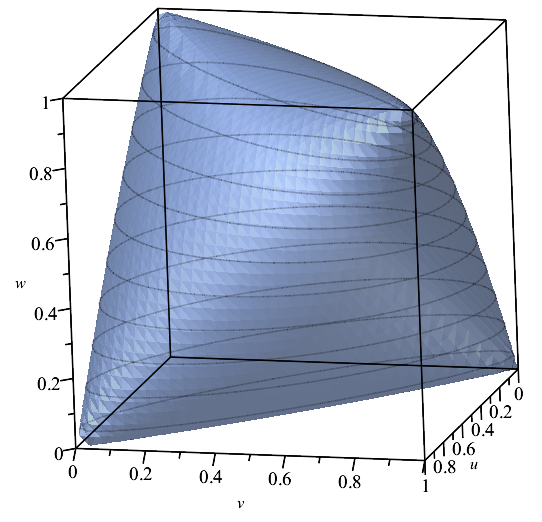}
\end{center}
\caption{The surface $\Delta=0$ inside the unit cube $0 < \uh,\vh,\wh < 1$.}
\label{fig:DeltaEq0}
\end{figure}

The reversed symbol can be written using the hexagon-function basis
constructed through weight 11~\cite{Caron-Huot:2019vjl,Caron-Huot:2019bsq}.
Because the eight-loop amplitude has weight 16, this means that
we express it in terms of its fifth derivatives, or more precisely,
its $\{11,1,1,1,1,1\}$ coproducts.  At this point, we know the symbol
of the MHV amplitude on the two-dimensional parity-preserving surface.
The next task is to lift it off that surface into the full
three-dimensional bulk kinematics, which is parametrized by
three dual-conformally invariant cross ratios,
$(\uh,\vh,\wh)$.\footnote{We put hats on the cross ratios in order
to distinguish them from similar ratios appearing in the three-particle
form factor.}
The nine-letter hexagon alphabet is
$\{\uh,\vh,\wh,1-\uh,1-\vh,1-\wh,y_u,y_v,y_w\}$.
The parity-preserving surface is $\Delta(\uh,\vh,\wh) = 0$,
where $\Delta(\uh,\vh,\wh) = (1-\uh-\vh-\wh)^2 - 4 \uh \vh \wh$.
The portion of this surface that lies within the unit cube
$0 < \uh,\vh,\wh < 1$ is shown in figure~\ref{fig:DeltaEq0}.
On this surface, $y_u = y_v = y_w = 1$.
Hence the symbol on the parity-preserving surface
has only six letters, $\{\uh,\vh,\wh,1-\uh,1-\vh,1-\wh\}$, and the lifting task
is to supply the dependence on the three additional parity-odd letters,
$\{y_u,y_v,y_w\}$.

An important question is: How well-defined is the lifting task? 
How many free parameters are left after constraining the symbol to
a definite value on the parity-preserving surface?
That is, suppose a parity-even
hexagon-function symbol is constrained to vanish on
$\Delta = 0$; how many independent functions satisfy this constraint?
Remarkably, through weight 10, the answer is zero --- the amplitude's
symbol is fixed {\it uniquely} by the antipodal information!
At 6 and 7 loops (weights 12 and 14, respectively), there are
1 and 3 such functions. Some of these function were encountered
in ref.~\cite{Caron-Huot:2019vjl}, and called $Z$ and $\tilde{Z}$.
They were encountered as ambiguities at a certain stage of the
traditional bootstrap procedure, because they vanish in the
near-collinear, or OPE limit
at the level of one flux-tube excitation
(${\cal O}(\sqrt{\vh}) = {\cal O}(\hat{T}^1)$ as
$\vh = \hat{T}^2/(1+\hat{T}^2) \to 0$, where $\hat{T}$ is a variable
in the OPE parametrization~\cite{Basso:2013vsa}).
The reason for the rapid vanishing of these functions in this limit
was traced to the vanishing of their parity-even first coproducts, 
$Z^{u_i} = Z^{1-u_i} = \tilde{Z}^{u_i} = \tilde{Z}^{1-u_i} = 0$.
Because only the parity-odd first coproducts,
$Z^{y_i}$ and $\tilde{Z}^{y_i}$, are nonvanishing,
their symbol's last-entry is always a $y_i$, and therefore
their symbol vanishes on the $\Delta = 0$ surface.

Dihedral symmetry was also applied in ref.~\cite{Caron-Huot:2019vjl};
when one relaxes that constraint, one still finds only 1 function ($Z$) at
weight 12, but there is a triplet of functions at weight 14.
The cyclically invariant sum of the three functions is the $\tilde{Z}$
function of ref.~\cite{Caron-Huot:2019vjl}.
In the present case, we need to determine how many such functions there
are at weight 16.  The answer is 9.  After imposing dihedral invariance,
3 functions are left.  We could impose the OPE constraints to further trim
this space, but at symbol level, since they all vanish at
${\cal O}(\hat{T}^1)$, it would require going to two flux-tube excitations,
${\cal O}(\hat{T}^2)$, which is computationally difficult.
It is simpler to go to the hexagon origin, the limit
$(\uh,\vh,\wh) \to (0,0,0)$,
where the behavior is now predicted to all
orders~\cite{Basso:2020xts,Basso:2022ruw}.  At symbol level,
the coefficients of 2 of the 3 remaining functions can be fixed at the origin,
but one linear combination vanishes there.  Fortunately, at function level,
the remaining ambiguity function does not actually vanish at the origin,
and so the remaining symbol-level parameter can be fixed in the process
of lifting the bulk symbol to a complete function.

There are also a few function-level ambiguities to fix at this stage.
We use a small amount of OPE data to do so; just the first term
in the series expansion of the one flux-tube OPE data is enough,
thanks to the power of symbol-level antipodal duality.

After fully fixing the eight-loop result, we
impose a number of cross-checks on it,
including the full one flux-tube OPE behavior, and the multi-Regge
limit~\cite{Basso:2014pla}.
We also check predictions of antipodal duality that go
beyond the symbol~\cite{Dixon:2021tdw}.  On the line
$(\uh,\vh,\wh)=(\uh,\uh,1)$,
the amplitude collapses to harmonic polylogarithms
(HPLs)~\cite{Remiddi:1999ew}.  Antipodal duality
predicts~\cite{Dixon:2021tdw} all terms
that do not have an explicit $\pi$ in them, including beyond-the-symbol
terms proportional to odd Riemann zeta values, and 
multiple zeta values (MZVs).
All the predictions of antipodal duality work perfectly.
In summary, although we used antipodal duality at symbol level to construct
the result, its beyond-the-symbol consequences are now tested at eight loops.

We plot the eight-loop result on a couple of lines. We also describe
what we can learn about the MHV amplitude's multi-final entry relations
at eight loops.  We provide the values of these multi-final entries at
the standard hexagon-function base-point $(\uh,\vh,\wh)=(1,1,1)$, where we
find that remarkably few are linearly independent, and where can
test the coaction principle (also known as Cosmic Galois invariance)
as in ref.~\cite{Caron-Huot:2019bsq}.
We are also able to use consistency of the eight-loop amplitude,
and of the $\Delta = 0$ ambiguity functions, to fix the final six remaining
constants of integration in the weight 11 hexagon function space,
which could not be fixed at seven loops~\cite{Caron-Huot:2019bsq}.

This paper is organized as follows. In section~\ref{sec:hexreview}
we provide a lightning review of hexagon functions and the coproduct
formalism.  In section~\ref{sec:symbolflip} we explain how to flip
a weight 16, 1.67 billion term symbol.
Section~\ref{sec:symbollift} describes how we lift the symbol off of the
parity-preserving slice, up to a one-parameter ambiguity.
In section~\ref{sec:fullfunction} we fix that ambiguity, as well as
all beyond-the-symbol ambiguities, using various boundary conditions.
Section~\ref{sec:checks} describes how the eight-loop amplitude passes
numerous checks involving four different kinematic limits.
In section~\ref{sec:multiplefinalentry} we describe empirical
linear relations among the multiple final entries of the amplitude,
which we expect to hold to all loop orders.
Section~\ref{sec:copparitylocking} describes how a few of the
beyond-the-symbol hexagon functions are not needed when the amplitude
is optimally (cosmically) normalized, because they always appear ``locked''
to other functions.
In section~\ref{sec:ampcoprods111} we study the values of the amplitudes
and their multiple coproducts at the hexagon-function base point $(1,1,1)$,
and test the coaction principle there.
Section~\ref{sec:wt16Zfunctions} describes a set of parity-even
``$Z$'' functions, which vanish on the entire parity-preserving surface,
but not off it; they are the essential ambiguities in lifting
off the surface. In section~\ref{sec:lines} we provide numerical
results at $(1,1,1)$ and along a pair of lines that
intersect at that point.  Finally, in section~\ref{sec:conclusions}
we present our conclusions.

A number of results are best presented as computer-readable ancillary
files.  We provide the following such files along with this paper: 
{\tt MHV8quintuples.txt} gives the weight 11 quintuple coproducts
of the eight-loop MHV amplitude in terms of the hexagon function basis;
{\tt EZMHVcoproducts111.txt} provides the constant values of the eight-loop
coproducts with higher weights (12 to 16) at $(1,1,1)$, along with
lower-loop values; {\tt RLncy.txt} provides the near-collinear (OPE) limits
of the remainder function; {\tt RLncyser.txt} gives a further series
expansion in $S$; {\tt MRKSigma.txt} provides the perturbative expansion
of the multi-Regge limit through eight loops;
{\tt EZsmallercoproductspace.txt} gives a reduced space of
hexagon functions revealed by the amplitudes' coproducts;
{\tt ftoMZV16.txt} converts between two representations of MZVs;
{\tt Zorigin.txt} gives the values of the weight 16 $Z$ functions
at the origin; and \texttt{EZMHVg\_uu1\_lin.txt} and
\texttt{EZMHVg\_u11\_lin.txt} give the values of the MHV amplitude
on the lines $(\uh,\uh,1)$ and $(\uh,1,1)$, respectively.
All these files are hosted at~\cite{CosmicWebsite}.

\section{Hexagon and form-factor function lightning review}
\label{sec:hexreview}

In this section we provide a lightning review of the space of
hexagon functions, which describes six-particle amplitudes
in planar ${\cal N}=4$ SYM.  For further details, the reader is
referred to refs.~\cite{Dixon:2013eka,Caron-Huot:2019vjl,Caron-Huot:2019bsq,%
  Caron-Huot:2020bkp}.  Then we briefly describe the space of form-factor
functions; see refs.~\cite{Dixon:2020bbt,Dixon:2022rse}
for further details.  We also introduce a new amplitude normalization
factor~\cite{Basso:2020xts},
and review the antipodal map between the three-particle form factor
and the MHV amplitude.

The coefficients in the perturbative expansion of six-point scattering
amplitudes in planar ${\cal N}=4$ SYM are hexagon functions,
a particular case of generalized polylogarithms, or iterated
integrals over logarithmic kernels.
Generalized polylogarithms can be defined iteratively by
\be \label{eq:G_func_def}
G_{a_1,\dots, a_n}(z)
= \int_0^z \frac{dt}{t-a_1} G_{a_2,\dots, a_n}(t)\,,
\qquad G_{\fwboxL{27pt}{{\underbrace{0,\dots,0}_{p}}}}(z) = \frac{\ln^p z}{p!} \, , 
\ee
with $G(z)\equiv1$.  The \emph{weight} of $G_{a_1,\dots, a_n}(z)$ is the
number of integrations $n$.

The differential of any such function $F$ has the form
\be \label{eq:A_differential}
d F
= \sum_{\phi \in {\cal L}} F^{\phi} \ d \ln \phi \, ,
\ee
where the \emph{symbol letters} $\phi$ belong to the \emph{alphabet} ${\cal L}$,
and the \emph{first coproduct} $F^{\phi}$ is also an iterated integral
with weight one lower than $F$.
For hexagon functions, the alphabet has nine letters.
The original alphabet for the hexagon function bootstrap
was~\cite{Dixon:2011pw}
\be
{\cal L}^u_\text{hex} =
\left\{ \uh, \vh, \wh, 1-\uh, 1-\vh, 1-\wh, y_u, y_v, y_w \right\} \,,
\label{hex_letters_uvw}
\ee
where the dual conformal cross ratios are
\be
\uh = \frac{x_{13}^2x_{46}^2}{x_{14}^2x_{36}^2}
    = \frac{s_{12}s_{45}}{s_{123}s_{345}} \,, \quad
\vh = \frac{x_{24}^2x_{51}^2}{x_{25}^2x_{41}^2} \,, \quad
\wh = \frac{x_{35}^2x_{62}^2}{x_{36}^2x_{52}^2} \,.
\label{hex_uvw_def}
\ee
Here we will mainly (but not always) use the equivalent
alphabet~\cite{Caron-Huot:2016owq},
\be
{\cal L}^a_\text{hex} =
\left\{ \ah, \bh, \ch, \dh, \eh, \fh, y_u, y_v, y_w \right\} \,,
\label{hex_letters_abc}
\ee
where
\be
\ah = \frac{\uh}{\vh\wh} \,, \quad
\bh = \frac{\vh}{\wh\uh} \,, \quad
\ch = \frac{\wh}{\uh\vh} \,, \quad
\dh = \frac{1-\uh}{\uh} \,, \quad
\eh = \frac{1-\vh}{\vh} \,, \quad
\fh = \frac{1-\wh}{\wh} \,.
\label{hatabcdef}
\ee
and $\uh,\vh,\wh$ are the usual dual conformal cross ratios.
We put hats on them to distinguish them from the ordinary ratios
$u,v,w$ for the form factor.

The parity-odd letters,
\be
y_u = \frac{\uh-z_+}{\uh-z_-}\,, \qquad y_v = \frac{\vh-z_+}{\vh-z_-}\,,
\qquad y_w = \frac{\wh - z_+}{\wh - z_-}\, ,
\label{yfromu}
\ee
are defined in terms of
\be
z_\pm = \frac{1}{2}\Bigl[-1+\uh+\vh+\wh \pm \sqrt{\Delta}\Bigr],
\qquad
\Delta = (1-\uh-\vh-\wh)^2 - 4 \uh \vh \wh .
\ee
The dihedral symmetry group $D_6$ contains an $S_3$ permuting
$(\uh_1,\uh_2,\uh_3)\equiv(\uh,\vh,\wh)$ and $(y_1,y_2,y_3)\equiv(y_u,y_v,y_w)$.
In terms of the alphabet ${\cal L}^a_\text{hex}$, the $S_3$ is
generated by:
\bea
&&\hbox{cycle:}\quad  \ah \to \bh \to \ch \to \ah,
  \quad \dh \to \eh \to \fh \to \dh,
  \quad y_u \to 1/y_v \to y_w \to 1/y_u,
\label{cycledef}\\
&&\hbox{flip:}\quad \ah \lr \bh, \quad \dh \lr \eh, \quad y_u \lr y_v,
\label{flipdef}
\eea
There is also the parity transformation, whose only effect is to
invert the parity-odd letters, $y_i \lr 1/y_i$.
On the parity-even surface,
\be
\Delta(\uh,\vh,\wh) = 0,
\label{Pevensurface}
\ee
all the $y_i \to 1$, which means that they drop out of the symbol,
reducing it to six letters, $\{\ah,\bh,\ch,\dh,\eh,\fh\}$,
the same number as for the three-particle form
factor~\cite{Brandhuber:2012vm,Dixon:2020bbt,Dixon:2022rse}.

For the form factor for three massless particles, momentum conservation
reads $p_1 + p_2 + p_3 = - q$, where $q$ is the operator momentum.
Squaring this relation and using $p_i^2=0$, we have
\be
q^2 = s_{123} = s_{12} + s_{23} + s_{31} \,,
\label{qsq_relation}
\ee
where $s_{ij} = (p_i+p_j)^2$.  The dimensionless ratios,
\be
 u = \frac{s_{12}}{q^2} \,, \quad 
 v = \frac{s_{23}}{q^2} \,, \quad 
 w = \frac{s_{31}}{q^2} \,,
\label{uvw_def}
\ee
obey
\be
u + v + w = 1,
\label{uvwsumto1}
\ee
as a consequence of \eqn{qsq_relation}.

The form-factor symbol alphabet
is~\cite{Brandhuber:2012vm,Dixon:2020bbt,Dixon:2022rse}
\be
{\cal L}^a_\text{F3} = \left\{ a, b, c, d, e, f \right\} \,,
\label{FF_letters_abc}
\ee
where
\be
a = \frac{u}{vw} \,, \quad
b = \frac{v}{wu} \,, \quad
c = \frac{w}{uv} \,, \quad
d = \frac{1-u}{u} \,, \quad
e = \frac{1-v}{v} \,, \quad
f = \frac{1-w}{w} \,.
\label{abcdef}
\ee
There is a dihedral symmetry $S_3$ generated by
\bea
&&\hbox{cycle:}\quad  a \to b \to c \to a,
  \quad d \to e \to f \to d,
\label{cycledefF}\\
&&\hbox{flip:}\quad a \lr b, \quad d \lr e.
\label{flipdefF}
\eea

Generalized polylogarithms are naturally equipped with a coaction
$\Delta$~\cite{Gonch2,Goncharov:2010jf,Brown:2011ik,Duhr:2012fh,%
Brown1102.1312,Brown:2015fyf} which maps weight $n$ functions to (sums of)
products of functions (roughly) of weight $(n-p)$ and $p$, for any integer $p$
between $0$ and $n$.  The $p=1$ case, or $\{n-1,1\}$ component
of the coaction, is equivalent to
the total differential~\eqref{eq:A_differential}:
\be
\Delta_{n-1,1} (F )
\ =\ \sum_{\phi \in {\cal L}} F^{\phi} \otimes \ln \phi \,.
\label{Deltanm11}
\ee
The map $\Delta_{\bullet,1}$ can be applied iteratively to each $F^\phi$, leading
to the \emph{second coproducts} $F^{\phi^\prime,\phi}$, etc.
Continuing on $n$ times, the weight $n$ polylogarithms entering $F$
are mapped to $n$-fold tensor products of logarithms,
the \emph{symbol} of $F$~\cite{Goncharov:2010jf}.
The arguments of the logarithms are the symbol alphabet.

The perturbative coefficients of the six-point amplitude at $L$ loops
can be expressed in terms of weight $2L$ polylogarithms.
However, it is not practical to write out such an expression explicitly
at high loop orders, because the number of $G$ functions required
grows roughly by a factor of 5 at each weight, so the seven loop
results might require around a billion terms.  Instead, we use a nested
description, the coproduct formalism~\cite{Dixon:2013eka,Caron-Huot:2019bsq}.
This formalism describes the first derivative
of any function $F$ (which might be an amplitude or a basis function)
in terms of its first coproducts $F^\phi$.  Those functions are in turn
described by their first coproducts, and so on down to logarithms.
Let $F^{(n)}_{i_n}$, $i_n=1,2,\ldots d_n$ denote a basis for the weight $n$
part of the hexagon function space $\Hhex$,
which has dimension $d_n$.
Also give the letters a discrete label $k$, so that $\phi_k \in {\cal L}$.
Then the coproducts are described by a $\mathbb{Q}$-valued three-index
tensor $T$, with dimension $d_n\times d_{n-1}\times |{\cal L}|$,
\be
\label{tensor_coproduct_single}
\Delta_{n-1,1} F^{(n)}_{i_n}
 = \sum_{i_{n-1},k} T^k_{i_n,i_{n-1}}F^{(n-1)}_{i_{n-1}} \otimes \ln \phi_k \,,
\ee
where $|{\cal L}_\text{hex}| = 9$ and $|{\cal L}_\text{F3}| = 6$.

We begin the construction of $\Hhex$ at weight 1 by imposing
the branch-cut conditions, which requires that the first letter
in the symbol is drawn from $\{ \uh, \vh, \wh \}$, or equivalently
$\{ \ah, \bh, \ch \}$.  In other words,
\be
\Hhex_1 = \{ \, \ln \ah,\ \ln \bh,\ \ln \ch \, \} \,.
\label{firstentry}
\ee
In constructing $\Hhex_n$ at higher weights $n$, we impose integrability,
$d^2F=0$, as well as the extended Steinmann
relations~\cite{Caron-Huot:2016owq,Caron-Huot:2019bsq},
that $\ah$ never appears adjacent to $\bh$,
\begin{align} 
\cancel{\ldots \otimes \ah \otimes \bh \otimes \ldots},
\label{eq:ExtSteinmannab}
\end{align}
plus the five other conditions generated by dihedral symmetry.

In terms of double coproducts of functions $F$, the extended
Steinmann conditions are
\be
F^{\ah,\bh} = F^{\bh,\ch} = F^{\ch,\ah} = F^{\bh,\ah} = F^{\ch,\bh} = F^{\ah,\ch} = 0.
\label{ExtSF}
\ee
We impose these conditions along with integrability and the first-entry
condition.
Together, these conditions lead to 41 relations among the $9\times 9 = 81$
adjacent pairs of letters, or equivalently 40 independent pairs of adjacent
entries~\cite{Caron-Huot:2018dsv,Caron-Huot:2019vjl,Caron-Huot:2019bsq},
which also corresponds to the $A_3$ cluster
adjacency conditions~\cite{Drummond:2017ssj,Drummond:2018caf}.
Included in these relations are the non-adjacency of symbol letters
$\ah$ and $\dh$, plus four other relations generated by dihedral symmetry:
\begin{align} 
F^{\ah,\dh} = F^{\bh,\eh} = F^{\ch,\fh} = F^{\dh,\ah} = F^{\eh,\bh} = F^{\fh,\ch} = 0.
\label{eq:ExtSteinmannad}
\end{align}

In addition, we only include constants (zeta values) as independent
functions when the amplitudes' coproducts dictate their presence.
As in refs.~\cite{Caron-Huot:2016owq,Caron-Huot:2019bsq},
this means that the only independent zeta-valued constants are
$\zeta_4$, $\zeta_6$, $\zeta_8$, $\zeta_{10}$, etc.

In some cases we use the old alphabet ${\cal L}^u_\text{hex}$,
for example to describe the multi-final entry relations in
section~\ref{sec:multiplefinalentry}.  One can convert between alphabets
using the following relations between coproducts:
\bea
F^{\uh} &=& F^{\ah} - F^{\bh} - F^{\ch} - F^{\dh} \,, \nn\\
F^{\vh} &=& - F^{\ah} + F^{\bh} - F^{\ch} - F^{\eh} \,, \nn\\
F^{\wh} &=& - F^{\ah} - F^{\bh} + F^{\ch} - F^{\fh} \,, \nn\\
F^{1-\uh} &=& F^{\dh} \,, \nn\\
F^{1-\vh} &=& F^{\eh} \,, \nn\\
F^{1-\wh} &=& F^{\fh} \,.
\label{oldnewcoproducts}
\eea

Besides giving all the first derivatives of the hexagon functions through
the coproduct tensors $T$, through weight 11,
we must also specify the values of the
basis functions at one point, as integration constants.
For hexagon functions, we use the base point $(\uh,\vh,\wh)=(1,1,1)$ in the
Euclidean region.  This point is dihedrally symmetric, all the functions
are finite there, and they all evaluate to multiple zeta values (MZVs).
The coproduct and base point data for $\Hhex$ is complete through weight 11
(up to a few weight 11 constants $n_i$, which are fixed below),
and it is provided in the ancillary files for
refs.~\cite{Caron-Huot:2019vjl,Caron-Huot:2019bsq},
which are stored at~\cite{CosmicWebsite}.

We remark that the
function-level part of the weight 11 basis in the parity-even sector
is sub-optimal; the coproduct tensors contain rational numbers with
large denominators in some cases.  This fact means that some care has to be
taken when reconstructing the rational-number coefficients of some of the
basis functions, if the constraint equations are initially solved over a
prime number field.

In this work, we use a new, all-orders
``cosmic'' BDS-like normalization of the MHV amplitude, inspired by
the amplitude's behavior at the origin~\cite{Basso:2020xts},
\be
\EE =
\frac{1}{\rho} \,\exp\biggl[ \frac{\Gcusp}{4} \EE^{(1)} + \RR \biggr]\,,
\label{EZMHVtoR6simplerho}
\ee
where $\RR$ is the remainder function, and
\bea
\rho = \det(\mathbb{I}+\mathbb{K})
  \,\exp\Bigl[ -\tfrac{1}{2} \zeta_2 \Gcusp \Bigr] \,.
\label{simplerho}
\eea
Here $\mathbb{K}$ is the semi-infinite matrix
kernel of the BES equation~\cite{Beisert:2006ez},
which provides the cusp anomalous dimension $\Gcusp$ to all orders in $g^2$,
in terms of the $1,1$ matrix element of the matrix inverse,
\be
\frac{\Gcusp}{4}
= g^2 \, \biggl[ (\mathbb{I}+\mathbb{K})^{-1} \biggr]_{1,1} \,,
\label{BEScuspformula}
\ee
and $\mathbb{I}$ is the identity matrix.

The formula~(\ref{simplerho}) has the perturbative expansion through
eight loops,
\bea
\rho(g^2) &=& 1 - \zeta_4 \, g^4
+ \biggl[ \frac{50}{3} \zeta_6 + 8 (\zeta_3)^2 \biggr] \, g^6
- \biggl[ \frac{2891}{12} \zeta_8 + 160 \zeta_3 \zeta_5 \biggr] \, g^8
\nonumber\\
&&\null\hskip0.0cm
+ \biggl[ \frac{10265}{3} \zeta_{10} - 40 \zeta_4 (\zeta_3)^2
       + 1680 \zeta_3 \zeta_7 + 912 (\zeta_5)^2 \biggr]
\, g^{10}
\nonumber\\
&&\null\hskip0.0cm
- \biggl[ \frac{4857061891}{99504} \zeta_{12}
   - \frac{1600}{3} \zeta_6 (\zeta_3)^2
   - 608 \zeta_4 \zeta_3 \zeta_5
   + 18816 \zeta_3 \zeta_9 + 20832 \zeta_5 \zeta_7 \biggr] \, g^{12}
\nonumber\\
&&\null\hskip0.0cm
+ \biggl[ \frac{50570065}{72} \zeta_{14}
   - 6370 \zeta_8 (\zeta_3)^2  - \frac{24320}{3} \zeta_6 \zeta_3 \zeta_5 
   - 5040 \zeta_4 \zeta_3 \zeta_7 - 2736 \zeta_4 (\zeta_5)^2
\nonumber\\
&&\null\hskip0.5cm
   + 221760 \zeta_3 \zeta_{11} + 247296 \zeta_5 \zeta_9 + 126240 (\zeta_7)^2
   \biggr] \, g^{14}
\nonumber\\
&&\null\hskip0.0cm
- \biggl[ \frac{63718004141707}{6250176} \zeta_{16}
   - \frac{230960}{3} \zeta_{10} (\zeta_3)^2 - 98056 \zeta_8 \zeta_3 \zeta_5
   - 67840 \zeta_6 \zeta_3 \zeta_7
\nonumber\\
&&\null\hskip0.5cm
   - 34880 \zeta_6 (\zeta_5)^2
   - 45696 \zeta_4 \zeta_3 \zeta_9 - 52128 \zeta_4 \zeta_5 \zeta_7
   - 320 \zeta_4 (\zeta_3)^4
\nonumber\\
&&\null\hskip0.5cm
   + 2718144 \zeta_3 \zeta_{13} + 3130560 \zeta_7 \zeta_9
   + 3041280 \zeta_5 \zeta_{11} \biggr] \, g^{16}
\ +\ {\cal O}(g^{18}).
\label{rho}
\eea
Previously, a different ``cosmic'' normalization factor $\rho_{\rm old}$
was used, namely eq.~(2.29) of ref.~\cite{Caron-Huot:2019vjl}.
It was determined by imposing the coaction principle
through weight 14, and a few other constraints.  It was inherently
ambiguous as to the pure $\zeta_{2L}$ terms ($\pi^{2L}$ terms),
because they have no nontrivial
terms in their coaction. Equation~(2.29) of ref.~\cite{Caron-Huot:2019vjl}
matches \eqn{rho} through seven loops, up to pure $\zeta_{2L}$ terms.
To convert from the normalizations used in
refs.~\cite{Caron-Huot:2019vjl,Caron-Huot:2019bsq} to the new,
all-orders normalization~(\ref{simplerho}),
one should multiply the old $\EE$ functions (and also the
NMHV components $E$ and $\Et$) by the factor
\bea
\frac{\rho_{\rm old}}{\rho} &=& 1 + \zeta_4 \, g^4
  - \frac{50}{3} \zeta_6 \, g^6 + \frac{2905}{12} \zeta_8 \, g^8
  - \frac{10375}{3} \zeta_{10} \, g^{10}
 + \frac{4937878055}{99504} \zeta_{12} \, g^{12}
\nonumber\\
&&\null\hskip0.0cm
 - \frac{51698725}{72} \zeta_{14} \, g^{14}\ +\ {\cal O}(g^{16}).
\label{XtoZfactor}
\eea

The precise statement of antipodal duality between the
three-particle form factor $\EE_c$ and the MHV six-particle amplitude
$\EE$ on the latter's parity-preserving surface
is\footnote{Note that $\EE$ here is called $A_6$ in
ref.~\cite{Dixon:2021tdw}, and $\EE_c$ here is called $F_3$ there.}
\begin{equation}
\label{antipodal_duality}
\EE(\uh,\vh,\wh)\Bigl|_{\Delta=0}\ =\ S\left( \EE_c(u,v,w) \right) \,.
\end{equation}
Here $S$ is the antipode map which reverses the order of letters in
every word in the symbol:
\begin{equation}
  S(x_1\otimes\dots\otimes x_m) = (-1)^m\ x_m\otimes\dots\otimes x_1\,.
\label{Sdef}
\end{equation}
The kinematic map in \eqn{antipodal_duality} can be written as
\be
\uh = \frac{vw}{(1-v)(1-w)} \,, \quad 
\vh = \frac{wu}{(1-w)(1-u)} \,, \quad
\wh = \frac{uv}{(1-u)(1-v)} \,.
\label{uvw_kin_map}
\ee
At the level of the symbol letters, going from the form-factor to
the amplitude, it acts as
\bea
a \to \hat{d}, \qquad b &\to& \hat{e}, \qquad c \to \hat{f},
\nonumber\\
d \to \sqrt{\hat{a}}, \qquad e &\to& \sqrt{\hat{b}}, \qquad
f \to \sqrt{\hat{c}}.
\label{abc_kin_map}
\eea
It is easy to verify that it maps the form-factor surface $u\,+\,v\,+\,w=1$
into the amplitude's parity-preserving surface $\Delta(\uh,\vh,\wh)=0$.
The dihedral symemtry $S_3$ of the form factor,
\eqns{cycledefF}{flipdefF}, maps into the dihedral symmetry
$D_6/Z_2 \equiv S_3$ of the amplitude, \eqns{cycledef}{flipdef}.
The antipode map can be defined beyond the symbol level, which
will play a role in checks later in the paper.  However,
mathematically it is only defined modulo $i\pi$
(see e.g.~ref.~\cite{DelDuca:2016lad}).  That is, any additive
term containing a factor of $i\pi$, including all even Riemann $\zeta$
values, which are powers of $\pi^2$, cannot be predicted at present.


\section{Flipping the symbol on \texorpdfstring{$\Delta=0$}{\Delta=0}}
\label{sec:symbolflip}

The first task in computing the eight-loop MHV six-particle
amplitude $\EE^{(8)}$ via antipodal duality is to construct its symbol
on the $\Delta=0$ surface.  The starting point is the symbol of
the eight-loop form-factor $\EE_c^{(8)}$,
which is provided in the ancillary file
{\tt Esymboct8.txt} for ref.~\cite{Dixon:2022rse}.
There are 279 linearly independent octuple final entries for the
form factor, which can be organized into 93 three-orbits under
the cyclic symmetry, and this is how the symbol is stored
in {\tt Esymboct8.txt}.  Additionally, instructions for
``integrating up'' the octuple representation to one based on
heptuples, hextuples, etc., using the multi-final-entry relations
obeyed by $\EE_c^{(8)}$, are given in the ancillary file
{\tt Esymb.txt} for ref.~\cite{Dixon:2022rse}.
See also section 4.2 of that reference for the first few multi-final
entry relations.
While the full symbol is too unwieldly, we choose to integrate it up
to the level of the triple final entries.  As mentioned in
ref.~\cite{Dixon:2022rse}, there are 12 independent final entries,
which can be organized into 4 three-orbits under
the cyclic symmetry.  Here we take as representatives
$\EE_c^{f,f,f}$, $\EE_c^{a,f,f}$, $\EE_c^{f,a,f}$ and
$\EE_c^{a,a,f}$.\footnote{For clarity,
we will drop the superscript $(8)$ for the
rest of this section, since everything is at eight loops.}
Using the alphabet ${\cal L}^a_\text{F3}$, the respective numbers of
terms in their weight 13 symbols are $n_{f,f,f} = \text{58,831,962}$,
$n_{a,f,f} = \text{95,178,164}$, $n_{f,a,f} = \text{58,826,293}$, and
$n_{a,a,f} = \text{36,362,651}$.

There is one other three-orbit of form-factor triple final entries,
which is not linearly independent, but which is useful to construct
in order to count all the terms in the symbol:
$\EE^{e,a,f} = \EE^{f,a,f} - \EE^{a,f,f}$, which has
$n_{e,a,f} = $\,58,826,293 terms in its symbol.
Using the form factor's double and triple final entry relations from
section 4.2 of ref.~\cite{Dixon:2022rse}, as well as dihedral
symmetry, we find that the number of terms in the independent
form-factor double final entries ($\EE_c^{f,f}$ and $\EE_c^{a,f}$) is
\bea
n_{f,f} &=& 2 \, n_{a,f,f} + n_{f,f,f} = \text{249,188,290},
\nonumber\\
n_{a,f} &=& n_{a,a,f} + n_{e,a,f} + n_{f,a,f} = \text{154,015,237}.
\label{FF_double_numbers}
\eea
The number of symbol terms for the one dihedrally independent
single final entry ($\EE_c^{f}$) is then
\be
n_f = 2 \, n_{a,f} + n_{f,f} = \text{557,218,764}.
\label{FF_single_numbers}
\ee
By cyclic symmetry,
the total number of terms in the form-factor symbol is just three times this,
\be
n_{\rm F3} = 3 \, n_{f} = \text{1,671,656,292}.
\label{FF_total_number}
\ee
This number will be useful as a check that we flipped the symbol properly.

On the amplitude side, we also work at the level of triple final entries.
These correspond to triple initial entries on the form-factor side.
We find that there are 62 independent triple final entries for the MHV
amplitude.  However, only 21 of them contain no $y_i$ index and
so can be obtained by flipping the form factor symbol.
(These 21 no-$y_i$ final entries
are in one-to-one correspondence with the number of independent
weight 3 functions in the form-factor space,
see Table 2 of ref.~\cite{Dixon:2022rse}.)
The 21 triple final entries
can be organized into 7 three-orbits under the amplitude cyclic symmetry.
We take as representatives
$\EE^{\ch,\ch,\dh}$, $\EE^{\bh,\dh,\dh}$, $\EE^{\dh,\ch,\dh}$, $\EE^{\eh,\ch,\dh}$,
$\EE^{\dh,\fh,\dh}$, $\EE^{\fh,\fh,\dh}$ and $\EE^{\dh,\dh,\dh}$.
Using the alphabet ${\cal L}^a_\text{hex}$, the respective numbers of
terms in their weight 13 symbols are
\bea
&&n_{\ch,\ch,\dh} = \text{20,632,545}, \quad
n_{\bh,\dh,\dh} = \text{20,627,110}, \quad
n_{\dh,\ch,\dh} = \text{33,380,402}, \quad
n_{\eh,\ch,\dh} = \text{33,380,402}, \nonumber\\
&&n_{\dh,\fh,\dh} = \text{33,340,897}, \quad
n_{\fh,\fh,\dh} = \text{33,340,897}, \quad
n_{\dh,\dh,\dh} = \text{33,338,326}.
\label{ntriplehat}
\eea
We obtained their symbols
from those of $\EE_c^{f,f,f}$, etc., by clipping off the appropriate
first three entries on the form-factor side
(e.g. $a,f,f$ for $\EE^{\ch,\ch,\dh}$), and then using the form-factor
triple, double and single final-entry relations to integrate up all
the way to the back of the form factor (front of the amplitude, after
applying the antipodal map).
It is straightforward to apply $S$ between the alphabets
${\cal L}^a_\text{F3}$ and ${\cal L}^a_\text{hex}$ because it is a simple
substitution, so that no re-factoring of the symbol is required.
(Re-factoring of such large symbols can be computationally expensive.)

There are 3 other three-orbits of amplitude triple final entries
which are not linearly independent,
but which are useful to construct in order to re-count
all the terms in the symbol:
$\EE^{\eh,\fh,\dh}$ ($n_{\eh,\fh,\dh} = \text{33,262,314}$ terms), 
$\EE^{\eh,\dh,\dh}$ ($n_{\eh,\dh,\dh} = \text{33,349,686}$ terms),
$\EE^{\bh,\fh,\dh}$ ($n_{\bh,\fh,\dh} = \text{20,625,966}$ terms).
In order to count the terms in the full symbol, we first use the
the triple final-entry enumeration to count the number of terms
in each double final entry.
Using the amplitude's double and triple final entry relations
(which include $\EE^{\ah,\fh,\dh}=0$) and dihedral symmetry, we find that
\bea
n_{\ch,\dh} &=& n_{\ch,\ch,\dh} + n_{\dh,\ch,\dh} + n_{\eh,\ch,\dh}
= \text{87,393,349},
\nonumber\\
n_{\fh,\dh} &=& n_{\bh,\fh,\dh} + n_{\dh,\fh,\dh} + n_{\eh,\fh,\dh} + n_{\fh,\fh,\dh}
= \text{120,570,074},
\nonumber\\
n_{\dh,\dh} &=& 2 \, n_{\bh,\dh,\dh}  + n_{\dh,\dh,\dh} + 2 \, n_{\eh,\dh,\dh}
= \text{141,291,918}.
\label{amp_double_numbers}
\eea
Finally we count the number of terms in the one dihedrally independent
single final entry ($\EE^{\dh}$):
\be
n_{\dh} = 2 \, n_{\ch,\dh} + n_{\dh,\dh} + 2 \, n_{\fh,\dh}  = \text{557,218,764}.
\label{amp_single_numbers}
\ee
The total number of terms in the eight-loop amplitude's
symbol on the parity-preserving surface is just three times this,
\be
n_{\Delta=0} = 3 \, n_{\dh} = \text{1,671,656,292}.
\label{amp_total_number}
\ee
This number matches the number of terms~(\ref{amp_total_number})
counted from the form-factor side,
which is a very useful cross-check that we flipped the symbol correctly.

The weight 13 triple final entries still have rather unwieldy symbols,
so the next step is to ``re-nest'' the information using the basis
for the hexagon function space $\Hhex$.  Since this basis (at least
a computationally useful one) only exists through weight 11,
we should organize the MHV amplitudes according to their
$\{11,1,1,1,1,1\}$ coproducts, or quintuple final entries.
We start with the MHV amplitude triple final entries in the bulk.
As discussed in \sect{sec:multiplefinalentry}, the number of such
entries has saturated by seven loops at 62, of which 31 are parity-even
and 31 are parity-odd.
We take the 62 MHV amplitude triple final entries, and put any of the
9 hexagon letters in front of them, in order to obtain $9\times 62$
potential quadruple final entries.
We then impose the 41 pair relations for $\Hhex$,
acting in the $4^{\rm th}$-$3^{\rm rd}$ slots from the back.
We find that there are 166 independent quadruple
final entries; 84 are parity-even and 82 are parity-odd.
48 of the 166 have no $y_i$ indices; they correspond to the 48 independent
weight 4 functions in the form-factor space --- see again
Table 2 of ref.~\cite{Dixon:2022rse}.

To get the independent quintuple final entries,
we repeat the exercise:
we put all 9 hexagon letters in front of the 166 independent quadruples
and impose the $\Hhex$ pair relations. We find 424 independent
quintuples, 211 parity-even and 213 parity-odd. Of the 424 quintuples,
108 have no $y_i$ indices, which matches the `108' in the weight 5 column
in Table 2 of ref.~\cite{Dixon:2022rse}.
We place these 108 quintuples into the (symbol-level part of)
the weight 11 hexagon function basis, using the above weight 13 symbols
$\EE^{\ch,\ch,\dh}$, $\ldots$.

The only slight catch is that we didn't write out the symbols for
all the weight 11 functions in $\Hhex$, because the table would have
been too large.  We only did so up to weight 9.  So in practice,
for each of the 108 no-$y_i$ quintuple final entries, we first generated
their weight 11 symbols by clipping the appropriate pairs of entries off
the back of the weight 13 symbols.  Then we clipped off pairs of
additional entries to get to weight 9, where we could map them into
$\Hhex$.  Then we found the unique weight 11 elements of $\Hhex$
with those weight 9 double coproducts.  These 108 weight 11 quintuple
final entries are uniquely determined in the bulk of the amplitude
kinematic space; that is because ambiguities in lifting off of the
$\Delta=0$ surface only begin at weight 12.

\section{Lifting the symbol off \texorpdfstring{$\Delta=0$}{\Delta=0}}
\label{sec:symbollift}

The next step in computing the eight-loop MHV six-particle
amplitude is to lift the symbol off of the $\Delta=0$ surface,
where all the $y_i = 1$, into the bulk three-dimensional kinematices.
To do this, we need to determine the $424-108 = 316$ quintuple final
entries that have one or more $y_i$ indices.  Of the 316 such quintuples,
$211 - 108 = 103$ are parity-even and 213 are parity odd.  The main tool
for determining them is the set of 41 pair relations for the
$6^{\rm th}$-$5^{\rm th}$ slots from the back.
One could in principle float unknowns for all 316 quintuple final entries
at once, and use the pair relations to relate them to the
coproducts of the 108 known quintuples.
At weight 11 and at symbol level, there are 1503 parity-even
and 382 parity-odd functions.  The total number of unknowns would be
$103 \times 1503 + 213 \otimes 382 = $\,236,175, which is a pretty
large linear system to solve.

In order to reduce the number of unknowns encountered at a single stage,
we solve for the missing quintuples in layers, organized by
the number and location of the $y_i$ entries, instead of solving for
all of them at once.  The 108 known quintuples have the generic form
$\EE^{e_{i_1},e_{i_2},e_{i_3},e_{i_4},e_{i_5}}$, where $e_i$ stands for any of the
6 parity-even hexagon letters.  Suppose we try to solve first for
quintuples of the form $\EE^{y_{j_1},e_{i_1},e_{i_2},e_{i_3},e_{i_4}}$,
with a $y$ in the first of the five slots.  We call these ``no-$y$ quads''.
For example, a pair relation of the form $F^{\uh,y_u} = F^{y_u,\uh}$,
applied in the $6^{\rm th}$-$5^{\rm th}$ slots from the back,
allows us to relate an even coproduct of a
$\EE^{y_{j_1},e_{i_1},e_{i_2},e_{i_3},e_{i_4}}$
to an odd coproduct of one of the known $\EE^{e_{i_1},e_{i_2},e_{i_3},e_{i_4},e_{i_5}}$.
After combining the no-$y$ quads with the known 108 quintuples,
and applying all of the quintuple relations, we find that there
are 99 more independent quintuples. We can determine them one at
a time using $6^{\rm th}$-$5^{\rm th}$ slot pair relations,
so we never need more than 1503 unknowns, making the relations
easy to solve.  The no-$y$ quads have weight 12, so we can expect
some unknown constants to crop up, which are related to the weight 12
function $Z$ discussed in the introduction.  Using dihedral symmetry,
we found just 11 constants in all, at quad level.  (These constant parameters
multiply weight 12 symbols, and are unrelated to zeta-valued constants.)

Next we add quintuples that come from the ``no-$y$ triples'',
i.e.~of the form $\EE^{e_{i_0},y_{j_1},e_{i_1},e_{i_2},e_{i_3}}$ or
$\EE^{y_{j_0},y_{j_1},e_{i_1},e_{i_2},e_{i_3}}$.
There are 69 additional quintuples of this type, and they are associated
with 21 no-$y$ triples.  Determining them is also not a major
computation.  When we solve for the weight 13 triples, we also determine
all but 2 of the 11 constants that were required at quad level,
and no additional triple-level constants have to be introduced.
So far we have fixed $108 + 99 + 69 = 276$ of the 424 quintuples, up to
these 2 free parameters.

Next we solve for the ``no-$y$ doubles'', which correspond to
54 additional quintuples, or $276 + 54 = 330$ in all.
Up to cyclic permutations, their last two entries are
$\{\ch,\dh\}$, $\{\dh,\dh\}$, and $\{\fh,\dh\}$.
We have to introduce $1 \times 1503 + 3 \times 382 = 2649$ unknowns
for 4 different quintuples associated with $\{\fh,\dh\}$,
$4 \times 1503 + 10 \times 382 = 9832$ unknowns for $\{\dh,\dh\}$,
and $4 \times 1503 + 9 \times 382 = 9450$ unknowns for $\{\fh,\dh\}$.
These equations could all be solved over the rational numbers,
up to just 3 undetermined constants.

Similarly, the ``no-$y$ singles'' correspond to 24 additional quintuples,
or 354 in all. Up to cyclic permutations, their last entry is $\dh$,
and we have to introduce $9 \times 1503 + 6 \times 382 = $\,15,819 unknown
parameters, which is still feasible to solve over the rationals.
The combined solution, for everything but the $y_i$ final entries,
has {\it no} undetermined constants at symbol level.

The final step in lifting the symbol to the bulk is to determine
the additional quintuples associated with an odd final entry, which
by cyclic symmetry can be taken to be $y_u$.
There are still $424 - 354 = 70$ such quintuples left to fix;
28 are parity-even and 42 are parity-odd.
The corresponding number of unknown parameters is
$28 \times 1503 + 42 \times 382 = $\,58,128.
This linear system is large enough that we had to first solve it over
prime number fields $\mathbb{Z}_p$ for a few different primes $p$,
and then reconstruct the rational number solution,
along the lines of e.g.~refs.~\cite{vonManteuffel:2014ixa,Peraro:2016wsq}.
We obtained a three-parameter, dihedrally symmetric, solution space.
(Later we relaxed the condition of dihedral symmetry on the homogeneous
part of the solution, i.e.~functions $F$ satisfying $F^{e_i}=0$ for
all parity-even letters $e_i$, and we found 9 unfixed parameters,
corresponding to a 9-dimensional space of ambiguities for lifting
weight 16 symbols off of the surface $\Delta=0$.)

Next we evaluated the 3-parameter solution (still over the primes)
at the kinematical origin, $(\uh,\vh,\wh) \to (0,0,0)$.
The remainder function $\RR$ in \eqn{EZMHVtoR6simplerho} is quadratic
in $\ln \uh_i$ at the origin, with zeta-valued
coefficients~\cite{Basso:2020xts}; thus its symbol vanishes there,
up to power-law corrections.
Also, at symbol level $\rho \to 1$ and $\Gcusp \to 4 g^2$.
Hence we require
\be
\EE \to \exp\Bigl[ g^2 \EE^{(1)} \Bigr]
        + {\cal O}(\uh^1,\vh^1,\wh^1)\,,
\label{EZMHVorigin}
\ee
as $(\uh,\vh,\wh) \to (0,0,0)$.
This condition fixes 2 of the 3 parameters, but one linear combination
vanishes (up to powers) at the origin, and so it cannot be fixed there.

At this point, we reconstructed the one-parameter symbol-level solution
for the 58,128 coefficients of the weight 11 basis functions
in terms of rational numbers.
The rational reconstruction is not too difficult
because, for the coefficient of the one-parameter ambiguity,
the denominator is $2^{26}$ (or a factor thereof)
for all but 766 of the 58,128 rational numbers.
For the coefficient of the inhomogeneous part, the denominator is
$3^2 \times 2^{26}$ (or a factor thereof) for all but 851 of the 58,128
rational numbers.  Furthermore, the rational numbers with ``bad''
denominators are all associated with just 33 of the 1503 weight 11
parity-even basis functions. We know in advance which 33 basis functions
they are, by inspecting previous seven-loop results expressed in the same basis.
So we can multiply through by $3^2 \times 2^{26}$, require the coefficients
to be integers, and reconstruct the rational coefficients for the 1470 other
parity-even basis functions, and all 382 parity-odd basis functions.
The remaining coefficients are few enough in number that we could solve
for them directly over the rational numbers, since we generated
all the equations over the rational numbers.

\section{The full function}
\label{sec:fullfunction}

With the symbol fixed up to one parameter, we proceed to fix all
the zeta-valued constants multiplying lower-weight functions,
i.e.~the ``beyond-the-symbol'' terms.
In the first stage, we do so at the level of the
weight 11 quintuple final entries.

\subsection{At quintuple level}
\label{sec:fullfunctionquintuples}

There are 215 parity-even and 36 parity-odd beyond-the-symbol functions
in $\Hhex$ at weight 11.  Since there are 211 parity-even and
213 parity-odd quintuples, there are
$211 \times 215 + 213 \times 36 = $\,53,033 parameters to determine.
This can be done in one stage, but it again requires using prime fields.
The workhorse for constraining these parameters is again the pair relations
in the  $6^{\rm th}$-$5^{\rm th}$ slots from the back. We also impose dihedral
symmetry.  The solution to the pair and dihedral constraints has 65 parameters.
Next we impose the vanishing of the remainder function $\RR$ in the
strict collinear limits, which leaves 17 parameters.

At this point, we return to the origin.  The value of the eight-loop
remainder function at the origin is~\cite{Basso:2020xts}
\be
\RR^{(8)} =
    c_1^{(8)} \sum_{i=1}^3 ( \ln^2 \uh_i + \ln \uh_i \ln \uh_{i+1} )
  + c_2^{(8)} \sum_{i=1}^3 \ln \uh_i \ln \uh_{i+1}
  + c_0^{(8)} + {\cal O}(\uh,\vh,\wh)\,,
\label{R8origin}
\ee
where
\bea
c_1^{(8)} &=& \frac{3622824549}{1280} \zeta_{14}
- \frac{69069}{4} \zeta_8 (\zeta_3)^2 - 33948 \zeta_6 \zeta_3 \zeta_5
- 34650 \zeta_4 \zeta_3 \zeta_7 - 16974 \zeta_4 (\zeta_5)^2
\nonumber\\
&&\null\hskip0.0cm
- 28224 \zeta_2 \zeta_3 \zeta_9 - 26208 \zeta_2 \zeta_5 \zeta_7
- 156 \zeta_2 (\zeta_3)^4
\nonumber\\
&&\null\hskip0.0cm
- 55440 \zeta_3 \zeta_{11} - 49728 \zeta_5 \zeta_9 - 23820 (\zeta_7)^2
- 800 (\zeta_3)^3 \zeta_5
\,, \label{R8originc1}\\
c_2^{(8)} &=& \frac{3730480571}{640} \zeta_{14}
+ \frac{29821}{2} \zeta_8 (\zeta_3)^2 + 28344 \zeta_6 \zeta_3 \zeta_5
+ 27300 \zeta_4 \zeta_3 \zeta_7 + 13356 \zeta_4 (\zeta_5)^2
\nonumber\\
&&\null\hskip0.0cm
+ 18816 \zeta_2 \zeta_3 \zeta_9 + 17472 \zeta_2 \zeta_5 \zeta_7
+ 120 \zeta_2 (\zeta_3)^4
+ 320 (\zeta_3)^3 \zeta_5
\,, \label{R8originc2}\\
c_0^{(8)} &=& \frac{157718118308821}{3703808} \zeta_{16}
- \frac{2180163}{16} \zeta_{10} (\zeta_3)^2
- 299677 \zeta_8 \zeta_3 \zeta_5
- \frac{688875}{2} \zeta_6 \zeta_3 \zeta_7
\nonumber\\
&&\null\hskip0.0cm
- \frac{337713}{2} \zeta_6 (\zeta_5)^2
- 441840 \zeta_4 \zeta_3 \zeta_9 - 406320 \zeta_4 \zeta_5 \zeta_7
- 2061 \zeta_4 (\zeta_3)^4
- 498960 \zeta_2 \zeta_3 \zeta_{11}
\nonumber\\
&&\null\hskip0.0cm
- 447552 \zeta_2 \zeta_5 \zeta_9
- 214380 \zeta_2 (\zeta_7)^2 - 6240 \zeta_2 (\zeta_3)^3 \zeta_5
\nonumber\\
&&\null\hskip0.0cm
- 679536 \zeta_3 \zeta_{13} - 760320 \zeta_5 \zeta_{11}
- 782640 \zeta_7 \zeta_9 - 8792 (\zeta_3)^2 (\zeta_5)^2
- 5880 (\zeta_3)^3 \zeta_7 
\,. \label{R8originc0}
\eea
We compute $\EE$ from this value for $\RR$, and the lower-loop values
given in ref.~\cite{Caron-Huot:2019vjl},
using \eqn{EZMHVtoR6simplerho}.
We then take its quintuple coproducts near the origin,
and compare them with those of our 17-parameter ansatz.
We find that all but 2 of the parameters are fixed.
Note that the precise values of the eight-loop coefficients $c_i^{(8)}$
in \eqn{R8origin} do not matter yet, because we are still performing
the analysis at weight 11, and the quintuple coproducts annihilate
anything not containing at least a weight 5 function,
which includes all of \eqn{R8origin}.
On the other hand, products of lower-loop terms from expanding the
exponential in \eqn{EZMHVtoR6simplerho}
can contribute at weight 11, because they can result in more than
four logarithms.

We fixed one of the two remaining beyond-the-symbol parameters
by observing that in the symbol-level solution, there were not actually
424 linearly-independent quintuple final entries, but only 384,
of which 199 are parity-even and 185 are parity-odd.
We required there to be only the same 384 quintuple final entries
at function level as well.

The final beyond-the-symbol parameter at the level of quintuples
multiplies a function which is
$\zeta_8$ times a weight 8 function. The weight 8 function
is dihedrally symmetric, vanishes in the strict collinear limit
and at the origin --- modulo terms with at least weight 4 in zeta values,
which vanish at the level of quintuple coproducts.
However, it is non-vanishing in the near-collinear limit, at the level
of one flux-tube excitation.
This limit can be accessed easily from the $(1,\vh,\vh)$ line
as $\vh\to0$, where the function has a nonvanishing $\vh^1 \ln^4 \vh$
behavior.  Thus we use a single term in the flux-tube OPE prediction
to fix the final quintuple-level parameter.  We provide the final set
of 384 linearly independent quintuple final entries in an ancillary file,
{\tt MHV8quintuples.txt}, along with a routine for constructing all
the other quintuples via the various linear relations.

\subsection{Constants beyond weight 11}
\label{sec:beyondweight11}

There are additional constants of integration above weight 11, which we
need to determine in order to completely specify the amplitude.
We specify the constants at the point $(1,1,1)$,
but we impose constraints determining them at the point $(1,0,0)$
--- strict collinear vanishing of $\RR$ and the one flux-tube excitation
OPE information --- as well as consistency between the $(1,\vh,\vh)$
and $(\uh,0,0)$ lines where they intersect.

At weight 12, we have 166 linearly independent quadruple final entries
to determine at the point $(1,1,1)$.
However, the 82 parity-odd quadruples all vanish at $(1,1,1)$,
since this point lies on the parity-preserving surface $\Delta=0$.
Also, using the branch-cut condition that $F^{1-\uh}=0$ for $\uh\to1$,
for any function $F$, we know that
\be
\EE^{\dh,x,y,z}(1,1,1) = \EE^{\dh,x,y,z}(1,1,1) = \EE^{\dh,x,y,z}(1,1,1) = 0,
\label{dhehfh111vanish}
\ee
for any letters $x,y,z$.  Combining this information with dihedral symmetry,
we find only 10 independent weight 12 quadruple constants at $(1,1,1)$.
Repeating the same exercise for the weight 13 triple constants yields
only 3 independent ones.  Similarly, there are
2 independent weight 14 double constants,
no nonvanishing weight 15 single final entries,
and one weight 16 constant value at $(1,1,1)$.
Given these $10+3+2+0+1=16$ constants,
we can integrate up the full function on
the line $(1,\vh,\vh)$, and then take $\vh\to0$ to access the soft endpoint
of the strict collinear limit (at leading power in $\vh$, namely $\vh^0$)
and a part of the one flux-tube excitation OPE information (at order $\vh^1$).
The result on the line $(1,\vh,\vh)$ can be expressed in terms
of HPLs~\cite{Remiddi:1999ew}
of the form $H_{\vec{a}}(\vh)$ with $a_i \in \{0,1\}$.
In carrying out this step, it is useful to be able to
evaluate rather high-weight MZVs; we use the program
{\sc HyperlogProcedures}~\cite{HyperlogProcedures} for this purpose.
Of the 16 constants, only 10 actually appear in $\EE^{(8)}(1,\vh,\vh)$.
Four of these are fixed by the strict collinear limit, and
three more are fixed by the OPE information at order $\vh^1$.

Next we integrate up along the collinear limit line, not just
the soft endpoint.  The results are HPLs, $H_{\vec{a}}(\uh=1-\wh)$
with $a_i \in \{0,1\}$. However, we find only one more constraint,
leaving 8 constants still to be fixed.

Finally, we construct the full function on the line $(\uh,0,0)$.
This means that $\uh$ has a generic value, while $\vh,\wh \ll 1$.
The results are HPLs $H_{\vec{a}}(\uh)$, $a_i \in \{0,1\}$, multiplied
by polynomials in $\ln \vh$ and $\ln \wh$.
There are consistency conditions from matching the results on the $(1,\vh,\vh)$
line, as $\vh\to0$, to the results on the $(\uh,0,0)$ line, as $\uh\to1$.
These consistency conditions suffice to fix all 8 of the
undetermined constants.  We then take the limit $\uh\to0$, to obtain
the value of $\EE^{(8)}$ at the origin.  It agrees perfectly with
the prediction~(\ref{R8origin}).

We provide all the values of the weight 12, 13, 14, and 16 constants
in the ancillary file {\tt EZMHVcoproducts111.txt},
along with some lower-loop values (at weight 12 and above)
and the $\rho$ factor from \eqn{rho}.


\section{Checks}
\label{sec:checks}

Besides checking the behavior at the origin, as just mentioned,
we checked several other limits where the amplitude's behavior
is well understood:
\begin{enumerate}
\item The self-crossing limit. 
\item Multiparticle factorization limit.
\item Near-collinear (OPE) limit.
\item Multi-Regge kinematics (MRK).
\end{enumerate}
In the remainder of this section we briefly describe these limits
and checks.
  
\subsection{Self-crossing limit}
\label{subsec:selfcross}

There is a limit of massless $2\to4$ scattering
that mimics double parton scattering,
in that the two incoming partons can each split into (almost on-shell)
pairs of partons, and then two separate $2\to2$ scatterings take place.
There is an analogous limit of $3\to3$ scattering, where one of the
three incoming partons splits, and one of the three outgoing partons
is a fusion of two almost on-shell outgoing partons.
These kinematic limits have an interpretation in the dual Wilson hexagon
as limits where two opposite sides of the hexagon almost cross
each other~\cite{Dixon:2016epj}.  (For earlier studies of the self-crossing
limit, see refs.~\cite{Georgiou:2009mp,Dorn:2011gf,Dorn:2011ec}.)
Logarithmic singularities are generated due to the exchange of virtual
gluons between the two nearby sides.
In ref.~\cite{Caron-Huot:2019vjl}, an all-orders formula for the singular
terms in this limit was presented, and it was checked against
perturbative results through seven loops.  Here we will check
it at eight loops.

The singular terms have a simpler structure in the $3\to3$ case than
in the $2\to4$ case because the hexagonal Wilson loop is quasi-Euclidean,
with sides alternating between incoming and outgoing. In self-crossing
kinematics, the cross ratios
$(\uh,\vh,\wh)$ approach $(1-\delta,\vh,\vh)$ with $\delta \to 0$,
after analytic continuation onto the correct sheet. Here $\delta$ is
a dual-conformally-invariant measure of the separation of the two sides
that are almost crossing.
In the $3\to3$ case, the analytic continuation is $\uh\to e^{+2\pi i}\uh$,
$\vh\to e^{+\pi i}\vh$, $\wh\to e^{+\pi i}\wh$. Also, $\delta \to 0$ from
the negative side, and $\vh$ is either negative
or greater than one~\cite{Dixon:2016epj}.
All the logarithmic singularities as $|\delta| \to 0$ appear
in the imaginary part of the amplitude, and they are independent of $\vh$.
The all-orders formula is~\cite{Caron-Huot:2019vjl}:
\bea
\frac{1}{2\pi i} \frac{d \EE_{3\to3}}{d\lnden} &=&
\frac{g^2}{\rho} \exp\Bigl[ \tfrac{1}{2} \zeta_2 \Gcusp + 2 \Gamma_3 \Bigr]
\nn\\
&&\hskip0.0cm\null
\times 2 \int_0^\infty d\nu J_1(2\nu) \exp\Bigl[
 - \frac{\Gcusp}{4} [\lambda(\nu)]^2 - \Gvirt \lambda(\nu) \Bigr]\,,
\label{gSimon}
\eea
where $J_1$ is the first Bessel function and
\be
\lambda(\nu) = 2 ( \ln\nu + \gamma_E ) - \lnden,
\label{lambdadef}
\ee
with $\gamma_E$ the Euler-Mascheroni constant.  The anomalous dimensions
$\Gamma_3$ and $\Gvirt$ are given in ref.~\cite{Caron-Huot:2019vjl}.

The perturbative expansions of \eqn{gSimon} were provided in
ref.~\cite{Caron-Huot:2019vjl} through seven loops (for the earlier
value of $\rho$).  Here we give the eight loop value, for $\rho$
in \eqn{rho}:
\bea
\frac{1}{2\pi i} \frac{d \EE_{3\to3}^{(8)}}{d\lnden} &=&
- \frac{1}{5040} \lndene{14} - \frac{\z2}{72} \lndene{12}
- \frac{17}{90} \z3 \lndene{11} - \frac{163}{120} \z4 \lndene{10} 
\nn\\ &&\null\hskip0cm
- \Bigl( \frac{236}{15} \z5 + \frac{77}{9} \z2 \z3 \Bigr) \lndene{9}
- \Bigl( \frac{2963}{48} \z6 + \frac{401}{9} (\z3)^2 \Bigr) \lndene{8}
\nn\\ &&\null\hskip0cm
- \Bigl( \frac{18142}{21} \z7 + \frac{1384}{3} \z2 \z5
       + \frac{1378}{3} \z4 \z3 \Bigr) \lndene{7}
\nn\\ &&\null\hskip0cm
- \Bigl( \frac{130889}{72} \z8 + \frac{56132}{15} \z3 \z5
       + \frac{9452}{9} \z2 (\z3)^2 \Bigr) \lndene{6}
\nn\\ &&\null\hskip0cm
- \Bigl( 28504 \z9 + 14860 \z2 \z7 + \frac{71672}{5} \z4 \z5
       + \frac{28507}{3} \z6 \z3 + 2432 (\z3)^3 \Bigr) \lndene{5}
\nn\\ &&\null\hskip0cm
- \Bigl( \frac{2725139}{80} \z{10} + \frac{260096}{3} \z3 \z7
       + 42920 (\z5)^2 + \frac{138584}{3} \z2 \z3 \z5
\nn\\ &&\null\hskip0.5cm
       + \frac{71152}{3} \z4 (\z3)^2 \Bigr) \lndene{4}
\nn\\ &&\null\hskip0cm
- \Bigl( 446544 \z{11} + 241696 \z2 \z9 + 218396 \z4 \z7 + 141348 \z6 \z5
       + \frac{842009}{9} \z8 \z3
\nn\\ &&\null\hskip0.5cm
       + \frac{318304}{3} (\z3)^2 \z5
       + \frac{64096}{3} \z2 (\z3)^3 \Bigr) \lndene{3}
\nn\\ &&\null\hskip0cm
- \Bigl( \frac{3772896325}{11056} \z{12} + 867552 \z3 \z9
       + 827936 \z5 \z7 + 412160 \z2 \z3 \z7 + 201696 \z2 (\z5)^2
\nn\\ &&\null\hskip0.5cm
       + 417272 \z4 \z3 \z5 + 138546 \z6 (\z3)^2 + 18848 (\z3)^4
       \Bigr) \lndene{2}
\nn\\ &&\null\hskip0cm
- \Bigl( 1755072 \z{13} + 1320480 \z2 \z{11} + 1048656 \z4 \z9
       + 652370 \z6 \z7 + \frac{1258672}{3} \z8 \z5
\nn\\ &&\null\hskip0.5cm
       + \frac{3033977}{10} \z{10} \z3 + 459744 (\z3)^2 \z7
       + 448960 \z3 (\z5)^2 +285632 \z2 (\z3)^2 \z5
\nn\\ &&\null\hskip0.5cm
       + 96800 \z4 (\z3)^3 \Bigr) \lnden
\nn\\ &&\null\hskip0cm
- \frac{622062547}{672} \z{14} - 1827072 \z3 \z{11} - 1724160 \z5 \z9
- 835360 (\z7)^2 - 565824 \z2 \z3 \z9
\nn\\ &&\null\hskip0cm
- 519232 \z2 \z5 \z7 - 633920 \z4 \z3 \z7
- 309616 \z4 (\z5)^2 - 405020 \z6 \z3 \z5
\nn\\ &&\null\hskip0cm
- \frac{398330}{3} \z8 (\z3)^2
- 115904 (\z3)^3 \z5 - 16000 \z2 (\z3)^4 \,.
\label{dEdlnde8}
\eea
It matches perfectly the result found directly from the
eight-loop amplitude.

\subsection{Factorization limit}
\label{subsec:fact}

Amplitudes generically have universal factorizing behavior near
multi-particle poles.  Using integrability in planar ${\cal N}=4$ SYM,
an all-orders formula has been found for this factorization limit
for the NMHV amplitude, which contains a pole at tree level.
The limit takes two of the three cross-ratios large, and the other
one ``small'', but in practice it can be taken to be of order 1.
The simplest way to take this limit is via the line $(\uh,1,\uh)$
by taking $\uh\to\infty$.
This NMHV limit was checked through 4 loops in ref.~\cite{Dixon:2015iva},
and it holds to 7 loops as well~\cite{DDToAppear}.

It was also realized~\cite{CaronHuotprivate} that the MHV
amplitude has a similar factorization behavior, even though
it does not have a tree-level pole; instead one should take a
kind of discontinuity of the limiting behavior.  More precisely,
it was found that the MHV and NMHV behavior is related by
\be
D^{(L)}(z)
\equiv \frac{z}{2} \frac{d^2\EE^{(L)}(1,1/z,1/z)}{dz^2} \biggl|_{z\to0}
= E^{(L-1)}(1/z,1,1/z) \biggl|_{z\to0} \ +\ {\cal O}(z)\,,
\label{Simon_fact}
\ee
where $E^{(L-1)}(\uh,\vh,\wh)$ is the parity-even component of the
NMHV amplitude at one-loop order lower, and we keep only the leading
power terms in the equation as $z\to0$.  Due to the final entry condition,
there are no logarithmic terms at leading power in
$\EE^{(L)}(1,1/z,1/z)$ as $z\to0$.  There is a constant, but it is
removed by the derivatives in \eqn{Simon_fact}.  The contributions to
\eqn{Simon_fact} come from terms of the form $z \ln^k z$ in
$\EE^{(L)}(1,1/z,1/z)$ with $k>0$ only, hence the relation to a discontinuity.
Note that the same definition of $\rho$ should be used for $\EE^{(L)}$
and $E^{(L-1)}$ in \eqn{Simon_fact}.

Here we give the values of $D^{(L)}(z)$ computed from the
MHV amplitude through eight loops:
\bea
D^{(1)}(z) &=& 1,
\label{D1z}\\
D^{(2)}(z) &=&  - 2 L^2 - 2 \z2 \,,
\label{D2z}\\
D^{(3)}(z) &=& 2 L^4 + 16 \z2 L^2 + 4 \z3 L + \frac{83}{2} \z4 \,,
\label{D3z}\\
D^{(4)}(z) &=&  - \frac{4}{3} L^6 - 28 \z2 L^4 - \frac{88}{3} \z3 L^3
 - 443 \z4 L^2 - (112 \z5 + 136 \z2 \z3) L
\nn\\ &&\null\hskip0cm
 - \frac{3177}{4} \z6 - 32 (\z3)^2 \,,
\label{D4z}\\
D^{(5)}(z) &=& \frac{2}{3} L^8 + \frac{80}{3} \z2 L^6 + \frac{152}{3} \z3 L^5
 + 1047 \z4 L^4 + \Bigl( 736 \z5 + \frac{2240}{3} \z2 \z3 \Bigr) L^3
\nn\\ &&\null\hskip0cm
 + ( 11048 \z6 + 408 (\z3)^2 ) L^2
 + ( 3140 \z7 + 3280 \z2 \z5 + 5934 \z4 \z3 ) L
\nn\\ &&\null\hskip0cm
 + \frac{916603}{48} \z8 + 1316 \z3 \z5 + 596 \z2 (\z3)^2 \,,
\label{D5z}
\eea
\bea
D^{(6)}(z) &=& - \frac{4}{15} L^{10} - \frac{52}{3} \z2 L^8 - 48 \z3 L^7
 - \frac{3790}{3} \z4 L^6
 - \Bigl( \frac{7264}{5} \z5 + \frac{4432}{3} \z2 \z3 \Bigr) L^5
\nn\\ &&\null\hskip0cm
 - \Bigl( \frac{65819}{2} \z6 + \frac{4048}{3} (\z3)^2 \Bigr) L^4
 - \Bigl( 21640 \z7 + 21216 \z2 \z5 + \frac{115924}{3} \z4 \z3 \Bigr) L^3
\nn\\ &&\null\hskip0cm
 - \Bigl( \frac{8062091}{24} \z8 + 23112 \z3 \z5 + 11928 \z2 (\z3)^2 \Bigr) L^2
\nn\\ &&\null\hskip0cm
 - ( 100464 \z9 + 95080 \z2 \z7 + 166328 \z4 \z5
    + 201269 \z6 \z3 + 2912 (\z3)^3 ) L
\nn\\ &&\null\hskip0cm
 - \frac{94215313}{160} \z{10} - 35792 \z3 \z7 - 17376 (\z5)^2
 - 33800 \z2 \z3 \z5 - 31268 \z4 (\z3)^2 \,,
\label{D6z}
\eea
\bea
D^{(7)}(z) &=& \frac{4}{45} L^{12} + \frac{128}{15} \z2 L^{10}
 + \frac{280}{9} \z3 L^9 + \frac{2987}{3} \z4 L^8
 + \Bigl( \frac{23744}{15} \z5 + \frac{4928}{3} \z2 \z3 \Bigr) L^7
\nn\\ &&\null\hskip0cm
 + \Bigl( \frac{144206}{3} \z6 + \frac{19376}{9} (\z3)^2 \Bigr) L^6
 + \Bigl( 49288 \z7 + \frac{242848}{5} \z2 \z5
    + \frac{269060}{3} \z4 \z3 \Bigr) L^5
\nn\\ &&\null\hskip0cm
 + \Bigl( \frac{9748865}{8} \z8 + \frac{270040}{3} \z3 \z5
    + \frac{142840}{3} \z2 (\z3)^2 \Bigr) L^4
\nn\\ &&\null\hskip0cm
 + \Bigl( 774368 \z9 + 719360 \z2 \z7 + 1261072 \z4 \z5
        + \frac{4638080}{3} \z6 \z3 + \frac{72224}{3} (\z3)^3 \Bigr) L^3
\nn\\ &&\null\hskip0cm
 + \Bigl( \frac{50905899}{4} \z{10} + 798224 \z3 \z7 + 380576 (\z5)^2
    + 793472 \z2 \z3 \z5 + 744132 \z4 (\z3)^2 \Bigr) L^2
\nn\\ &&\null\hskip0cm
 + \Bigl( 3793104 \z{11} + 3393600 \z2 \z9 + 5602166 \z4 \z7 + 6522082 \z6 \z5
    + \frac{31319745}{4} \z8 \z3
\nn\\ &&\null\hskip0.5cm
    + 297552 (\z3)^2 \z5 + 107088 \z2 (\z3)^3
     \Bigr) L
\nn\\ &&\null\hskip0cm
 + \frac{2017876455195}{88448} \z{12} + 1269296 \z3 \z9 + 1144752 \z5 \z7
 + 1170064 \z2 \z3 \z7
\nn\\ &&\null\hskip0cm
 + 556304 \z2 (\z5)^2 + 2064582 \z4 \z3 \z5
 + \frac{2565617}{2} \z6 (\z3)^2
 + 10352 (\z3)^4 \,,
\label{D7z}
\eea
\bea
D^{(8)}(z) &=& - \frac{8}{315} L^{14} - \frac{152}{45} \z2 L^{12}
 - \frac{688}{45} \z3 L^{11} - \frac{2882}{5} \z4 L^{10}
 - \Bigl( \frac{17504}{15} \z5 + \frac{11120}{9} \z2 \z3 \Bigr) L^9
\nn\\ &&\null\hskip0cm
 - \Bigl( \frac{89039}{2} \z6 + \frac{19168}{9} (\z3)^2 \Bigr) L^8
 - \Bigl( \frac{1298288}{21} \z7 + \frac{923072}{15} \z2 \z5
        + 115208 \z4 \z3 \Bigr) L^7
\nn\\ &&\null\hskip0cm
 - \Bigl( \frac{25200161}{12} \z8 + \frac{818416}{5} \z3 \z5
   + \frac{791824}{9} \z2 (\z3)^2 \Bigr) L^6
\nn\\ &&\null\hskip0cm
 - \Bigl( 2035936 \z9 + 1899824 \z2 \z7 + \frac{16717616}{5} \z4 \z5
 + 4152190 \z6 \z3 + 68096 (\z3)^3 \Bigr) L^5
\nn\\ &&\null\hskip0cm
 - \Bigl( \frac{4378007151}{80} \z{10} + \frac{10796672}{3} \z3 \z7
 + 1697152 (\z5)^2 + \frac{10838224}{3} \z2 \z3 \z5
\nn\\ &&\null\hskip0.5cm
 + 3423104 \z4 (\z3)^2 \Bigr) L^4
\nn\\ &&\null\hskip0cm
 - \Bigl( 33391008 \z{11} + 29783488 \z2 \z9 + 49137612 \z4 \z7
 + 57381324 \z6 \z5
\nn\\ &&\null\hskip0.5cm
 + \frac{419455771}{6} \z8 \z3
  + \frac{8332640}{3} (\z3)^2 \z5 + \frac{3025184}{3} \z2 (\z3)^3 \Bigr) L^3
\nn\\ &&\null\hskip0cm
 - \Bigl( \frac{25976759893203}{44224} \z{12}
  + 33830944 \z3 \z9 + 29944736 \z5 \z7
  + 31742912 \z2 \z3 \z7
\nn\\ &&\null\hskip0.5cm
  + 14939104 \z2 (\z5)^2 + 56171500 \z4 \z3 \z5
  + 35249431 \z6 (\z3)^2 + 295392 (\z3)^4 \Bigr) L^2
\nn\\ &&\null\hskip0cm
 - \Bigl( 169405632 \z{13} + 146577312 \z2 \z{11} + 231088872 \z4 \z9
 + 253004589 \z6 \z7
\nn\\ &&\null\hskip0.5cm
 + 289917579 \z8 \z5 + \frac{14566638681}{40} \z{10} \z3
 + 12164928 (\z3)^2 \z7 + 11391680 \z3 (\z5)^2
\nn\\ &&\null\hskip0.5cm
  + 12311648 \z2 (\z3)^2 \z5 + 7857888 \z4 (\z3)^3 \Bigr) L
\nn\\ &&\null\hskip0cm
 - \frac{1374127004947}{1280} \z{14} - 55664064 \z3 \z{11}
 - 47148096 \z5 \z9 - 22132160 (\z7)^2
\nn\\ &&\null\hskip0cm
 - 49702240 \z2 \z3 \z9 - 43868896 \z2 \z5 \z7
 - 82235944 \z4 \z3 \z7 - 38672640 \z4 (\z5)^2
\nn\\ &&\null\hskip0cm
 - 96406097 \z6 \z3 \z5 - \frac{177982387}{3} \z8 (\z3)^2
 - 1596160 (\z3)^3 \z5 - 438880 \z2 (\z3)^4
 \,,
\label{D8z}
\eea
where $L = \ln z$.

These results agree with the limit~(\ref{Simon_fact}) of the NMHV amplitude
through six loops as given in ref.~\cite{Caron-Huot:2019vjl}, after
taking into account the different choice of $\rho$ via \eqn{XtoZfactor}.
The eight-loop MHV result agrees with the seven-loop NMHV result
found in ref.~\cite{DDToAppear}.  Notice the
strict sign alternation for all terms in $D^{(L)}(z)$
through eight loops, and also in \eqn{dEdlnde8}.

\subsection{Near-collinear (OPE) limit}
\label{subsec:ope}

Another powerful check of the eight-loop MHV amplitude is provided by
its behavior in the near-collinear limit in the Euclidean region.
This behavior is governed, to any order in the coupling, by the
Wilson loop (or Pentagon) Operator Product
Expansion~\cite{Alday:2010ku,Basso:2013vsa,Basso:2013aha,Basso:2014koa,%
Basso:2014jfa,Basso:2014nra,Belitsky:2014sla,Belitsky:2014lta,Basso:2014hfa,%
Belitsky:2015efa,Basso:2015rta,Basso:2015uxa,Belitsky:2016vyq}.
The limit is usually described by the variables
$T=e^{-\tau}$, $S=e^{\sigma}$, and $F=e^{i\phi}$, where $\vh=T^2/(1+T^2)$,
and it is similar to the collinear limit $\vh\to0$, $\uh+\wh\to1$,
but now keeping power-suppressed terms in $T$ (or $v$).
There is a straightforward recipe for computing the first couple of terms
in the $T$ expansion, which was carried out through seven loops
for MHV in ref.~\cite{Caron-Huot:2019vjl}, so we will not repeat it here.

In the ancillary file {\tt RLncy.txt}, we provide the $T^1$ terms in
the near-collinear limit of $\RR^{(L)}$ through eight loops.
They have the form
$\RR^{(L)} \sim T (F+1/F) f^{(L)}(S)$, where $S^2 = (1-u)/u \equiv y$.
We write $f(S)$ in terms of iterated integrals
Iy$(\vec{w}) \equiv G_{\vec{w}}(y)$.
The file is large because there are tens of thousands of such iterated
integrals in the eight loop expression.  The OPE recipe at order $T^1$
involves a sum over residues in the rapidity of the single flux-tube
excitation, which can be difficult to resum exactly.  Instead we
can expand the expressions in {\tt RLncy.txt} around $S=0$, which
corresponds to truncating the residue sum.  We provide the (much shorter)
series expansions through $S^{41}$ and through eight loops in the
ancillary file {\tt RLncyser.txt}.

They agree perfectly with all the OPE predictions we have computed.
At eight loops, we evaluated the full residue sum out to $S^{11}$,
and the $\ln^k T$ terms with $k \geq 4$ out to $S^{41}$.

\subsection{Multi-Regge kinematics}
\label{sec:MRK}

The limit of high-energy $2\to4$ scattering with large rapidity
separation between the four outgoing partons is referred to
as multi-Regge kinematics (MRK).  There is an analogous limit
of $3\to3$ scattering which is slightly simpler
(as in the self-crossing case).
In this limit the Fourier-Mellin transform of the
amplitude factorizes~\cite{Bartels:2008ce,Bartels:2009vkz,Fadin:2011we}.
The all-orders behavior of the BFKL eigenvalue and impact factor
that enter the factorization formula is now understood to all orders via
integrability and analytic continuation from the near-collinear
limit~\cite{Basso:2014pla}. 

To take the MRK limit in $2\to4$ scattering kinematics,
the cross ratio $\uh$ is first analytically continued out of the Euclidean
region, $\uh \to \uh e^{-2\pi i}$, and then we send $\uh\to1$
while $\vh,\wh\to0$, holding fixed the ratios
\be
\frac{\vh}{1-\uh} \equiv \frac{1}{|1-z|^2} \,, \qquad
\frac{\wh}{1-\uh} \equiv \frac{|z|^2}{|1-z|^2} \,.
\label{zzbardef}
\ee
In this limit the parity-odd variables become
\be
y_u = 1,\qquad
y_v = \frac{1-\bar{z}}{1-z} \,, \qquad
y_w = \frac{(1-z)\bar{z}}{(1-\bar{z})z} \,.
\label{yMRK}
\ee
At each order in perturbation theory, large logarithms are developed
in $(1-\uh)$, or alternatively in
$\tau \equiv \sqrt{\vh\wh} = (1-\uh)|z|/|1+z|^2$.
The coefficients of each power of $\ln\tau$ are single-valued
(real analytic) functions of $z \in \mathbb{C}$, in fact they are
single-valued HPLs (SVHPLs)~\cite{BrownSVHPLs,Dixon:2012yy}
${\cal L}_{\vec{a}}(z,\bar{z})$, $a_i \in \{0,1\}$.

In particular, in the ancillary file {\tt hexMRKL1-7.m}~\cite{CosmicWebsite}
to ref.~\cite{Caron-Huot:2019vjl} the limiting behavior of the MHV
amplitudes and NMHV amplitudes were provided in terms of
a certain Fourier-Mellin integral.  Here we use the coupling normalization
and other conventions in ref.~\cite{DelDuca:2022skz},
where the remainder function in the $3\to3$ MRK limit is given by,
\be
\exp(\RR-i\pi\delta_6)|_{{\rm MRK}, \, 3\to3}
\ =\ \cos\Bigl(\pi\frac{\Gcusp}{4} \ln|z|^2\Bigr) - i\pi \Sigma(L_\tau),
\label{MHVMRK33review}
\ee
where
\be
\delta_6\ =\ \frac{\pi\Gcusp}{4} \ln\biggl( \frac{|z|^2}{|1-z|^4} \biggr) \,,
\label{delta6def}
\ee
and
\be
L_\tau = \ln\tau, \qquad \tau = \sqrt{\vh \wh} \,.
\label{Ltaudef}
\ee
The Fourier-Mellin representation of $\Sigma$ is
\be
\Sigma(L_\tau)\ =\
\frac{g^2}{\pi} \sum_{m=-\infty}^{\infty} \left(\frac{z}{\zb}\right)^{{m\over 2}}
{\cal P} \int_{-\infty}^{\infty}
{d\nu \,|z|^{2i\nu} \over \nu^2+{n^2\over 4}}
\Phi_{\textrm{reg}}(\nu,m)
e^{- L_\tau \omega(\nu,m)} \,,
\label{SigmaFM}
\ee
where $\omega(\nu,m)$ is the BFKL eigenvalues, $\Phi_{\textrm{reg}}(\nu,m)$
is the impact factor, and ${\cal P}$ stands for the principal part.
The $2\to4$ MRK limit is expressed in terms of $\Sigma(L_\tau+i\pi)$:
\be
\exp(\RR+i\pi\delta_6)|_{{\rm MRK}, \, 2\to4}
\ =\ \cos\Bigl(\pi\frac{\Gcusp}{4} \ln|z|^2\Bigr)
  + i\pi \Sigma(L_\tau+i\pi),
\label{MHVMRK24review}
\ee

The perturbative expansion of $\Sigma$ is
\be
\Sigma(L_\tau)\ =\ \sum_{L=1}^\infty g^{2L}
           \sum_{n=0}^{L-1} \Sigma_{n}^{(L)} \, (L_\tau)^n \,.
\label{MRKSigma}
\ee
In the ancillary file {\tt MRKSigma.txt},
we provide the values of $\Sigma_{n}^{(L)}$ through $L = 8$ loops.
Note that $\Sigma_{n}^{(8)}$ is a weight $15-n$ SVHPL, because the
amplitude has weight 16, of which weight 1 goes to the $2\pi i$
from analytic continuation, and $n$ to $(L_\tau)^n$.

The values through seven loops are taken from
ref.~\cite{Caron-Huot:2019vjl}, re-expressed in terms of $\Sigma_{n}^{(L)}$.
To get the eight-loop values, we used tables giving the
$2\to4$ MRK behavior of all elements of $\Hhex$ at weight 11, in order to
specify the behavior of all the MHV quintuples in that limit.
Then we integrated up the results from the quintuples to get
$\EE^{(8)}|_{{\rm MRK}, \, 2\to4}$.
The integration is straightforward using the definitions of the SVHPLs
and the relations between
coproducts on the MRK surface (holding $(1-\uh)$ fixed)
and those in the bulk:
\be
F^z = F^w - F^{y_w} \,, \qquad
F^{1-z} = - F^v - F^w - F^{y_v} + F^{y_w} \,.
\label{zcop}
\ee
However, the integration
also requires specifying boundary conditions in the MRK limit
at each weight from 12 to 16.

We transported the boundary conditions
from the base point $(\uh,\vh,\wh)=(1,1,1)$ along two different routes.
One route was to construct the answer on the line $(\uh,1,1)$,
then take $\uh \to \uh e^{-2\pi i}$ to get on the $2\to4$ sheet,
then return to the point $(1,1,1)$, or rather $(1-\delta,1,1)$.
Then we move down the $2\to4$ self-crossing line $(1-\delta,\vh,\vh)$
until $\vh\to0^+$, which approaches the limit of MRK in which
$z\to1, \bar{z}\to1$.  This route also gave us the higher weight constants
for the self-crossing line.
For the second route, we moved down the line $(1,\vh,\vh)$ on the Euclidean
sheet.  Then we moved along the line $(\uh,0,0)$ (i.e.~where $\vh,\wh$
are infinitesimal) to the origin. Next we took $\uh \to \uh e^{-2\pi i}$
to get on the $2\to4$ sheet, and then returned to the MRK point $(1,0,0)$
along the line $(\uh,0,0)$.  In this case we approach the limit of MRK
from a different direction, $z\to0, \bar{z}\to\infty$.
We used the program {\sc HyperlogProcedures}~\cite{HyperlogProcedures}
to extract the $z\to0, \bar{z}\to\infty$ limits of the SVHPLs.
This route also gave us the higher weight constants at the origin,
as a byproduct.

We got the same result via both routes, which is a useful cross-check
of the integration procedure.  Then we converted the result
from $\EE$ to $\RR$ using \eqn{EZMHVtoR6simplerho}
and extracted the perturbative coefficients of $\Sigma(L_\tau+i\pi)$.
We computed $\Sigma_{n}^{(8)}$
via the Fourier-Mellin integral for $n=4,5,6,7$, and the results agreed
perfectly with the results obtained from the amplitude.

\section{Multiple final entry relations}
\label{sec:multiplefinalentry}

In this section we describe relations between the $k^{\rm th}$ final entries
of the MHV and NMHV amplitudes that are independent of the loop order $L$.
Such relations are very useful for bootstrapping in the coproduct formalism,
because they can greatly reduce the number of initial parameters in an
ansatz.  With the help of parity decompositions, we
will find that many of the relations have saturated or
stabilized by seven loops, and we can use this information to
find ``bonus'' final entry relations.
We use the old alphabet ${\cal L}^u_\text{hex}$ to describe the relations
because they seem to be somewhat simpler in that alphabet.

A useful table for understanding the saturation of the final entries
with loop order is Table~\ref{tab:MHVdim}.  This table gives 
the number of independent $\{n,1,1,\ldots,1\}$ coproducts of the
MHV amplitudes. The numbers through $L=7$ are from ref.~\cite{DDToAppear}.
The eight loop numbers only became available after the eight-loop
computation was completed, of course.
A green color is used when the $(L+1,n)$ entry is the same as the
$(L,n)$ entry; it indicates saturation of the hexagon function space $\Hhex$
at weight $n$.
The numbers at weights $n=6,7$ and $L = 5,6,7$ are slightly smaller than
the numbers in the corresponding
Table 8 of ref.~\cite{Caron-Huot:2019bsq} because of the new
all-orders cosmic normalization $\rho$.  The smaller numbers indicate
that $\rho$ is a more optimal normalization than the previous $\rho_{\rm old}$.
A blue color is used when the $(L+1,n+2)$ entry is the same as
the $(L,n)$ entry; it indicates saturation of the space of
$k^{\rm th}$-final entries, where $k=2L-n$.

\renewcommand{\arraystretch}{1.25}
\begin{table}[!t]
\centering
\begin{tabular}[t]{l c c c c c c c c c c c c c c c c c}
\hline\hline
weight $n$
& 0 & 1 & 2 & 3 & 4 &  5 &  6 &  7 &  8 &  9 & 10 & 11 & 12 & 13 & 14
& 15 & 16
\\\hline\hline
$L=1$
& \green{1} & \green{3} & \blue{1} &  &  &  &  &  &  &  &  &  &  &  & 
\\\hline
$L=2$
& \green{1} & \green{3} & \green{6} & 4 & \blue{1}
  &  &  &  &  &  &  &  &  &  & 
\\\hline
$L=3$
& \green{1} & \green{3} & \green{6} & \green{13} & 14 & \blue{6} & \blue{1}
  &  &  &  &  &  &  &  & 
\\\hline
$L=4$
& \green{1} & \green{3} & \green{6} & \green{13} & \green{27} & 35
& 20 & \blue{6} & \blue{1} &  &  &  &  &  & 
\\\hline
$L=5$
& \green{1} & \green{3} & \green{6} & \green{13} & \green{27} & \green{54}
& 77 & 51 & \blue{21} & \blue{6} & \blue{1} &  &  &  & 
\\\hline
$L=6$
& \green{1} & \green{3} & \green{6} & \green{13} & \green{27} & \green{54}
& \green{102} & 163 & 126 & 58 & \blue{21} & \blue{6} & \blue{1} &  & 
\\\hline
$L=7$
& \green{1} & \green{3} & \green{6} & \green{13} & \green{27} & \green{54}
& \green{102} & \green{190} & 318 & 293 & 159 & \blue{62} & \blue{21} &
\blue{6} & \blue{1}
\\\hline
$L=8$
& \green{1} & \green{3} & \green{6} & \green{13} & \green{27} & \green{54}
& \green{102} & \green{190} & 343 & 579 & 630 & 384 & 162
& \blue{62} & \blue{21} & \blue{6} & \blue{1}
\\\hline\hline
\end{tabular}
\caption{The number of independent $\{n,1,1,\ldots,1\}$ coproducts
  of the MHV amplitudes $\EE^{(L)}$ through $L=8$ loops, at function level.
  The green and blue entries indicate, respectively, saturation of
  $\Hhex$ at weight $n$ and of the $(2L-n)^{\rm th}$-final entries,
  as explained in the text.}
\label{tab:MHVdim}
\end{table}

\subsection{MHV single final entries}
\label{subsec:MHVsinglefinalentry}

The number of linearly
independent $k^{\rm th}$ final entries generally stabilizes at a sufficiently
high number of loops, for small enough $k$.  For example, in
Table~\ref{tab:MHVdim} we see that there are 6 final entries
($k=1$, or weight $n=2L-1$) for the MHV amplitude, not 9.
This is not surprising, because the $\bar{Q}$
equation~\cite{Bullimore:2011kg,CaronHuot:2011kk} leads to the three
final-entry relations,
\be
\EE^{1-\uh_i} = - \EE^{\uh_i} \,, \quad {\rm or}
\quad \EE^{\ah} = \EE^{\bh} = \EE^{\ch} = 0.
\label{MHVfe}
\ee
%

\subsection{MHV double final entries}
\label{subsec:MHVdoublefinalentry}

Table~\ref{tab:MHVdim} also indicates that there are 21 MHV double final
entries ($k=2$, or weight $n=2L-2$).  Inspecting them more closely,
12 are parity-even and 9 are parity-odd.
How many MHV double final entries should we expect?
In refs.~\cite{Caron-Huot:2018dsv,Caron-Huot:2019bsq} it
was remarked that for the hexagon function space as a whole,
imposing the branch-cut conditions and the extended Steinmann relations
iteratively leads to 40 independent pairs of adjacent symbol entries;
that is, there are 41 adjacency relations among the $9\times9 = 81$
possible pairs of 9 letters.
(Integrability alone provides only 26 relations.)
Now let us also impose the MHV final-entry conditions~(\ref{MHVfe}),
which are really $9\times3=27$ conditions, $\EE^{x,1-\uh_i} = -\EE^{x,\uh_i}$
for any of the 9 letters $x$.  The 40 adjacency relations
and the 27 MHV final-entry conditions together constitute 56 independent
relations, and they reduce the expected number of pairs to 25,
15 parity-even and 10 parity-odd.

The $\bar{Q}$ equations also can be used to constrain the
MHV double final entries~\cite{CaronHuotprivate}.
There are six such relations, three even and three odd.
We give them in the old alphabet ${\cal L}^u_\text{hex}$:
\bea
\EE^{y_v,y_u} &=& \EE^{y_w,y_u} - \EE^{y_v,y_w}  + \EE^{y_v,y_v} + \EE^{\wh,\uh} \,,
\label{MHVQbarEven1}\\
\EE^{y_w,y_v} &=& \EE^{y_u,y_v} - \EE^{y_w,y_u}  + \EE^{y_w,y_w} + \EE^{\uh,\vh} \,,
\label{MHVQbarEven2}\\
\EE^{y_u,y_w} &=& \EE^{y_v,y_w} - \EE^{y_u,y_v} + \EE^{y_u,y_u} + \EE^{\vh,\wh}  \,,
\label{MHVQbarEven3}\\
\EE^{1-\uh,y_u} &=& \EE^{y_v,\vh} - \EE^{y_w,\uh} - \EE^{\vh,y_w} \,,
\label{MHVQbarOdd1}\\
\EE^{1-\vh,y_v} &=& \EE^{y_w,\wh} - \EE^{y_u,\vh} - \EE^{\wh,y_u} \,,
\label{MHVQbarOdd2}\\
\EE^{1-\wh,y_w} &=& \EE^{y_u,\uh} - \EE^{y_v,\wh} - \EE^{\uh,y_v} \,.
\label{MHVQbarOdd3}
\eea
However, these relations are automatically satisfied by the 25 independent
pairs.

Since the number of independent functions in Table~\ref{tab:MHVdim}
has stabilized at 12 parity-even, and 9 parity-odd, there must be
three parity-even and one parity-odd ``bonus'' relations.
They are found to be, in the old alphabet,
\bea
\EE^{y_v,y_u} &=& \EE^{y_u,y_v} + \EE^{\uh,\uh} + \EE^{1-\uh,\uh}
                - \EE^{\vh,\vh} - \EE^{1-\vh,\vh} \,,
\label{MHVBonusEven1}\\
\EE^{y_w,y_v} &=& \EE^{y_v,y_w} + \EE^{\vh,\vh} + \EE^{1-\vh,\vh}
                - \EE^{\wh,\wh} - \EE^{1-\wh,\wh} \,,
\label{MHVBonusEven2}\\
\EE^{y_u,y_w} &=& \EE^{y_w,y_u} + \EE^{\wh,\wh} + \EE^{1-\wh,\wh}
                - \EE^{\uh,\uh} - \EE^{1-\uh,\uh} \,,
\label{MHVBonusEven3}\\
\EE^{1-\uh,y_u} &=&  - \EE^{y_u,\uh} - \EE^{\vh,y_w} + \EE^{y_w,\vh} \,.
\label{MHVBonusOdd}
\eea
The first three (even) equations permute into each other under cyclic
permutations, and flips do not given anything new.  The last (odd)
equation~(\ref{MHVBonusOdd}) appears to be asymmetric,
and dihedral permutations of it would naively seem to generate more equations,
but they turn out to all be equivalent to this relation when taking into
account the other 26 odd relations.  

Using all these relations, we can take the 12 independent parity-even
MHV double final entries to be
\be
\{ \EE^{\uh_i,\uh_i},\ \EE^{1-\uh_i,\uh_i},\ \EE^{y_i,y_i},\ \EE^{y_i,y_{i+1}} \},
\quad i = 1,2,3.
\label{MHVEvenDoubleIndep_uvw}
\ee
The remaining 33 even double final entries are given by
\bea 
\EE^{y_v,y_u} &=& \EE^{\uh,\uh} + \EE^{1-\uh,\uh}
   - \EE^{\vh,\vh} - \EE^{1-\vh,\vh} + \EE^{y_u,y_v} \,,
\label{MHVEvenDoubleRelA}\\
\EE^{\uh,\vh} &=& \EE^{\vh,\vh} + \EE^{1-\vh,\vh}
   - \EE^{\wh,\wh} - \EE^{1-\wh,\wh}
   - \EE^{y_w,y_w} - \EE^{y_u,y_v} + \EE^{y_v,y_w} + \EE^{y_w,y_u} \,,
\label{MHVEvenDoubleRelB}\\
\EE^{1-\uh,\vh} &=& \EE^{\uh,\uh} + \EE^{1-\uh,\uh}
   - \EE^{\vh,\vh} - \EE^{1-\vh,\vh}
   + \EE^{y_w,y_w} + \EE^{y_u,y_v} - \EE^{y_v,y_w} - \EE^{y_w,y_u} \,,
\label{MHVEvenDoubleRelC}\\
\EE^{\uh_i,1-\uh_j} &=& - \EE^{\uh_i,\uh_j} \,,
\label{MHVEvenDoubleRelD}\\
\EE^{1-\uh_i,1-\uh_j} &=& - \EE^{1-\uh_i,\uh_j} \,,
\label{MHVEvenDoubleRelE}
\eea
plus the dihedral images of these relations.

Similarly, we can take the 9 independent parity-odd
MHV double final entries to be
\be
\{ \EE^{\uh_i,y_j} \},\quad i,j = 1,2,3.
\label{MHVOddDoubleIndep_uvw}
\ee
The remaining 27 odd double final entries follow from the relations,
\bea
\EE^{1-\uh,y_u} &=& \frac{1}{2} \Bigl[
\EE^{\vh,y_u} + \EE^{\wh,y_u} - \EE^{\uh,y_v} - \EE^{\wh,y_v}
                       - \EE^{\uh,y_w} - \EE^{\vh,y_w} \Bigr] \,,
\label{MHVOddDoubleRelA}\\
\EE^{1-\uh,y_v} &=& \EE^{\uh,y_u} + \EE^{\vh,y_u} + \EE^{\wh,y_u}
  - \EE^{\uh,y_v} - \EE^{\uh,y_w} - \EE^{\vh,y_w} - \EE^{\wh,y_w} \,,
\label{MHVOddDoubleRelB}\\
\EE^{y_u,\uh} &=& \EE^{\uh,y_u} \,,
\label{MHVOddDoubleRelC}\\
\EE^{y_v,\uh} &=& \frac{1}{2} \Bigl[
- \EE^{\vh,y_u} - \EE^{\wh,y_u} + \EE^{\uh,y_v} - \EE^{\wh,y_v}
+ \EE^{\uh,y_w} + \EE^{\vh,y_w} \Bigr]
+ \EE^{\wh,y_w} \,,
\label{MHVOddDoubleRelD}\\
\EE^{y_i,1-\uh_j} &=& - \EE^{y_i,\uh_j} \,.
\label{MHVOddDoubleRelE}
\eea
and their dihedral images.

\subsection{MHV triple final entries}
\label{subsec:MHVtriplefinalentry}

We can perform a similar analysis for the MHV triple final entries,
$\EE^{x,y,z}$, where $x,y,z$ are generic letters.
We require the last two slots ($y,z$) to be in the 21-dimensional space
of MHV double final entries, and the first two slots ($x,y$) to
be in the generic 40-dimensional space of adjacent pairs.
The first requirement gives $9 \times 60 = 540$ relations,
and the second one $41 \times 9 = 369$ relations.  Solving the equations,
we find that there are 65 independent triple final entries,
34 parity-even and 31 parity-odd.
Now we take the 58 and 62 triple final entries at six and seven loops,
shown in Table~\ref{tab:MHVdim}, and determine their parity.
There are 31 even and 27 odd at six loops,
and 31 even and 31 odd at seven loops.
Thus the even number has saturated, while the odd number already agrees
with the analysis based on the double final entries and the general 40-pair
restriction.  Hence the number of MHV triple final entries shown
in Table~\ref{tab:MHVdim} has stabilized at \blue{62},
31 even and 31 odd.

We also conclude that there must be three ``bonus'' triple final
entry relations in the parity-even sector.  By comparing the 34 parity-even
functions inferred from the double-final-entry analysis with the
actual 31 functions at six and seven loops,
we find that the bonus relations can be written as,
\bea
\EE^{1-\uh,\uh,\uh} &=& \EE^{\uh,1-\uh,\uh}
+ 2 \, \EE^{y_u,\uh,y_u} + \EE^{y_u,\uh,y_v} + \EE^{y_u,\uh,y_w}
+ 5 \, ( \EE^{y_u,\vh,y_u} + \EE^{y_u,\wh,y_u} )
\nonumber\\ &&\hskip0.0cm\null
- 2 \, ( \EE^{y_v,\uh,y_v} + \EE^{y_w,\uh,y_w} - \EE^{y_v,\vh,y_w} - \EE^{y_w,\wh,y_v}
      + \EE^{y_v,\wh,y_v} +\EE^{y_w,\vh,y_w} )
\nonumber\\ &&\hskip0.0cm\null
- 4 \, ( \EE^{y_v,\vh,y_u} + \EE^{y_w,\wh,y_u} )
+ \EE^{y_u,\vh,y_v} + \EE^{y_u,\wh,y_w}
- \EE^{y_v,\vh,y_v} - \EE^{y_w,\wh,y_w}
\nonumber\\ &&\hskip0.0cm\null
- \EE^{y_w,\vh,y_v} - \EE^{y_v,\wh,y_w} \,,
\label{BonusTripleEquation1}
\eea
plus the two equations obtained by cyclic permutations of this one.
There is some arbitrariness in how the bonus relations are written,
since they are modulo a large number of other relations.
We have checked that \eqn{BonusTripleEquation1} holds for
all loop orders through seven loops.

\section{Parity decomposition of amplitude coproducts and locking}
\label{sec:copparitylocking}

\renewcommand{\arraystretch}{1.25}
\begin{table}[!t]
\centering
\begin{tabular}[t]{l c c c c c c c c c c c c c c c c c}
\hline\hline
weight $n$
& 0 & 1 & 2 & 3 & 4 &  5 &  6 &  7 &  8 &  9 & 10 & 11 & 12 & 13 & 14
& 15 & 16
\\\hline\hline
$L \leq 7$
& \green{0} & \green{0} & \green{0} & \green{1} & \green{2}
& \green{6} & \green{13} & \green{29} & \green{57} & \green{113} & 161
& 112 & \blue{39} & \blue{12} & \blue{2} & $-$ & $-$
\\\hline
$L = 8$
& \green{0} & \green{0} & \green{0} & \green{1} & \green{2}
& \green{6} & \green{13} & \green{29} & \green{57} & \green{113} & 193
& 185 & 78 & \blue{31} & \blue{9} & \blue{3} & \blue{0}
\\\hline\hline
\end{tabular}
\caption{The number of parity odd $\{n,1,1,\ldots,1\}$ coproducts of the
MHV and NMHV amplitudes through 7 loops, followed by the
number for the 8 loop MHV amplitude alone.
The color coding is the same as in Table~\ref{tab:MHVdim}.}
\label{tab:ALLodddim}
\end{table}

\renewcommand{\arraystretch}{1.25}
\begin{table}[!t]
\centering
\begin{tabular}[t]{l c c c c c c c c c c c c c c c c c}
\hline\hline
weight $n$
& 0 & 1 & 2 & 3 & 4 &  5 &  6 &  7 &  8 &  9 & 10 & 11 & 12 & 13 & 14
& 15 & 16
\\\hline\hline
$L \leq 7$
& \green{1} & \green{3} & \green{6} & \green{12} & \green{25}
& \green{48} & \green{89} & \green{161} & 280 & 377 & 255
& 107 & \blue{43} & \blue{12} & \blue{4} & $-$ & $-$
\\\hline
$L = 8$
& \green{1} & \green{3} & \green{6} & \green{12} & \green{25}
& \green{48} & \green{89} & \green{161} & 286 & 466 & 437
& 199 & 84 & \blue{31} & \blue{12} & \blue{3} & \blue{1}
\\\hline\hline
\end{tabular}
\caption{The number of parity even $\{n,1,1,\ldots,1\}$ coproducts of the
MHV and NMHV amplitudes through 7 loops, followed by the
number for the 8 loop MHV amplitude alone.
The color coding is the same as in Table~\ref{tab:MHVdim}.}
\label{tab:ALLevendim}
\end{table}

In Table~\ref{tab:ALLodddim} we provide the number of 
independent parity-odd $\{n,1,1,\ldots,1\}$ coproducts for the
combined system of MHV and NMHV amplitudes through seven loops,
followed by the number for the eight loop MHV amplitude alone.
With the addition of this last amplitude,
saturation of the odd functions is now achieved all the way
through weight 9.
These data allow us to see cleanly that the space of hexagon
functions used in refs.~\cite{Caron-Huot:2019vjl,Caron-Huot:2019bsq}
can be reduced further in size.

Starting at weight 7, the saturated number of parity-odd functions is lower
than with the previous normalization factor $\rho$, due to the locking
phenomenon mentioned in ref.~\cite{DDToAppear}.  At weight 7 odd,
only one function is removed:  $\zeta_4\,\tilde{\Phi}_6$,
where $\zeta_4=\pi^4/90$ and $\tilde{\Phi}_6$ is the unique
weight 3 parity-odd function, namely the $D=6$ scalar hexagon integral.

This function must be added to the other 29, symbol-level weight 7 odd
functions with fixed coefficients.  In terms of the basis used in
refs.~\cite{Caron-Huot:2019vjl,Caron-Huot:2019bsq},
which is called $\{ {\rm YO}[7,i] \}$, $i=1,2,\ldots,30$
in the ancillary files, the smaller 29-dimension space is
\be
\{ {\rm YO}[7,i] + c^{7{\rm o}}_{i} \, {\rm YO}[7,30] \}, \quad i=1,2,\ldots,29,
\label{new7o}
\ee
where ${\rm YO}[7,30] = \zeta_4 \, \tilde{\Phi}_6$, and
\bea
c^{7{\rm o}}_{i} &=& [
9, -12, -12, -12, -9, -9, 9, 48, -9, 3, -39, -18, -51, -6, 9,
\nonumber\\&&\hskip0.1cm
-30, 12, -21, 27, 15, 6, 3, -27, -6, 24, -30, 12, -33, -30 ]\,.
\label{shift7o}
\eea
Notice that the coefficients are all a multiple of 3.  This property holds
also for all the analogous $\zeta_4$-associated
coefficients for weight 8 and 9 odd and
weight 6 and 7 even.  The coefficients of
${\rm YO}[9,120] = \zeta_6 \, \tilde{\Phi}_6$ encountered at weight 9 odd
are always integers, but not always a multiple of 3.
The coefficients for all these linear combinations are provided in the
ancillary file {\tt EZsmallercoproductspace.txt}.

At weight 8 odd, two functions are removed, corresponding to
$\zeta_4$ times the two weight 4 odd functions.
At weight 9 odd, seven functions are removed, corresponding to
$\zeta_4$ times the six weight 5 odd functions, plus $\zeta_6$
times the one weight 3 odd function.
In other words, the number of independent odd functions for weights 7,8,9
in the new normalization is exactly equal to the number of consistent symbols.

Table~\ref{tab:ALLevendim} provides the corresponding numbers
in the parity-even sector.  Comparing them with the numbers
for the old amplitude normalization reveals the following:
At weight 6 even, three functions are removed, corresponding to
$\zeta_4 \, {\rm Li}_2(1-1/u_i)$.  However, the three pure logarithms
$\zeta_4 (\ln^2 a_i + 4\zeta_2)$ are still required to be independent
functions.
At weight 7 even, nine functions are removed, corresponding to
$\zeta_4$ multiplied by the weight 3 even functions containing $1-u_i$ in
their symbols, while the six functions
$\zeta_4 \ln^3 a_i + \ldots$ and $\zeta_6 \ln a_i$
are still independent functions.
We don't have quite enough information yet to confirm this pattern
past weight 7 in the parity-even sector.  The number of weight 8 even
functions obtained through seven loops is 280, while for the eight loop
MHV amplitude there are 286, so the number of functions may not have
stabilized yet.


\section{Amplitudes and multiple coproducts at
   \texorpdfstring{$(1,1,1)$}{(1,1,1)}}
\label{sec:ampcoprods111}

In this section we examine the values of the amplitudes and their
multiple coproducts at the dihedrally symmetric, finite base point,
$(\uh,\vh,\wh)=(1,1,1)$, where they evaluate to MZVs.
They belong to a restricted set of MZVs, $\Hhex(1,1,1)$, which
obeys a coaction principle~\cite{Caron-Huot:2019bsq}.
The MHV 8-loop amplitude allows us to test the coaction principle
further and search for ``dropouts'' that imply further constraints
at higher weights.
In addition, there are interesting relations between the multiple coproducts
of the amplitudes at $(1,1,1)$, that hold at every loop order
through $L=8$, for which we do not yet have a complete explanation.

\subsection{Coaction principle at \texorpdfstring{$(1,1,1)$}{(1,1,1)}}
\label{subsec:coaction111}

Multiple polylogarithms have motivic versions which are subject to
a coaction~\cite{Gonch2,Brown:2011ik,2011arXiv1101.4497D,%
  Duhr:2011zq,Duhr:2012fh}.
The coaction $\Delta$ (not to be confused with the polynomial $\Delta$
whose vanishing defines the parity-preserving surface!)
maps a general space of polylogarithms ${\cal G}$ essentially into two
copies of itself,
\be
\Delta({\cal G}) = {\cal G} \otimes {\cal G}^{\rm \mathfrak{dR}} \,.
\label{DeltaG}
\ee
The right ``de Rham'' space ${\cal G}^{\rm \mathfrak{dR}}$ loses
some information about contours of integration and is therefore
only defined modulo $i\pi$ (roughly; for a more detailed discussion see
e.g.~ref.~\cite{Caron-Huot:2019bsq}).
Since ${\cal G}$ is graded by the weight,
\begin{equation}
{\cal G}\ =\ \bigoplus_{n=0}^\infty \, {\cal G}_n \, ,
\end{equation}
$\Delta$ acting on ${\cal G}_n$ can be split into components
$\Delta_{n-p,p}$ according to the grading, where $p\in\mathbb{Z}$.
The case $p=1$ is the total differential, as discussed around \eqn{Deltanm11}.
Iterations of $\Delta_{n-1,1}$ lead to the symbol.

The coaction principle is a statement about the stability of the left-hand
side of the coaction, for a subspace of a space of multiple polylogarithms
(or perhaps MZVs, which are multiple polylogarithms evaluated at a particular
point) that is picked out by a given physical problem.
For the space of hexagon functions $\Hhex$, we ask whether it obeys
\be
\Delta \Hhex \subset \Hhex \otimes {\cal K}^\pi \,.
\label{eq:coaction_principle_intro}
\ee
Here ${\cal K}^\pi$ involves iterated integrals whose symbols have
the same pair-adjacency relations as in $\Hhex$, but they lack the first
entry condition, so the space ${\cal K}_n^\pi$ is much larger than $\Hhex_n$
for a given weight $n$.

Because we construct the space of hexagon functions $\Hhex$ iteratively,
by requiring their derivatives to be in the space, the part of the coaction
principle that involves only $\Delta_{n-1,1}$ is automatically obeyed.
The interesting question has to do with $\Delta_{n-p,p}$ for $p>1$,
and in particular with constants on the right-hand side of the coaction,
since such constants are invisible at the level of differentials
($\Delta_{n-1,1}$).  Such constants can be seen by evaluating all
the hexagon functions at the point $(1,1,1)$,
and we refer to the space of MZVs there as $\Hhex(1,1,1)$.
In ref.~\cite{Caron-Huot:2019bsq} it was found that only a restricted
set of MZVs appears in $\Hhex(1,1,1)$, and that this set
was stable under the coaction.

A way to represent MZVs which respects the coaction is to use
an $f$-alphabet~\cite{Brown:2011ik,Schnetz:2013hqa}.
In this description, each odd Riemann zeta value $\zeta_{2k+1}$
is mapped to a letter $f_{2k+1}$, for $k=1,2,3,\ldots$.
If we take a free algebra $\mathbb{Q}(f_{2k+1})$ over the rational numbers,
and supplement it with powers of $\pi^2$, then that space is isomorphic to
the vector space of the MZVs over the rationals.
A free algebra means that the letters $f_{2k+1}$ do not commute with
each other; the different orderings allow for irreducible MZVs to be
encoded.
There is a derivation operation $\del_{2k+1}$ associated with every letter
$f_{2k+1}$. It acts to remove any $f_{2k+1}$ in the de Rham (right) factor of
the coaction; if another $f_{2k'+1}$ is there it returns 0.

The antipode map associated with the coaction has a particularly simple
action on MZVs in the $f$-alphabet:
it simply reverses the ordering of all the $f$'s.
Since it is only defined modulo the $i\pi$ ambiguity on the right-hand
side of the coaction, one should ignore all $\pi$'s
(and all even Riemann zeta values) in computing the antipode.

We use the $f$-alphabet version available in the program
{\sc HyperlogProcedures}~\cite{HyperlogProcedures}.
We write $f_{2k_1+1,2k_2+1,\ldots} \equiv f_{2k_1+1}f_{2k_2+1}\ldots$
as a shorthand.
We use an ordering convention of the $f$'s from
refs.~\cite{Panzer:2016snt,HyperlogProcedures} which
is reversed with respect to our convention for the symbol;
thus, the derivation $\del_{2k+1}$ acts on the {\it left} side of a
string of $f$'s.

The following relations allow the conversion of the $f$-alphabet to
more conventional MZV notation through weight 10:
\bea
f_{3,3} &=& \frac{1}{2} (\zeta_3)^2 \,, \\
f_{5,3} &=& -\frac{1}{5} \zeta_{5,3} \,, \\
f_{3,3,3} &=& \frac{1}{6} (\zeta_3)^3 \,, \\
f_{3,7} &=& \zeta_3 \zeta_7
       + \frac{1}{14} \Bigl[ 3 (\zeta_5)^2 + \zeta_{7,3} \Bigr] \,, \\
f_{7,3} &=& - \frac{1}{14} \Bigl[ 3 (\zeta_5)^2 + \zeta_{7,3} \Bigr] \,, \\
f_{5,5} &=& \frac{1}{2} (\zeta_5)^2 \,.
\label{ftozeta}
\eea
The analogous conversions through weight 16 are given in the ancillary file
{\tt ftoMZV16.txt}. (See ref.~\cite{Caron-Huot:2019bsq} for conversions
through weight 14.)

In the $f$ alphabet, the vector space $\Hhex(1,1,1)$
was shown~\cite{Caron-Huot:2019bsq} to have the following elements,
through weight 12:
\bea
&&1\label{finalHhex111}\\
&& - \nonumber\\
&& \zeta_2 \nonumber\\
&& - \nonumber\\
&& \zeta_4 \nonumber\\
&& 5 f_5 - 2\zeta_2 f_3 \nonumber\\
&& \zeta_6 \nonumber\\
&& 7 f_7 - \zeta_2 f_5 - 3\zeta_4 f_3 \nonumber\\
&& \zeta_8 \,,\ \ 5 f_{3,5} - 2\zeta_2 f_{3,3} \nonumber\\
&& 7 f_9 - 6\zeta_4f_5 \,,\ \ 5 f_9 - 3\zeta_6 f_3,\ \ \zeta_2 f_7-\zeta_6 f_3
\nonumber\\
&& \zeta_{10} \,,\ \ 7 f_{3,7}-\zeta_2 f_{3,5}-3\zeta_4 f_{3,3}  \,,\ \
5 f_{5,5}-2\zeta_2f_{5,3}
\nonumber\\ 
&& 33 f_{11} - 20 \zeta_8 f_3\,,
\ \zeta_2 f_9 - \zeta_8 f_3\,,
\ 3 \zeta_4 f_7 - 2 \zeta_8 f_3\,,
\ 3 \zeta_6 f_5 - 2 \zeta_8 f_3\,,
\ 5 f_{3,3,5} - 2 \zeta_2 f_{3,3,3} + \frac{5611}{132} \zeta_8 f_3
\nonumber\\
&&\zeta_{12} \,,
\ 7 f_{3,9} \! - \! 6 \zeta_4 f_{3,5}\,,
\ 5 f_{3,9} \! - \! 3 \zeta_6 f_{3,3}\,,
\ \zeta_2 f_{3,7} \! - \! \zeta_6 f_{3,3}\,,
\ 7 f_{5,7} \! - \! \zeta_2 f_{5,5} \! - \! 3 \zeta_4 f_{5,3}\,,
\ 5 f_{7,5} \! - \! 2 \zeta_2 f_{7,3} \,.
\nonumber
\eea

Notice that $\zeta_3 = f_3$ does not appear, and only one linear combination
appears out of the two possible at weight 5, $5 f_5 - 2\zeta_2 f_3$.
These two facts, and the coaction principle, dictate that at weight 8,
only $\zeta_8$ and $5 f_{3,5} - 2\zeta_2 f_{3,3}$ can appear.
The first element, $\zeta_8$,
like any even Riemann zeta value $\zeta_{2k}$,
gives nothing nontrivial (lower weight) under the coaction. For this
reason, it is always allowed by the coaction principle.
The second weight 8 MZV, $5 f_{3,5} - 2\zeta_2 f_{3,3}$, is allowed because
\begin{enumerate}
\item $\del_5$ annihilates it, acting on the left, which is necessary
  because $f_3=\zeta_3$ is absent at weight 3 in \eqn{finalHhex111}).
\item $\del_3 (5 f_{3,5} - 2\zeta_2 f_{3,3}) = 5 f_{5} - 2\zeta_2 f_{3}$, which
  is proportional to the one linear combination in $\Hhex_5$.
\end{enumerate}
Similary, it is easy to see that the weight 10 basis is consistent
with the coaction principle because, other than $\zeta_{10}$, the
basis elements are obtained by adding a ``3'' to the left of the one weight 7
basis element, and a ``5'' to the left of the one weight 5 basis element.
In general one can add odd indices to the left of lower-weight basis elements,
and also add either $\zeta_{2k}$ or $f_{2k+1}$ to the basis, in order to get
a candidate basis at the next weight that is consistent with the
coaction principle.

We refer to the absence of $f_3$, and of the other linear combination
of $f_5$ and $\zeta_2 f_3$, as {\it dropouts}.  These are missing zeta values
whose absence is not required by the coaction principle. Without such dropouts,
there would be no consequences of the coaction principle at $(1,1,1)$.
The number of dropouts at different weights in the list~(\ref{finalHhex111})
is easily counted to be:
1 at weight 3, 1 at weight 5, 2 at weight 7, 1 at weight 9, 1 at weight 11.

It becomes increasingly difficult to establish the existence of dropouts
at high weight, because as we will see later in this section,
there can be relatively few independent amplitude coproducts at $(1,1,1)$.
This is particularly true for odd weights, because the single coproducts
for both MHV and NMHV amplitudes all vanish at $(1,1,1)$; and for MHV
there are very few independent triple coproducts at $(1,1,1)$.
However, with the benefit of all the eight-loop MHV amplitude coproducts,
we can establish that all of the basis elements in the
list~(\ref{finalHhex111}) are present, with the exception of weight 12,
where there is one dropout.  Thus, the last line of \eqn{finalHhex111}
should have only 5 entries instead of 6:
\bea
&&\biggl\{ \zeta_{12}\,,\ \ 
 7 f_{3,9} - 6 \zeta_4 f_{3,5}
   + \frac{1}{3} ( 7 f_{5,7} - \zeta_2 f_{5,5} - 3 \zeta_4 f_{5,3} )\,,
\nonumber\\
&& 7 (5 f_{3,9} - 3 \zeta_6 f_{3,3})
   + \frac{5}{3} (7 f_{5,7} - \zeta_2 f_{5,5} - 3 \zeta_4 f_{5,3} )\,,\ \ 
\zeta_2 f_{3,7} - \zeta_6 f_{3,3}\,,
\nonumber\\
&& 3 (5 f_{3,9} - 3 \zeta_6 f_{3,3}) - (5 f_{7,5} - 2 \zeta_2 f_{7,3}) \biggr\}
\label{wt12new}\\
&=& \biggl\{ \zeta_{12}\,,\ \
7 f_{3,9} + \frac{7}{3} f_{5,7} - \frac{1}{3} \zeta_2 f_{5,5}
         - \zeta_4 (f_{5,3}+6 f_{3,5})\,,
\nonumber\\
&& 35 f_{3,9} + \frac{35}{3} f_{5,7} - \frac{5}{3} \zeta_2 f_{5,5}
         - 5 \zeta_4 f_{5,3} - 21 \zeta_6 f_{3,3}\,,\ \ 
\zeta_2 f_{3,7} - \zeta_6 f_{3,3}\,,
\nonumber\\
&& 15 f_{3,9} - 5 f_{7,5} + 2 \zeta_2 f_{7,3} - 9 \zeta_6 f_{3,3} \biggr\} \,.
\label{wt12newALT}
\eea
The first form makes clear the linear combinations of the previous
basis elements, and that it obeys the coaction principle.
The 6-loop amplitudes $\EE^{(6)}(1,1,1)$ and (NMHV) $E^{(6)}(1,1,1)$
are linear combinations of these 5 basis elements,
as well as the 7-loop double coproducts of $\EE^{(7)}$, $E^{(7)}$ and
(NMHV parity-odd) $\Et^{(7)}$.

The independent quadruple coproducts of $\EE^{(8)}$ at $(1,1,1)$ furnish
a more stringent test of \eqn{wt12new}.
As mentioned in \sect{sec:beyondweight11}, there are 10
such eight-loop constants before imposing more detailed constraints.
The fact that they all live in the same 5-dimensional space provides
convincing evidence of the first dropout to appear at an even weight.

At weight 13, the coaction principle allows for 9 possible basis elements.
However, there is only one independent triple coproduct of $\EE^{(8)}$
at $(1,1,1)$, and no single coproduct of $\EE^{(7)}$,
so we are unable to search for dropouts at weight 13 (or higher).

\subsection{Amplitude at \texorpdfstring{$(1,1,1)$}{(1,1,1)}}
\label{subsec:amps111}

The value of the eight-loop MHV 6-point amplitude at $u=v=w=1$
provides us with one weight 16 MZV.
In terms of the $f$-alphabet, it is:
\bea
\EE^{(8)}(1,1,1) &=&
\blue{9122624} \, f_{9,7} + \blue{11543472} \, f_{7,9}
+ \blue{5153280} \, f_{11,5} + \blue{19603536} \, f_{5,11}
+ \blue{23915376} \, f_{3,13}\nonumber\\
&&\hskip-0.5cm\null
+ \blue{371520} \, f_{5,3,3,5}
+ \blue{400320} \, f_{3,3,5,5} + \blue{400320} \, f_{3,5,3,5}
+ \blue{825216} \, f_{3,3,3,7} \nonumber\\
&&\hskip-0.5cm\null
- \zeta_2  \,  ( 701856 \, f_{7,7} + 1303232 \, f_{9,5}
+ 430656 \, f_{5,9} + 2061312 \, f_{11,3}
- 309696 \, f_{3,11}\nonumber\\
&&\hskip0.5cm\null
+ 160128 \, f_{3,5,3,3} + 160128 \, f_{3,3,5,3}
+117888 \, f_{3,3,3,5}+148608 \, f_{5,3,3,3} ) \nonumber\\
&&\hskip-0.5cm\null
- \zeta_4  \,  ( 3243888 \, f_{5,7} + 3475296 \, f_{7,5}
+ 3909696 \, f_{9,3} + 3215472 \, f_{3,9} + 353664 \, f_{3,3,3,3} ) \nonumber\\
&&\hskip-0.5cm\null
- \zeta_6 ( 3612804 \, f_{5,5} + 3791520 \, f_{7,3} + 3409152 \, f_{3,7} )
- \zeta_8 ( 3720664 \, f_{5,3} \! + \! 3456614 \, f_{3,5} ) \nonumber\\
&&\hskip-0.5cm\null
- \frac{19560489}{5}  \,  \zeta_{10}  \,  f_{3,3}
- \frac{512193667550809}{7639104} \,  \zeta_{16} \,.
\label{EZMHVfg8_111}
\eea
It is straightforward to check that application of $\del_{2k+1}$
to this results lands in the basis~(\ref{finalHhex111})
for $k=1,2,3,4,5,6,7$, as required by the coaction principle.

As mentioned above, to apply the antipode map to this expression,
one only has to reverse the ordering of the $f$ indices,
and ignore any term with an $i\pi$ (none here) or a $\pi^2$ or
an even Riemann zeta value $\zeta_{2k}$. 
That means focusing on the integers shown in \blue{blue}
in \eqn{EZMHVfg8_111}.
Reversing the ordering of the $f$ subscripts in these terms,
we recover the appropriate value of the eight-loop form factor given in the
ancillary file {\tt AntipodePointsSummary.txt} for ref.~\cite{Dixon:2021tdw}
(modulo $\pi^2$ terms).  Thus we confirm that antipodal duality works at
eight loops beyond symbol level.  Although antipodal duality was used in the
construction of the amplitude, it was only used at symbol level,
so this is quite a nice confirmation of its full action.
We also confirmed antipodal duality at eight loops beyond symbol level
on the entire line $(1,\vh,\vh)$, using the prediction in
the ancillary file {\tt A6line1vv.dat} for ref.~\cite{Dixon:2021tdw},
and comparing it with the results in the ancillary file
\texttt{EZMHVg\_uu1\_lin.txt}.

In terms of conventional MZVs, the eight-loop value at $(1,1,1)$ is
\bea
\EE^{(8)}(1,1,1) &=&
\frac{4901904}{11} \zeta_{11,5} - 2764512 \zeta_{13,3}
+ 58944 \zeta_{7,3,3,3} - 54720 \zeta_{5,5,3,3}
+ 576 (\zeta_{5,3})^2 \nn\\
&&\hskip-0.5cm\null
+ 54720 \zeta_3 \zeta_{5,5,3} - 58944 \zeta_3 \zeta_{7,3,3}
+ 5760 \zeta_5 \zeta_{5,3,3} + 29472 (\zeta_3)^2 \zeta_{7,3}
- \frac{57222368}{11} \zeta_7 \zeta_9  \nn\\
&&\hskip-0.5cm\null
+ 3016464 \zeta_5 \zeta_{11}
+ 23915376 \zeta_3 \zeta_{13}
+ 188496 (\zeta_3)^2 (\zeta_5)^2 + 137536 (\zeta_3)^3 \zeta_7 \nn\\
&&\hskip-0.5cm\null
+ \zeta_2 \Bigl( 1799904 \zeta_{11,3} - 411968 \zeta_{9,5} + 6144 \zeta_{5,3,3,3}
       - 8448 \zeta_3 \zeta_{5,3,3}
       + 4224 (\zeta_3)^2 \zeta_{5,3} \nn\\
&&\hskip0.8cm\null
       + 5485728 (\zeta_7)^2 + 7035328 \zeta_5 \zeta_9
       - 4161408 \zeta_3 \zeta_{11} - 19648 (\zeta_3)^3 \zeta_5 \Bigr) \nn\\
&&\hskip-0.5cm\null
+ \zeta_4 \Bigl(
  259312 \zeta_{9,3} - 1910928 \zeta_5 \zeta_7 - 4448304 \zeta_3 \zeta_9
       - 14736 (\zeta_3)^4 \Bigr)  \nn\\
&&\hskip-0.5cm\null
+ \zeta_6 \Bigl( 56784 \zeta_{7,3} - 1550370 (\zeta_5)^2
              - 3217728 \zeta_3 \zeta_7 \Bigr)
+ \zeta_8 \Bigl( 57930 \zeta_{5,3} - 3421414 \zeta_3 \zeta_5 \Bigr)
\nn\\
&&\hskip-0.5cm\null
- \frac{19560489}{10} \zeta_{10} (\zeta_3)^2
- \frac{512193667550809}{7639104} \zeta_{16} \,.
\label{EZMHVg8_111}
\eea
We give this value (and its $f$-alphabet form)
in the ancillary file {\tt EZMHVcoproducts111.txt}.
We give its numerical value in \eqn{EZMHVg8_111_num} below.

\subsection{Amplitude coproducts at \texorpdfstring{$(1,1,1)$}{(1,1,1)}}
\label{subsec:single111}

Now we turn to relations among the MHV amplitude's multiple coproducts
at $(1,1,1)$. In ref.~\cite{DDToAppear}, the subspace of MZVs encountered
by evaluating the single, double and triple coproducts for
both MHV and NMHV 6-particle amplitudes at $(1,1,1)$ is explored
through 7 loops.   The MHV structure is particularly simple,
and we verify here that it continues to be obeyed through 8 loops.
In this discussion, we use the old alphabet ${\cal L}^u_\text{hex}$
because the equations are a little shorter.

For MHV, the $\{2L-1,1\}$ first coproducts of the amplitudes must all
vanish at $(1,1,1)$.  This result follows from parity, the branch-cut
condition that $\EE^{1-\uh_i}$ vanishes at
$\uh_i=1$~\cite{Dixon:2013eka,Caron-Huot:2019bsq},
and the final-entry condition $\EE^{\uh_i} = -\EE^{1-\uh_i}$.

In ref.~\cite{DDToAppear} it is shown that all of the double coproducts
of the MHV amplitudes at $(1,1,1)$ either vanish or can be expressed in terms
of $\EE^{\uh,\uh}(1,1,1)$ and $\EE^{y_u,y_u}(1,1,1)$:
\bea
\EE^{\uh,1-\vh}(1,1,1) &=& \EE^{\uh,\vh}(1,1,1)
= \EE^{1-\uh,\vh}(1,1,1) = \EE^{1-\uh,\uh}(1,1,1) = 0,
\label{MHVDoubleCoproducts111A}\\
\EE^{\uh,1-\uh}(1,1,1) &=& - \EE^{\uh,\uh}(1,1,1), \label{MHVDoubleCoproducts111B}\\
\EE^{y_u,y_v}(1,1,1) &=& \EE^{y_u,y_u}(1,1,1),
\label{MHVDoubleCoproducts111C}
\eea
including also the dihedral images of these relations,
and the vanishing of parity-odd double coproducts.
Hence to specify all MHV double coproducts at $(1,1,1)$ it is enough
to tabulate $\EE^{\uh,\uh}(1,1,1)$ and $\EE^{y_u,y_u}(1,1,1)$
through 8 loops. The tabulation through 7 loops is provided in
ref.~\cite{DDToAppear}; here we give the 8 loop values:
\bea
\EE^{(8)\, \uh,\uh}(1,1,1) &=& 
- 2246816 \Bigl[f_{3,11} - \frac{20}{33} \zeta_8 f_{3,3}\Bigr]
+ \frac{151360}{3} \Bigl[\zeta_2 f_{3,9} - \zeta_8 f_{3,3}\Bigr]
\nn\\ &&\hskip-0.5cm\null
+ 120224   \Bigl[3 \zeta_4 f_{3,7} - 2 \zeta_8 f_{3,3}\Bigr]
+ \frac{370232}{3} \Bigl[3 \zeta_6 f_{3,5} - 2 \zeta_8 f_{3,3}\Bigr]
\nn\\ &&\hskip-0.5cm\null
- 11264 \Bigl[5 f_{3,3,3,5} - 2 \zeta_2 f_{3,3,3,3}
             +\frac{5611}{132} \zeta_8 f_{3,3}\Bigr]
- 65056 \Bigl[7 f_{5,9} - 6 \zeta_4 f_{5,5}\Bigr]
\nn\\ &&\hskip-0.5cm\null
- 146176 \Bigl[5 f_{5,9} - 3 \zeta_6 f_{5,3}\Bigr]
+ 73536 \Bigl[\zeta_2 f_{5,7} -\zeta_6 f_{5,3}\Bigr]
\nn\\ &&\hskip-0.5cm\null
- 130304 \Bigl[7 f_{7,7} - \zeta_2 f_{7,5} - 3 \zeta_4 f_{7,3}\Bigr]
\nn\\ &&\hskip-0.5cm\null
- \frac{331136}{3} \Bigl[5 f_{9,5} - 2 \zeta_2 f_{9,3}\Bigr]
- \frac{42437345879}{30240} \zeta_{14}
\label{EZMHVfg8uu}\\
&=&
- 197536 \zeta_{9,5} + 693536 \zeta_{11,3} 
- 11264 \zeta_{5,3,3,3} + 11264 \zeta_3 \zeta_{5,3,3}
- 5632 (\zeta_3)^2 \zeta_{5,3}
\nn\\ &&\hskip-0.5cm\null
+ 2976704 (\zeta_7)^2 + \frac{8868512}{3} \zeta_5 \zeta_9
- 2246816 \zeta_3 \zeta_{11}
- \frac{28160}{3} (\zeta_3)^3 \zeta_5
\nn\\ &&\hskip-0.5cm\null
- \zeta_2 \Bigl[ \frac{394688}{9} \zeta_{9,3} + \frac{568768}{3} \zeta_5 \zeta_7
  - \frac{1672000}{3} \zeta_3 \zeta_9
  - \frac{2816}{3} (\zeta_3)^4 \Bigr]
\nn\\ &&\hskip-0.5cm\null
- \zeta_4 ( 2160 \zeta_{7,3} - 188688 (\zeta_5)^2 - 394464 \zeta_3 \zeta_7 )
- \zeta_6 ( 4584 \zeta_{5,3} - 342072 \zeta_3 \zeta_5 )
\nn\\ &&\hskip-0.5cm\null
+ \frac{517768}{3} \zeta_8 (\zeta_3)^2
- \frac{42437345879}{30240}   \zeta_{14} \,,
\label{EZMHVg8uu}
\eea
and
\bea
\EE^{(8)\, y_u,y_u}(1,1,1) &=&
5593216   \Bigl[f_{3,11} - \frac{20}{33} \zeta_8 f_{3,3}\Bigr]
- \frac{374816}{3}   \Bigl[\zeta_2 f_{3,9} - \zeta_8 f_{3,3}\Bigr]
\nn\\ &&\hskip-0.5cm\null
- 318400   \Bigl[3 \zeta_4 f_{3,7} - 2 \zeta_8 f_{3,3}\Bigr]
- \frac{993784}{3}   \Bigl[3 \zeta_6 f_{3,5}-2 \zeta_8 f_{3,3}\Bigr]
\nn\\ &&\hskip-0.5cm\null
  + 24448   \Bigl[5 f_{3,3,3,5} - 2 \zeta_2 f_{3,3,3,3}
                 + \frac{5611}{132} \zeta_8 f_{3,3}\Bigr]
+ 208292 \Bigl[7 f_{5,9} - 6 \zeta_4 f_{5,5}\Bigr]
\nn\\ &&\hskip-0.5cm\null
+ 484724   \Bigl[5 f_{5,9} - 3 \zeta_6 f_{5,3}\Bigr]
- 235488   \Bigl[\zeta_2 f_{5,7} - \zeta_6 f_{5,3}\Bigr]
\nn\\ &&\hskip-0.5cm\null
+ 483712   \Bigl[7 f_{7,7} - \zeta_2 f_{7,5} - 3 \zeta_4 f_{7,3}\Bigr]
\nn\\ &&\hskip-0.5cm\null
+ \frac{1304416}{3}   \Bigl[5 f_{9,5} - 2 \zeta_2 f_{9,3}\Bigr]
- \frac{323971645187}{30240} \zeta_{14} \label{EZMHVfg8yuyu}\\
&=&
430784 \zeta_{9,5} - 1496656 \zeta_{11,3}
+ 24448 \zeta_{5,3,3,3} - 24448 \zeta_3 \zeta_{5,3,3}
+ 12224 (\zeta_3)^2 \zeta_{5,3}
\nn\\ &&\hskip-0.5cm\null
- 5677624 (\zeta_7)^2 - \frac{15172576}{3} \zeta_5 \zeta_9
+ 5593216 \zeta_3 \zeta_{11}
+ \frac{61120}{3} (\zeta_3)^3 \zeta_5
\nn\\ &&\hskip-0.5cm\null
+ \zeta_2 \Bigl[ \frac{981664}{9} \zeta_{9,3} + \frac{1256864}{3} \zeta_5 \zeta_7
       - \frac{3675296}{3} \zeta_3 \zeta_9 - \frac{6112}{3} (\zeta_3)^4 \Bigr]
\nn\\ &&\hskip-0.5cm\null
+ \zeta_4 ( 35424 \zeta_{7,3} - 518604 (\zeta_5)^2 - 1028544 \zeta_3 \zeta_7 )
+ \zeta_6 ( 57204 \zeta_{5,3} - 932664 \zeta_3 \zeta_5 )
\nn\\ &&\hskip-0.5cm\null
- \frac{1389512}{3}   \zeta_8 (\zeta_3)^2
- \frac{323971645187}{30240} \zeta_{14} \,. \label{EZMHVg8yuyu}
\eea
The weight 14 values (\ref{EZMHVfg8uu}) and (\ref{EZMHVfg8yuyu})
also obey the coaction principle.  In the ancillary file
{\tt EZMHVcoproducts111.txt} we give the double coproducts in the
alphabet ${\cal L}^a_\text{hex}$.

Finally we discuss the values of the MHV
triple coproducts at $(1,1,1)$. Empirically, they obey
the following relations:
\bea
\EE^{\uh,\uh,\uh}(1,1,1) &=& \EE^{\vh,\vh,\uh}(1,1,1) = - \EE^{\vh,\wh,\uh}(1,1,1)
= \EE^{y_u,y_u,\uh}(1,1,1) = \EE^{y_u,y_v,\uh}(1,1,1) \nn\\
&=& \EE^{y_v,y_u,\uh}(1,1,1) = \EE^{y_v,y_v,\uh}(1,1,1) = \EE^{y_v,y_w,\uh}(1,1,1)
= \EE^{y_u,\uh,y_u}(1,1,1) \nn\\
&=& \EE^{y_v,\uh,y_u}(1,1,1) = \EE^{\uh,y_u,y_u}(1,1,1) = \EE^{\uh,y_v,y_u}(1,1,1),
\label{MHVtriplerelations}
\eea
where we omitted giving: values with $1-\uh$ in the final entry,
which are of course related by $\EE^{1-\uh}=-\EE^{\uh}$;
values related by the total dihedral symmetry of the MHV amplitude;
and vanishing values.  Thus, somewhat remarkably,
all MHV triple coproducts are either $0$ or $\pm \EE^{\uh,\uh,\uh}(1,1,1)$.
The values of $\EE^{\uh,\uh,\uh}(1,1,1)$
through seven loops are given in ref.~\cite{DDToAppear}.
The eight loop value is:
\bea
\EE^{(8)\, \uh,\uh,\uh}(1,1,1) &=&
  6672   \Bigl[ 5 f_{3,5,5} - 2 \zeta_2 f_{3,5,3} \Bigr]
+ 6672   \Bigl[ 5 f_{5,3,5} - 2 \zeta_2 f_{5,3,3} \Bigr]
\nn\\ &&\hskip-0.5cm\null
+ 9824   \Bigl[ 7 f_{3,3,7} - \zeta_2 f_{3,3,5} - 3 \zeta_4 f_{3,3,3} \Bigr]
+ 1992948 f_{13}
+ 25808 \zeta_2 f_{11}
\nn\\ &&\hskip-0.5cm\null
- 267956 \zeta_4 f_{9} - 284096 \zeta_6 f_{7}
- \frac{1728307}{6} \zeta_8 f_{5}
- \frac{6520163}{20} \zeta_{10} f_{3} \label{EZMHVfg8uuu}\\
&=&
4560 \zeta_{5,5,3}
- 4912 \zeta_{7,3,3} + 4912 \zeta_3 \zeta_{7,3} + 1992948 \zeta_{13}
+ 31416 \zeta_3 (\zeta_5)^2
\nn\\ &&\hskip-0.5cm\null
+ 34384 (\zeta_3)^2 \zeta_7
- \zeta_2   ( 704   \zeta_{5,3,3} - 704 \zeta_3 \zeta_{5,3}
+ 346784 \zeta_{11} + 4912 (\zeta_3)^2 \zeta_5 )
\nn\\ &&\hskip-0.5cm\null
- \zeta_4 ( 370692 \zeta_9 + 4912 (\zeta_3)^3 )
- 268144 \zeta_6 \zeta_7 
- \frac{1710707}{6} \zeta_8 \zeta_5
\nn\\ &&\hskip-0.5cm\null
- \frac{6520163}{20} \zeta_{10} \zeta_3 
\,. \label{EZMHVg8uuu}
\eea
The value $\EE^{(8)\, \ah,\ah,\dh}(1,1,1)$
in the alphabet ${\cal L}^a_\text{hex}$
is given in {\tt EZMHVcoproducts111.txt};
they are related by
$\EE^{(8)\, \uh,\uh,\uh}(1,1,1) = -2 \, \EE^{(8)\, \ah,\ah,\dh}(1,1,1)$

The values of the 166 (mostly) linearly independent
weight 12 quadruple coproducts at $(1,1,1)$ are also provided
in the ancillary file
{\tt EZMHVcoproducts111.txt}. We find that there are 5 independent
values, which span the basis~(\ref{wt12new}), indicating that there
is a weight 12 dropout, as discussed in \sect{subsec:coaction111}.


\section{Weight 16 \texorpdfstring{$Z$}{Z} functions}
\label{sec:wt16Zfunctions}

In the context of using antipodal duality to determine the MHV
amplitude, it is of general interest to know how many independent
parity-even hexagon functions vanish identically on the parity-preserving
surface, and what other properties they might have.
All the examples we have found so far, through weight 16,
have the property that their parity-even first coproducts $Z^{e_i}$
vanish identically, in the entire $(\uh,\vh,\wh)$ space.
We call such functions $Z$ functions.
They are completely characterized by their odd first coproducts, $Z^{y_i}$.
They only seem to appear at even weight.
At weight 12, there is a single $Z$ function, mentioned
in ref.~\cite{Caron-Huot:2019vjl}.  At weight 14, there is a triplet
of such functions~\footnote{We thank \"Omer G\"urdo\u{g}an for
discussions on this subject.}, which permute into each other under
cyclic symmetry, and whose cyclic sum is the function $\tilde{Z}$
mentioned in ref.~\cite{Caron-Huot:2019vjl}.

At weight 16 we found 9 such functions at symbol level.
They form 3 separate triplets under the dihedral symmetry.
(One linear combination of the 3 dihedrally-invariant symbols
vanishes at the origin, at symbol level.)
All the $Z$ functions obey the same final entry conditions
as the MHV amplitude $\EE$, and addition relations that follow
from setting $Z^{e_i} \to 0$, see Table~\ref{tab:countfinalentries}
for the number of independent multi-final entries.
We integrated up all of the weight 12, 14 and 16 $Z$ functions
from symbol-level to functions.  We could do this uniquely,
up to the existence of one beyond-the-symbol parameter at weight 16,
which is none other than the weight 12 $Z$ function multiplied
by the zeta value $\zeta_4$, which has a free coefficient in $\Hhex$.
(On the other hand, $\zeta_2$ and $\zeta_3$ do {\it not} have
free coefficients, and the next free one is $\zeta_6$.)

\begin{table}[!t]
\begin{center}
\begin{tabular}{|c|c|c|c|c|c|}
\hline\hline
function & singles & doubles & triples & quadruples & quintuples \\
\hline\hline
${\cal E}$ & 6 & 21 & 62 & 166 & 424 \\\hline
$Z$        & 3 &  6 & 14 &  31 &  70 \\
\hline\hline
\end{tabular}
\caption{\label{tab:countfinalentries} The number of independent
multi-final entries for the MHV amplitude ${\cal E}$, compared with
those for the $Z$ functions. The latter are a subspace of the former.}
\end{center}
\end{table}

\subsection{Behavior at \texorpdfstring{$(1,0,0)$}{(1,0,0)}}
\label{sec:Z100}

In ref.~\cite{Caron-Huot:2019bsq} it was mentioned that the space $\Hhex$
constructed there was slightly over-complete, starting with the
parity-even weight 8 functions, where 3 functions should be removed
because they lead to irreducible MZVs, $\zeta_{5,3}$ in this case,
at the points $(1,0,0)$, $(0,1,0)$ and $(0,0,1)$.  If the space $\Hhex$
is defined to be the minimal space containing all the coproducts
of the MHV and NMHV amplitudes to all loop orders, then these
3 functions should not be in the space.  That's because the
OPE approach~\cite{Basso:2013vsa} implies that these limits
can only contain Riemann zeta values, not irreducible MZVs such as
$\zeta_{5,3}$. It is worth examining the multiple coproducts of the
$Z$ functions at $(1,0,0)$, etc., to see if any of these ambiguity functions
can be discarded based on this function-level information.

In fact, we find that all the independent single, double, and triple
coproducts of the $Z$ functions at weights 12, 14 and 16 vanish completely
at $(\uh,\vh,\wh)=(1,0,0)$ (and at its cyclic images)!
As shown in Table~\ref{tab:countfinalentries}, there are 31 independent
quadruple final entries, 15 parity-even and 16 parity-odd.
The 16 odd quadruples have to vanish at $(1,0,0)$, but so do 14 of the 15 even
ones.  The 15 even quadruples can be taken to be
\bea
&&[\dh,y_v,\eh,y_v],\ [\dh,y_v,\fh,y_w],\ [\fh,y_w,\eh,y_v],\
[y_u,\eh,\fh,y_w],\ [y_v,\ch,\eh,y_v],\
\nonumber\\
&&[y_v,\dh,\eh,y_v],\ [y_v,\eh,\eh,y_v],\ [y_v,\eh,\fh,y_w],\
[y_u,y_v,y_w,y_w],\ [y_u,y_w,y_v,y_v],\
\nonumber\\
&&[y_u,y_w,y_w,y_w],\ [y_v,y_v,y_w,y_w],\ [y_v,y_w,y_v,y_v],\
[y_v,y_w,y_w,y_w],\ [y_w,y_w,y_u,y_u].~~~~
\label{Wquadindepeven15}
\eea
Then the only nonvanishing quadruple at $(1,0,0)$ is $[\fh,y_w,\eh,y_v]$,
and this is true for the weight 12, 14 and 16 $Z$ functions.

For example, the relevant quadruple coproduct of the
suitably normalized weight 12 $Z$ function is
\bea
Z_{\text{wt.~12}}^{\fh,y_w,\eh,y_v}(1,0,0) &=& - \frac{1}{576} \ln^4\vh \ln^4\wh
- \frac{\zeta_2}{48} \ln^2\vh \ln^2\wh ( \ln^2\vh + \ln^2\wh )
\nonumber\\
&&\null
+ \frac{5}{96} \zeta_4 ( \ln^4\vh + \ln^4\wh - 12 \ln^2\vh \ln^2\wh )
+\Bigl[ \frac{35}{32} \zeta_6 - (\zeta_3)^2 \Bigr] ( \ln^2\vh + \ln^2\wh )
\nonumber\\
&&\null
- \frac{175}{96} \zeta_8 + 8 \zeta_3 \zeta_5 - 4 \zeta_2 (\zeta_3)^2 \,,
\label{J_mw_yw_mv_yv}
\eea
which contains no irreducible MZV; i.e.~no $\zeta_{5,3}$.
The quadruples $Z_{\text{wt.~14}}^{\fh,y_w,\eh,y_v}$ and
$Z_{\text{wt.~16}}^{\fh,y_w,\eh,y_v}$ also contain no irreducible MZVs
at weight 10 and 12, respectively. (Lower-weight irreducible MZVs could also
appear in principle, multiplied by $\ln\vh$ and/or $\ln\wh$, but they do not.)

Similarly, very few of the 70 independent quintuples are nonvanishing
at $(1,0,0)$.  With the basis we use, only four quintuples
are nonvanishing, and three of them are always identical to each other:
\be
Z^{\eh,y_w,y_w,y_u,y_u} \,; \qquad
Z^{\fh,\dh,y_v,\fh,y_w} = Z^{\fh,y_v,\dh,\eh,y_v} = Z^{\fh,y_v,\eh,\fh,y_w}
\label{nonvanishingWquints}
\ee
None of these quantities contain irreducible MZVs either.
We worked out the hextuples at $(1,0,0)$ as well, for the
weight 14 and weight 16 $Z$ functions for which they have
weights 8 and 10, respectively, and we found no irreducible MZVs.
Since there are no irreducible MZVs below weight 8, that completes the
search for the weight 12 and 14 $Z$ functions,
and makes it implausible that there are any for weight 16 either.
The conclusion is that the $Z$ functions seem to be genuine function-level
ambiguities in $\Hhex$, for lifting off of the $\Delta=0$ surface.

\subsection{Behavior at the origin}
\label{sec:Z000}

As mentioned earlier, one of the dihedrally symmetric weight 16 $Z$
functions vanishes at the origin at symbol level.  However,
it is non-vanishing at function level.  In fact all of the weight 16
$Z$ function ambiguities can be fixed using only coefficients of $\ln^k u$ for
$k=4,6,8$. This statement is consistent with our earlier analysis,
where the higher-weight constants were all fixed by consistency between
the $(1,\vh,\vh)$ and $(\uh,0,0)$ lines, before ever going to the origin.
Again, it means that the 8-loop value at the origin~(\ref{R8origin})
is a pure cross check.  Interestingly, the maximal degree of the
$Z$ functions in any individual $\ln \uh_i$ at the origin is equal to the
weight minus 8;  i.e.~degree 4 at weight 12, degree 6 at weight 14,
and degree 8 at weight 16.
In fact, on the entire line $(\uh,0,0)$, the $Z$ functions
have this same maximal degree in $\ln\vh$ and in $\ln\wh$.
In the ancillary file {\tt Zorigin.txt}, we give the behavior at
the origin for the 9 true weight 16 $Z$ functions, as well as the
tenth which is $\z{4}$ times the weight 12 $Z$ function.

\subsection{Fixing the last weight 11 constants}
\label{sec:wt11constants}

Refs.~\cite{Caron-Huot:2019vjl,Caron-Huot:2019bsq} gave a description of
hexagon functions through weight 11.  At weight 11, however, there were six
missing constants, $n_i$, $i=1,2,\ldots,6$,
associated with the constant values of the hexagon functions
at the reference point $(\uh,\vh,\wh)=(1,1,1)$ that were provided in
the ancillary file {\tt SixGluonAmpsAndCops}.
The six $n_i$ were expected to be integers, but there was not enough
information to fix them all at that stage.

In the process of integrating up the full eight-loop MHV function,
consistency conditions on the integration completely fix all but one of
these constants.  (In contrast, the seven-loop NMHV
amplitude~\cite{DDToAppear}
still leaves some of the six constants unfixed.)
The final constant can be fixed by examining the weight 16
ambiguity functions and requiring the following branch-cut condition
to hold,
\be
  Z_{\text{wt.~16}}^{y_v,x_1,x_2,x_3,x_4}(1,0,0)
  = Z_{\text{wt.~16}}^{y_w,x_1,x_2,x_3,x_4}(1,0,0) = 0,
\label{ambigvanishcondition}  
\ee
for any letters $x_i$.
The resulting values of the $n_i$ are all indeed integers,
\bea
n_1 &=& -337920, \quad n_2 = -16896, \quad n_3 = -608256,
\nonumber\\
n_4 &=& 236544, \quad n_5 = -1317888, \quad n_6 = 9934848.
\label{fixni}
\eea
This result completes the specification of the hexagon function space
$\Hhex$ through weight 11.  As seen in section~\ref{sec:copparitylocking},
we now know that a bit fewer functions are actually required, but
at least their behavior is now totally determined.


\section{The lines \texorpdfstring{$(\uh,\uh,1)$}{(\uh,\uh,1)} and
    \texorpdfstring{$(\uh,1,1)$}{(\uh,1,1)}}
\label{sec:lines}

The $L$-loop MHV amplitudes
are very complicated for general $(\uh,\vh,\wh)$, but they simplify
drastically on two lines, where we will plot them.  
On the lines $(\uh,\vh,\wh)=(\uh,\uh,1)$ and $(\uh,\vh,\wh)=(\uh,1,1)$,
the hexagon symbol alphabet
${\cal L}_\text{hex}$ collapses to just two letters, $\{\uh,1-\uh\}$.
All hexagon functions become HPLs~\cite{Remiddi:1999ew} $H_{\vec{a}}(1-\uh)$,
with $a_i \in \{0,1\}$. By dihedral symmetry, it is enough to provide and
plot $\EE(\uh,\uh,1)$ and $\EE(\uh,1,1)$.  In the ancillary files
\texttt{EZMHVg\_uu1\_lin.txt} and \texttt{EZMHVg\_u11\_lin.txt},
we provide these functions through eight loops,
in four different linearized HPL representations, which can be used
to series expand around the points $\uh=0,1,\infty$.
(Results in the old normalization through seven loops
were provided in the ancillary file \texttt{SixGluonHPLLines.m}
for refs.~\cite{Caron-Huot:2019vjl,Caron-Huot:2019bsq}.)
These series expansions have overlapping regions of convergence,
which makes it possible to plot the amplitudes on the full lines.

Before moving outward on the two lines, we discuss the behavior
at two points.  The first point is the Euclidean base point $(1,1,1)$
which lies at the intersection of the two lines.  The analytical
value is given in \eqn{EZMHVg8_111}, and the \blue{blue} coefficients
are predicted by antipodal duality from the eight-loop form factor,
by reversing the order of the indices of the $f$'s.
The numerical value of the eight-loop amplitude
at that point is
\be
\EE^{(8)}(1,1,1)\ =\
-47748904.85576496624997891660892743663\ldots
\label{EZMHVg8_111_num}
\ee
In Table~\ref{tab:MHV111numerics8loops} we give the numerical
values of $\EE^{(L)}(1,1,1)$ through $L=8$ loops, as well as ratios of
successive loop orders, which appear to tend toward the ratio $-16$
obeyed by the cusp anomalous dimension~\cite{Beisert:2006ez},
$\Gamma_{\rm cusp}^{(L)}/\Gamma_{\rm cusp}^{(L-1)} \to -16$ as $L\to\infty$.
In Table~\ref{tab:RV111numerics8loops} we provide the values of the
successive loop-order ratios for the remainder function
at $(1,1,1)$.  This ratio approaches $-16$
somewhat faster than the one for $\EE$.

\begin{table}[!t]
\begin{center}
\begin{tabular}{|l|c|c|}
\hline\hline
$L$ & $\EE^{(L)}(1,1,1)$
    & $\EE^{(L)}(1,1,1)/\EE^{(L-1)}(1,1,1)$ \\
\hline\hline
1  &     0             & -- \\
2  &   $-$9.740909108  & -- \\
3  &      123.0985106  & $-$12.63727124 \\
4  &   $-$1508.319856  & $-$12.25294968 \\
5  &      19196.41479  & $-$12.72701855 \\
6  &   $-$253379.3991  & $-$13.19930841 \\
7  &      3440841.652  & $-$13.57980035 \\
8  &   $-$47748904.85  & $-$13.87710034 \\
\hline\hline
\end{tabular}
\caption{\label{tab:MHV111numerics8loops} The value of the $L$-loop MHV
amplitude at $(\uh,\vh,\wh)=(1,1,1)$ through eight loops,
in the new cosmic normalization, and the ratio to the previous loop order.}
\end{center}
\end{table}

\begin{table}[!t]
\begin{center}
\begin{tabular}{|l|c|c|}
\hline\hline
$L$ & $\RR^{(L)}(1,1,1)$ & $\RR^{(L)}(1,1,1)/\RR^{(L-1)}(1,1,1)$ \\
\hline\hline
2  & $-$10.823232337111 &  --              \\
3  &    151.61375458732 & $-$14.00817702   \\
4  & $-$1997.6785778253 & $-$13.17610386   \\
5  &    26805.799395885 & $-$13.41847467   \\
6  & $-$368506.63803885 & $-$13.74727284   \\
7  &    5169794.4229268 & $-$14.02904015   \\
8  & $-$73696028.745912 & $-$14.25511784   \\
\hline\hline
\end{tabular}
\caption{\label{tab:RV111numerics8loops} MHV remainder function
and successive loop order ratios at $(1,1,1)$ through eight loops.}
\end{center}
\end{table}

The second point we discuss is the limit as $\uh\to\infty$
along the line $(\uh,\uh,1)$, or $(\infty,\infty,1)$ for short.
Values at this point, like at $(1,1,1)$
are MZVs, and antipodal duality relates this point to
the form factor at the point $(u,v)=(1,+\infty)$,
i.e.~the limit $v\to\infty$ of the line $u=1$.
The value is
\bea 
\EE^{(8)}(\infty,\infty,1) &=&
-\blue{\frac{5259063125}{18}} \, f_{9,7}
- \blue{\frac{40774437223}{144}} \, f_{7,9}
- \blue{\frac{49527178109}{120}} \, f_{11,5}
\nn\\ &&\null\hskip0.0cm
- \blue{\frac{76882560437}{240}} \, f_{5,11}
- \blue{\frac{589315976143}{1680}} \, f_{13,3}
- \blue{\frac{3806228668547}{16800}} \, f_{3,13}
\nn\\ &&\null\hskip0.0cm
- \blue{19525796} \, f_{5,5,3,3} - \blue{19216724} \, f_{5,3,5,3}
- \blue{21675228} \, f_{5,3,3,5} - \blue{13812532} \, f_{3,5,5,3}
\nn\\ &&\null\hskip0.0cm
- \blue{15382068} \, f_{3,5,3,5} - \blue{15703712} \, f_{3,3,5,5}
- \blue{20001939} \, f_{7,3,3,3} - \blue{12334055} \, f_{3,7,3,3}
\nn\\ &&\null\hskip0.0cm
- \blue{12638203} \, f_{3,3,7,3} - \blue{13916245} \, f_{3,3,3,7}
\nn
\eea
\bea
&&\null
- \z{2} \Bigl( \frac{5163836129}{16} f_{7,7}
         + \frac{2956391617}{9} f_{9,5} + \frac{5638517153}{18} f_{5,9}
         + \frac{49275366491}{120} f_{11,3}
\nn\\ &&\null\hskip0.7cm
         + \frac{54190442797}{240} f_{3,11}
         + 21672656 f_{5,3,3,3} + 14950352 f_{3,5,3,3}
\nn\\ &&\null\hskip0.7cm
	 + 15412504 f_{3,3,5,3} + 15167512 f_{3,3,3,5} \Bigr)
\nn\\ &&\null\hskip0.0cm
- \z{4} \Bigl( 417940090 f_{7,5} + \frac{799758063}{2} f_{5,7}
             + 384992568 f_{9,3} + \frac{5187282665}{18} f_{3,9}
\nn\\ &&\null\hskip0.7cm
             + 21691976 f_{3,3,3,3} \Bigr)
\nn\\ &&\null\hskip0.0cm
- \z{6} \Bigl( \frac{1608784821}{4} f_{5,5}
             + \frac{9165154933}{24} f_{7,3}
             + \frac{1126390233}{4} f_{3,7} \Bigr)
\nn\\ &&\null\hskip0.0cm
- \z{8} \Bigl( \frac{11463894607}{30} f_{5,3}
             + \frac{53213485753}{180} f_{3,5} \Bigr)
- \frac{50699228086}{175} \z{10} f_{3,3}
\nn\\ &&\null\hskip0.0cm
- \frac{537331216774792288437829}{82584825523200} \z{16} \,.
\label{EZMHVfg8_ii1}
\eea
The coefficients in \blue{blue} agree perfectly with the predictions of
antipodal duality from the eight-loop form factor.

Note that \emph{all} coefficients entering
$\EE^{(8)}(\infty,\infty,1)$ are negative, whereas for $\EE^{(8)}(1,1,1)$
in \eqn{EZMHVfg8_111} they are all positive for the pure $f$ terms,
but have opposite sign for all $\zeta_{2n}$ ($\pi^{2n}$) terms except one.
Remarkably,
these sign patterns are also true at all lower loops (with no exceptions),
after multiplying by an overall sign for odd loops.
(See the ancillary file {\tt AntipodePointsSummary.txt} for
ref.~\cite{Dixon:2021tdw} for the lower loop formulae.)

\begin{figure}[t]
\centering
 \begin{subfigure}[b]{0.49\textwidth}
         \centering
         \includegraphics[height=0.24\paperwidth]{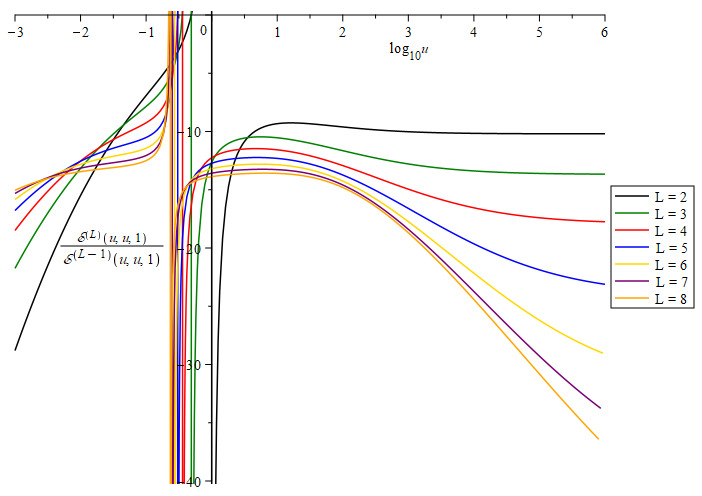}
         \caption{\phantom{.}}
         \label{fig:eeuu1}
     \end{subfigure}
     \hfill
      \begin{subfigure}[b]{0.49\textwidth}
         \centering
         \includegraphics[height=0.24\paperwidth]{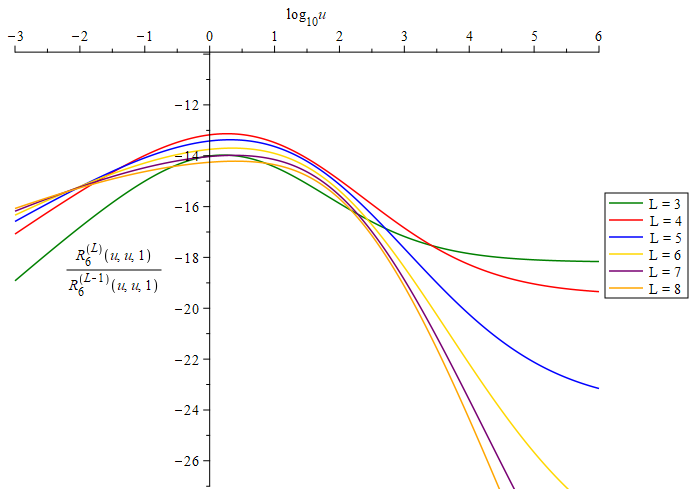}
         \caption{\phantom{.}}
         \label{fig:ruu1}
     \end{subfigure}
\caption{(a) $\EE^{(L)}(\uh,\uh,1)/\EE^{(L-1)}(\uh,\uh,1)$
  evaluated at successive loop orders $L-1$ and $L$.
  As there are points where $\EE^{(L)}(\uh,\uh,1)=0$ in this interval,
  the plot has many dip/spike features.
  (b) $\RR^{(L)}(\uh,\uh,1)/\RR^{(L-1)}(\uh,\uh,1)$ evaluated at successive
  loop orders.}
\label{fig:eeruu1}
\end{figure}

Next, we plot $\EE(\uh,\uh,1)$ in figure~\ref{fig:eeruu1}(a) 
through eight loops, as ratios of successive
loop orders.
For $\uh$ between $10^{-2}$ and $10^2$, remarkably, the ratios flatten out
more and more with each additional loop, and they appear to be steadily
approaching the cusp asymptotic ratio of $-16$.
There is a dip/spike feature in all the plots, simply because each function
crosses zero at a slightly different value of $\uh$.
For $\uh\rightarrow 0$ and $\uh\rightarrow\infty$, the ratios no longer
display the expected radius of convergence, either diverging logarithmically
at different rates (for $\uh\to0$)
or approaching constant values (as $\uh \to \infty$) that do not have
the same ratio of $-16$ between loop orders.

We plot the remainder function ratios
$\RR^{(L)}(\uh,\uh,1)/\RR^{(L-1)}(\uh,\uh,1)$
on the same line in figure~\ref{fig:eeruu1}(b).
These ratios do not exhibit the same degree
of flattening in $\uh$ as the ratios
$\EE^{(L)}(\uh,\uh,1)/\EE^{(L-1)}(\uh,\uh,1)$.

\begin{figure}[t]
\centering
 \begin{subfigure}[b]{0.46\textwidth}
         \centering
         \includegraphics[height=0.22\paperwidth]{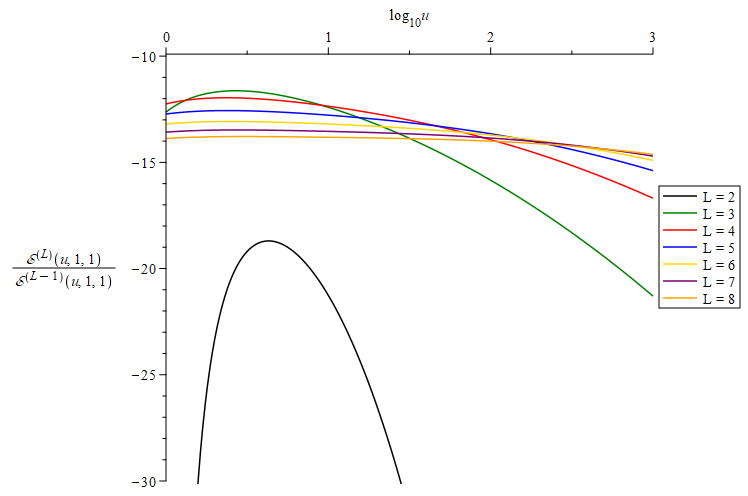}
         \caption{\phantom{.}}
         \label{fig:eeu11}
     \end{subfigure}
     \hfill
      \begin{subfigure}[b]{0.46\textwidth}
         \centering
         \includegraphics[height=0.22\paperwidth]{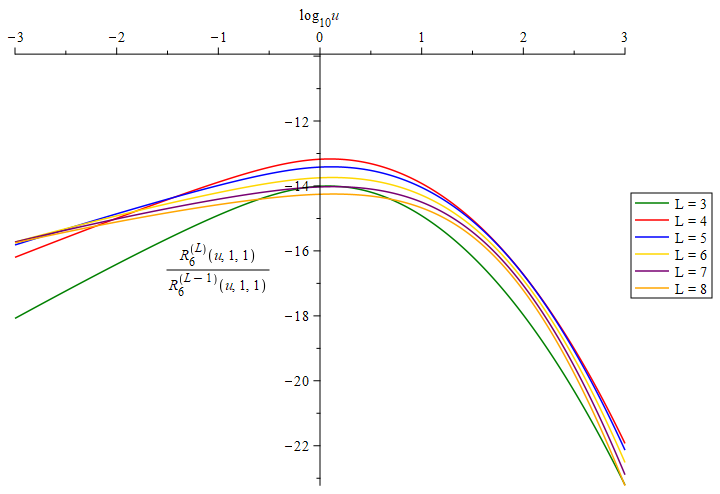}
         \caption{\phantom{.}}
         \label{fig:ru11}
     \end{subfigure}
\caption{(a) The ratio $\EE^{(L)}(\uh,1,1)/\EE^{(L-1)}(\uh,1,1)$
  evaluated at successive loop orders, $L-1$ and $L$.
  (b) $\RR^{(L)}(\uh,1,1)/\RR^{(L-1)}(\uh,1,1)$ evaluated at successive
  loop orders.}
\label{fig:eeru11}
\end{figure}

Now we turn to the line $(\uh,1,1)$, which leaves the $\Delta=0$ surface.
We plot $\EE(\uh,1,1)$ in figure~\ref{fig:eeru11}(a) through eight loops,
as ratios between successive loop orders, for $1 \leq \uh \leq 1000$.
There is a pretty striking flattening of the ratios for $1 \leq \uh \leq 100$,
again generally approaching the radius of convergence suggested by the cusp
anomalous dimension.  In figure~\ref{fig:eeru11}(b) the same ratio is plotted
for the remainder function; as was the case for the line $(\uh,\uh,1)$,
the flattening is much less pronounced for the remainder function.

\section{Conclusions}
\label{sec:conclusions}

In this paper, we computed the eight-loop MHV six-particle amplitude
in planar ${\cal N}=4$ SYM, using the novel approach of determining
it by assuming antipodal duality, and relying on the
recently-computed~\cite{Dixon:2022rse}
eight-loop three-point form factor of the chiral stress-energy tensor.
This approach is also a bootstrap, making heavy use of the hexagon function
space~\cite{Caron-Huot:2019vjl,Caron-Huot:2019bsq}, but the number
of linear equations that have to be solved is far fewer with
the form factor information.  The number of equations that have to be solved
at any one time could also be minimized by grading the quintuple
final entries in the number of parity-odd $y_i$ letters they contain.
The amount of OPE information required was truly minimal, just enough
to fix one beyond-the-symbol constant.

The fully fixed amplitude is in turn a rich source of information
about the hexagon-function space, revealing additional final-entry
relations, both in the bulk and at the base point $(1,1,1)$,
as well as functions and constants that ``drop out'', i.e.~are not needed,
particularly when the new cosmic normalization of the amplitude is
employed.  That there is a solution at all, and some of its
beyond-the-symbol properties, is a validation of antipodal duality
at eight loops.

The general idea of first fixing an amplitude or form factor on a suitable
lower-dimensional surface, where the symbol alphabet simplifies,
and then lifting it off the surface to full kinematics,
should be applicable more broadly, even when antipodal duality
is not there as a crutch.  Parity-preserving surfaces also exist
for higher-point amplitudes and form factors, and would be a natural
place to try out such methods in the future.

\vskip0.5cm
\noindent {\large\bf Acknowledgments}
\vskip0.3cm

\noindent We are grateful to \"Omer G\"urdo\u{g}an, Andrew McLeod and 
Matthias Wilhelm for collaboration on earlier related projects
and for stimulating discussions.
We also thank Benjamin Basso for useful conversations.
This research was supported by the US Department of Energy under contracts
DE--AC02--76SF00515 and DE--FOA--0002705, KA/OR55/22 (AIHEP),
and by the Munich Institute for Astro-, Particle and BioPhysics (MIAPbP)
which is funded by the Deutsche
Forschungsgemeinschaft (DFG, German Research Foundation)
under Germany's Excellence Strategy – EXC-2094 – 390783311.
LD thanks MIAPbP for hospitality during part of the writing of
this paper.


\noindent

\bibliographystyle{JHEP}

\providecommand{\href}[2]{#2}\begingroup\raggedright\endgroup


\begin{thebibliography}{10}

\bibitem{Chen}
K.-T. Chen, \emph{Iterated path integrals}, {\emph{Bull. Amer. Math. Soc.}
  {\bfseries 83} (1977) 831}.

\bibitem{G91b}
A.~B. Goncharov, \emph{Geometry of configurations, polylogarithms, and motivic
  cohomology},
  \href{https://doi.org/http://dx.doi.org/10.1006/aima.1995.1045}{\emph{Adv.
  Math.} {\bfseries 114} (1995) 197}.

\bibitem{Goncharov:1998kja}
A.~B. Goncharov, \emph{{Multiple polylogarithms, cyclotomy and modular
  complexes}}, \href{https://doi.org/10.4310/MRL.1998.v5.n4.a7}{\emph{Math.
  Res. Lett.} {\bfseries 5} (1998) 497}
  [\href{https://arxiv.org/abs/1105.2076}{{\ttfamily 1105.2076}}].

\bibitem{Remiddi:1999ew}
E.~Remiddi and J.~Vermaseren, \emph{{Harmonic polylogarithms}},
  \href{https://doi.org/10.1142/S0217751X00000367}{\emph{Int. J. Mod. Phys. A}
  {\bfseries 15} (2000) 725}
  [\href{https://arxiv.org/abs/hep-ph/9905237}{{\ttfamily hep-ph/9905237}}].

\bibitem{Borwein:1999js}
J.~M. Borwein, D.~M. Bradley, D.~J. Broadhurst and P.~Lisonek, \emph{{Special
  values of multiple polylogarithms}},
  \href{https://doi.org/10.1090/S0002-9947-00-02616-7}{\emph{Trans. Am. Math.
  Soc.} {\bfseries 353} (2001) 907}
  [\href{https://arxiv.org/abs/math/9910045}{{\ttfamily math/9910045}}].

\bibitem{Moch:2001zr}
S.~Moch, P.~Uwer and S.~Weinzierl, \emph{{Nested sums, expansion of
  transcendental functions and multiscale multiloop integrals}},
  \href{https://doi.org/10.1063/1.1471366}{\emph{J.Math.Phys.} {\bfseries 43}
  (2002) 3363} [\href{https://arxiv.org/abs/hep-ph/0110083}{{\ttfamily
  hep-ph/0110083}}].

\bibitem{Goncharov:2010jf}
A.~B. Goncharov, M.~Spradlin, C.~Vergu and A.~Volovich, \emph{{Classical
  Polylogarithms for Amplitudes and Wilson Loops}},
  \href{https://doi.org/10.1103/PhysRevLett.105.151605}{\emph{Phys. Rev. Lett.}
  {\bfseries 105} (2010) 151605}
  [\href{https://arxiv.org/abs/1006.5703}{{\ttfamily 1006.5703}}].

\bibitem{Drummond:2006rz}
J.~M. Drummond, J.~Henn, V.~A. Smirnov and E.~Sokatchev, \emph{{Magic
  identities for conformal four-point integrals}},
  \href{https://doi.org/10.1088/1126-6708/2007/01/064}{\emph{JHEP} {\bfseries
  01} (2007) 064} [\href{https://arxiv.org/abs/hep-th/0607160}{{\ttfamily
  hep-th/0607160}}].

\bibitem{Bern:2006ew}
Z.~Bern, M.~Czakon, L.~J. Dixon, D.~A. Kosower and V.~A. Smirnov, \emph{{The
  Four-Loop Planar Amplitude and Cusp Anomalous Dimension in Maximally
  Supersymmetric Yang-Mills Theory}},
  \href{https://doi.org/10.1103/PhysRevD.75.085010}{\emph{Phys. Rev.}
  {\bfseries D75} (2007) 085010}
  [\href{https://arxiv.org/abs/hep-th/0610248}{{\ttfamily hep-th/0610248}}].

\bibitem{Bern:2007ct}
Z.~Bern, J.~Carrasco, H.~Johansson and D.~Kosower, \emph{{Maximally
  supersymmetric planar Yang-Mills amplitudes at five loops}},
  \href{https://doi.org/10.1103/PhysRevD.76.125020}{\emph{Phys.Rev.} {\bfseries
  D76} (2007) 125020} [\href{https://arxiv.org/abs/0705.1864}{{\ttfamily
  0705.1864}}].

\bibitem{Alday:2007hr}
L.~F. Alday and J.~M. Maldacena, \emph{{Gluon scattering amplitudes at strong
  coupling}}, \href{https://doi.org/10.1088/1126-6708/2007/06/064}{\emph{JHEP}
  {\bfseries 06} (2007) 064} [\href{https://arxiv.org/abs/0705.0303}{{\ttfamily
  0705.0303}}].

\bibitem{Drummond:2008vq}
J.~M. Drummond, J.~Henn, G.~P. Korchemsky and E.~Sokatchev, \emph{{Dual
  superconformal symmetry of scattering amplitudes in $\mathcal{N}=4$
  super-Yang-Mills theory}},
  \href{https://doi.org/10.1016/j.nuclphysb.2009.11.022}{\emph{Nucl. Phys.}
  {\bfseries B828} (2010) 317}
  [\href{https://arxiv.org/abs/0807.1095}{{\ttfamily 0807.1095}}].

\bibitem{Bern:2008ap}
Z.~Bern, L.~J. Dixon, D.~A. Kosower, R.~Roiban, M.~Spradlin, C.~Vergu et~al.,
  \emph{{The Two-Loop Six-Gluon MHV Amplitude in Maximally Supersymmetric
  Yang-Mills Theory}},
  \href{https://doi.org/10.1103/PhysRevD.78.045007}{\emph{Phys. Rev.}
  {\bfseries D78} (2008) 045007}
  [\href{https://arxiv.org/abs/0803.1465}{{\ttfamily 0803.1465}}].

\bibitem{Drummond:2008aq}
J.~Drummond, J.~Henn, G.~Korchemsky and E.~Sokatchev, \emph{{Hexagon Wilson
  loop = six-gluon MHV amplitude}},
  \href{https://doi.org/10.1016/j.nuclphysb.2009.02.015}{\emph{Nucl.Phys.}
  {\bfseries B815} (2009) 142}
  [\href{https://arxiv.org/abs/0803.1466}{{\ttfamily 0803.1466}}].

\bibitem{DelDuca:2009au}
V.~Del~Duca, C.~Duhr and V.~A. Smirnov, \emph{{An Analytic Result for the
  Two-Loop Hexagon Wilson Loop in N = 4 SYM}},
  \href{https://doi.org/10.1007/JHEP03(2010)099}{\emph{JHEP} {\bfseries 03}
  (2010) 099} [\href{https://arxiv.org/abs/0911.5332}{{\ttfamily 0911.5332}}].

\bibitem{DelDuca:2010zg}
V.~Del~Duca, C.~Duhr and V.~A. Smirnov, \emph{{The Two-Loop Hexagon Wilson Loop
  in N = 4 SYM}}, \href{https://doi.org/10.1007/JHEP05(2010)084}{\emph{JHEP}
  {\bfseries 05} (2010) 084} [\href{https://arxiv.org/abs/1003.1702}{{\ttfamily
  1003.1702}}].

\bibitem{Golden:2013xva}
J.~Golden, A.~B. Goncharov, M.~Spradlin, C.~Vergu and A.~Volovich,
  \emph{{Motivic Amplitudes and Cluster Coordinates}},
  \href{https://doi.org/10.1007/JHEP01(2014)091}{\emph{JHEP} {\bfseries 1401}
  (2014) 091} [\href{https://arxiv.org/abs/1305.1617}{{\ttfamily 1305.1617}}].

\bibitem{Golden:2014pua}
J.~Golden and M.~Spradlin, \emph{{A Cluster Bootstrap for Two-Loop MHV
  Amplitudes}}, \href{https://doi.org/10.1007/JHEP02(2015)002}{\emph{JHEP}
  {\bfseries 02} (2015) 002} [\href{https://arxiv.org/abs/1411.3289}{{\ttfamily
  1411.3289}}].

\bibitem{Dixon:2011pw}
L.~J. Dixon, J.~M. Drummond and J.~M. Henn, \emph{{Bootstrapping the three-loop
  hexagon}}, \href{https://doi.org/10.1007/JHEP11(2011)023}{\emph{JHEP}
  {\bfseries 1111} (2011) 023}
  [\href{https://arxiv.org/abs/1108.4461}{{\ttfamily 1108.4461}}].

\bibitem{Dixon:2011nj}
L.~J. Dixon, J.~M. Drummond and J.~M. Henn, \emph{{Analytic result for the
  two-loop six-point NMHV amplitude in $\mathcal{N}=4$ super Yang-Mills
  theory}}, \href{https://doi.org/10.1007/JHEP01(2012)024}{\emph{JHEP}
  {\bfseries 1201} (2012) 024}
  [\href{https://arxiv.org/abs/1111.1704}{{\ttfamily 1111.1704}}].

\bibitem{Dixon:2013eka}
L.~J. Dixon, J.~M. Drummond, M.~von Hippel and J.~Pennington, \emph{{Hexagon
  functions and the three-loop remainder function}},
  \href{https://doi.org/10.1007/JHEP12(2013)049}{\emph{JHEP} {\bfseries 1312}
  (2013) 049} [\href{https://arxiv.org/abs/1308.2276}{{\ttfamily 1308.2276}}].

\bibitem{Dixon:2014iba}
L.~J. Dixon and M.~von Hippel, \emph{{Bootstrapping an NMHV amplitude through
  three loops}}, \href{https://doi.org/10.1007/JHEP10(2014)065}{\emph{JHEP}
  {\bfseries 1410} (2014) 65}
  [\href{https://arxiv.org/abs/1408.1505}{{\ttfamily 1408.1505}}].

\bibitem{Dixon:2014voa}
L.~J. Dixon, J.~M. Drummond, C.~Duhr and J.~Pennington, \emph{{The four-loop
  remainder function and multi-Regge behavior at NNLLA in planar $\mathcal{N} =
  4$ super-Yang-Mills theory}},
  \href{https://doi.org/10.1007/JHEP06(2014)116}{\emph{JHEP} {\bfseries 1406}
  (2014) 116} [\href{https://arxiv.org/abs/1402.3300}{{\ttfamily 1402.3300}}].

\bibitem{Dixon:2015iva}
L.~J. Dixon, M.~von Hippel and A.~J. McLeod, \emph{{The four-loop six-gluon
  NMHV ratio function}},
  \href{https://doi.org/10.1007/JHEP01(2016)053}{\emph{JHEP} {\bfseries 01}
  (2016) 053} [\href{https://arxiv.org/abs/1509.08127}{{\ttfamily
  1509.08127}}].

\bibitem{Caron-Huot:2016owq}
S.~Caron-Huot, L.~J. Dixon, A.~McLeod and M.~von Hippel, \emph{{Bootstrapping a
  Five-Loop Amplitude Using Steinmann Relations}},
  \href{https://doi.org/10.1103/PhysRevLett.117.241601}{\emph{Phys. Rev. Lett.}
  {\bfseries 117} (2016) 241601}
  [\href{https://arxiv.org/abs/1609.00669}{{\ttfamily 1609.00669}}].

\bibitem{Caron-Huot:2019vjl}
S.~Caron-Huot, L.~J. Dixon, F.~Dulat, M.~von Hippel, A.~J. McLeod and
  G.~Papathanasiou, \emph{{Six-Gluon amplitudes in planar $ \mathcal{N} $ = 4
  super-Yang-Mills theory at six and seven loops}},
  \href{https://doi.org/10.1007/JHEP08(2019)016}{\emph{JHEP} {\bfseries 08}
  (2019) 016} [\href{https://arxiv.org/abs/1903.10890}{{\ttfamily
  1903.10890}}].

\bibitem{Caron-Huot:2020bkp}
S.~Caron-Huot, L.~J. Dixon, J.~M. Drummond, F.~Dulat, J.~Foster,
  O.~G\"urdo\u{g}an et~al., \emph{{The Steinmann Cluster Bootstrap for
  $\mathcal{N}$ = 4 Super Yang-Mills Amplitudes}},
  \href{https://doi.org/10.22323/1.376.0003}{\emph{PoS} {\bfseries CORFU2019}
  (2020) 003} [\href{https://arxiv.org/abs/2005.06735}{{\ttfamily
  2005.06735}}].

\bibitem{DDToAppear}
L.~Dixon and F.~Dulat, ``{The Seven-Loop Six-Gluon NMHV Amplitude in Planar
  ${\cal N}=4$ Super-Yang-Mills Theory}.'' to appear.

\bibitem{Gaiotto:2011dt}
D.~Gaiotto, J.~Maldacena, A.~Sever and P.~Vieira, \emph{{Pulling the straps of
  polygons}}, \href{https://doi.org/10.1007/JHEP12(2011)011}{\emph{JHEP}
  {\bfseries 12} (2011) 011} [\href{https://arxiv.org/abs/1102.0062}{{\ttfamily
  1102.0062}}].

\bibitem{Caron-Huot:2018dsv}
S.~Caron-Huot, L.~J. Dixon, M.~von Hippel, A.~J. McLeod and G.~Papathanasiou,
  \emph{{The Double Pentaladder Integral to All Orders}},
  \href{https://doi.org/10.1007/JHEP07(2018)170}{\emph{JHEP} {\bfseries 07}
  (2018) 170} [\href{https://arxiv.org/abs/1806.01361}{{\ttfamily
  1806.01361}}].

\bibitem{Caron-Huot:2019bsq}
S.~Caron-Huot, L.~J. Dixon, F.~Dulat, M.~Von~Hippel, A.~J. McLeod and
  G.~Papathanasiou, \emph{{The Cosmic Galois Group and Extended Steinmann
  Relations for Planar $\mathcal{N} = 4$ SYM Amplitudes}},
  \href{https://doi.org/10.1007/JHEP09(2019)061}{\emph{JHEP} {\bfseries 09}
  (2019) 061} [\href{https://arxiv.org/abs/1906.07116}{{\ttfamily
  1906.07116}}].

\bibitem{He:2021mme}
S.~He, Z.~Li and Q.~Yang, \emph{{Comments on all-loop constraints for
  scattering amplitudes and Feynman integrals}},
  \href{https://doi.org/10.1007/JHEP01(2022)073}{\emph{JHEP} {\bfseries 01}
  (2022) 073} [\href{https://arxiv.org/abs/2108.07959}{{\ttfamily
  2108.07959}}].

\bibitem{Drummond:2017ssj}
J.~Drummond, J.~Foster and {\"O}.~G{\"u}rdo{\u{g}}an, \emph{{Cluster Adjacency
  Properties of Scattering Amplitudes in $\mathcal{N}=4$ Supersymmetric
  Yang-Mills Theory}},
  \href{https://doi.org/10.1103/PhysRevLett.120.161601}{\emph{Phys. Rev. Lett.}
  {\bfseries 120} (2018) 161601}
  [\href{https://arxiv.org/abs/1710.10953}{{\ttfamily 1710.10953}}].

\bibitem{Drummond:2018caf}
J.~Drummond, J.~Foster, {\"O}.~G{\"u}rdo{\u{g}}an and G.~Papathanasiou,
  \emph{{Cluster adjacency and the four-loop NMHV heptagon}},
  \href{https://doi.org/10.1007/JHEP03(2019)087}{\emph{JHEP} {\bfseries 03}
  (2019) 087} [\href{https://arxiv.org/abs/1812.04640}{{\ttfamily
  1812.04640}}].

\bibitem{Alday:2010ku}
L.~F. Alday, D.~Gaiotto, J.~Maldacena, A.~Sever and P.~Vieira, \emph{{An
  Operator Product Expansion for Polygonal null Wilson Loops}},
  \href{https://doi.org/10.1007/JHEP04(2011)088}{\emph{JHEP} {\bfseries 1104}
  (2011) 088} [\href{https://arxiv.org/abs/1006.2788}{{\ttfamily 1006.2788}}].

\bibitem{Basso:2013vsa}
B.~Basso, A.~Sever and P.~Vieira, \emph{{Spacetime and Flux Tube S-Matrices at
  Finite Coupling for $\mathcal{N}=4$ Supersymmetric Yang-Mills Theory}},
  \href{https://doi.org/10.1103/PhysRevLett.111.091602}{\emph{Phys. Rev. Lett.}
  {\bfseries 111} (2013) 091602}
  [\href{https://arxiv.org/abs/1303.1396}{{\ttfamily 1303.1396}}].

\bibitem{Basso:2013aha}
B.~Basso, A.~Sever and P.~Vieira, \emph{{Space-time S-matrix and Flux tube
  S-matrix II. Extracting and Matching Data}},
  \href{https://doi.org/10.1007/JHEP01(2014)008}{\emph{JHEP} {\bfseries 1401}
  (2014) 008} [\href{https://arxiv.org/abs/1306.2058}{{\ttfamily 1306.2058}}].

\bibitem{Basso:2014koa}
B.~Basso, A.~Sever and P.~Vieira, \emph{{Space-time S-matrix and Flux-tube
  S-matrix III. The two-particle contributions}},
  \href{https://doi.org/10.1007/JHEP08(2014)085}{\emph{JHEP} {\bfseries 08}
  (2014) 085} [\href{https://arxiv.org/abs/1402.3307}{{\ttfamily 1402.3307}}].

\bibitem{Drummond:2007aua}
J.~Drummond, G.~Korchemsky and E.~Sokatchev, \emph{{Conformal properties of
  four-gluon planar amplitudes and Wilson loops}},
  \href{https://doi.org/10.1016/j.nuclphysb.2007.11.041}{\emph{Nucl. Phys. B}
  {\bfseries 795} (2008) 385}
  [\href{https://arxiv.org/abs/0707.0243}{{\ttfamily 0707.0243}}].

\bibitem{Brandhuber:2007yx}
A.~Brandhuber, P.~Heslop and G.~Travaglini, \emph{{MHV amplitudes in
  $\mathcal{N}=4$ super Yang-Mills and Wilson loops}},
  \href{https://doi.org/10.1016/j.nuclphysb.2007.11.002}{\emph{Nucl. Phys. B}
  {\bfseries 794} (2008) 231}
  [\href{https://arxiv.org/abs/0707.1153}{{\ttfamily 0707.1153}}].

\bibitem{Alday:2007he}
L.~F. Alday and J.~Maldacena, \emph{{Comments on gluon scattering amplitudes
  via AdS/CFT}},
  \href{https://doi.org/10.1088/1126-6708/2007/11/068}{\emph{JHEP} {\bfseries
  0711} (2007) 068} [\href{https://arxiv.org/abs/0710.1060}{{\ttfamily
  0710.1060}}].

\bibitem{Drummond:2007au}
J.~Drummond, J.~Henn, G.~Korchemsky and E.~Sokatchev, \emph{{Conformal Ward
  identities for Wilson loops and a test of the duality with gluon
  amplitudes}},
  \href{https://doi.org/10.1016/j.nuclphysb.2009.10.013}{\emph{Nucl.Phys.}
  {\bfseries B826} (2010) 337}
  [\href{https://arxiv.org/abs/0712.1223}{{\ttfamily 0712.1223}}].

\bibitem{Alday:2008yw}
L.~F. Alday and R.~Roiban, \emph{{Scattering Amplitudes, Wilson Loops and the
  String/Gauge Theory Correspondence}},
  \href{https://doi.org/10.1016/j.physrep.2008.08.002}{\emph{Phys. Rept.}
  {\bfseries 468} (2008) 153}
  [\href{https://arxiv.org/abs/0807.1889}{{\ttfamily 0807.1889}}].

\bibitem{Adamo:2011pv}
T.~Adamo, M.~Bullimore, L.~Mason and D.~Skinner, \emph{{Scattering Amplitudes
  and Wilson Loops in Twistor Space}},
  \href{https://doi.org/10.1088/1751-8113/44/45/454008}{\emph{J. Phys. A}
  {\bfseries 44} (2011) 454008}
  [\href{https://arxiv.org/abs/1104.2890}{{\ttfamily 1104.2890}}].

\bibitem{Ben-Israel:2018ckc}
R.~Ben-Israel, A.~G. Tumanov and A.~Sever, \emph{{Scattering amplitudes
  \textemdash{} Wilson loops duality for the first non-planar correction}},
  \href{https://doi.org/10.1007/JHEP08(2018)122}{\emph{JHEP} {\bfseries 08}
  (2018) 122} [\href{https://arxiv.org/abs/1802.09395}{{\ttfamily
  1802.09395}}].

\bibitem{Dixon:2020bbt}
L.~J. Dixon, A.~J. McLeod and M.~Wilhelm, \emph{{A Three-Point Form Factor
  Through Five Loops}},
  \href{https://doi.org/10.1007/JHEP04(2021)147}{\emph{JHEP} {\bfseries 04}
  (2021) 147} [\href{https://arxiv.org/abs/2012.12286}{{\ttfamily
  2012.12286}}].

\bibitem{Dixon:2022rse}
L.~J. Dixon, O.~Gurdogan, A.~J. McLeod and M.~Wilhelm, \emph{{Bootstrapping a
  stress-tensor form factor through eight loops}},
  \href{https://doi.org/10.1007/JHEP07(2022)153}{\emph{JHEP} {\bfseries 07}
  (2022) 153} [\href{https://arxiv.org/abs/2204.11901}{{\ttfamily
  2204.11901}}].

\bibitem{Brandhuber:2010ad}
A.~Brandhuber, B.~Spence, G.~Travaglini and G.~Yang, \emph{{Form Factors in
  $\mathcal{N}=4$ Super Yang-Mills and Periodic Wilson Loops}},
  \href{https://doi.org/10.1007/JHEP01(2011)134}{\emph{JHEP} {\bfseries 01}
  (2011) 134} [\href{https://arxiv.org/abs/1011.1899}{{\ttfamily 1011.1899}}].

\bibitem{Brandhuber:2012vm}
A.~Brandhuber, G.~Travaglini and G.~Yang, \emph{{Analytic two-loop form factors
  in $\mathcal{N}=4$ SYM}},
  \href{https://doi.org/10.1007/JHEP05(2012)082}{\emph{JHEP} {\bfseries 05}
  (2012) 082} [\href{https://arxiv.org/abs/1201.4170}{{\ttfamily 1201.4170}}].

\bibitem{Sever:2020jjx}
A.~Sever, A.~G. Tumanov and M.~Wilhelm, \emph{{Operator Product Expansion for
  Form Factors}},
  \href{https://doi.org/10.1103/PhysRevLett.126.031602}{\emph{Phys. Rev. Lett.}
  {\bfseries 126} (2021) 031602}
  [\href{https://arxiv.org/abs/2009.11297}{{\ttfamily 2009.11297}}].

\bibitem{Sever:2021nsq}
A.~Sever, A.~G. Tumanov and M.~Wilhelm, \emph{{An Operator Product Expansion
  for Form Factors II. Born level}},
  \href{https://doi.org/10.1007/JHEP10(2021)071}{\emph{JHEP} {\bfseries 10}
  (2021) 071} [\href{https://arxiv.org/abs/2105.13367}{{\ttfamily
  2105.13367}}].

\bibitem{Sever:2021xga}
A.~Sever, A.~G. Tumanov and M.~Wilhelm, \emph{{An Operator Product Expansion
  for Form Factors III. Finite Coupling and Multi-Particle Contributions}},
  \href{https://doi.org/10.1007/JHEP03(2022)128}{\emph{JHEP} {\bfseries 03}
  (2022) 128} [\href{https://arxiv.org/abs/2112.10569}{{\ttfamily
  2112.10569}}].

\bibitem{Dixon:2021tdw}
L.~J. Dixon, {\"O}.~G{\"u}rdo{\u{g}}an, A.~J. McLeod and M.~Wilhelm,
  \emph{{Folding Amplitudes into Form Factors: An Antipodal Duality}},
  \href{https://doi.org/10.1103/PhysRevLett.128.111602}{\emph{Phys. Rev. Lett.}
  {\bfseries 128} (2022) 111602}
  [\href{https://arxiv.org/abs/2112.06243}{{\ttfamily 2112.06243}}].

\bibitem{Dixon:2022xqh}
L.~J. Dixon, O.~G\"urdo\u{g}an, Y.-T. Liu, A.~J. McLeod and M.~Wilhelm,
  \emph{{Antipodal Self-Duality for a Four-Particle Form Factor}},
  \href{https://doi.org/10.1103/PhysRevLett.130.111601}{\emph{Phys. Rev. Lett.}
  {\bfseries 130} (2023) 111601}
  [\href{https://arxiv.org/abs/2212.02410}{{\ttfamily 2212.02410}}].

\bibitem{MoreASDToAppear}
L.~J. Dixon, O.~G\"urdo\u{g}an, Y.-T. Liu, A.~J. McLeod and M.~Wilhelm, ``{More
  Antipodal Self-Duality}.'' to appear.

\bibitem{Basso:2020xts}
B.~Basso, L.~J. Dixon and G.~Papathanasiou, \emph{{The Origin of the Six-Gluon
  Amplitude in Planar $\mathcal{N}=4$ SYM}},
  \href{https://doi.org/10.1103/PhysRevLett.124.161603}{\emph{Phys. Rev. Lett.}
  {\bfseries 124} (2020) 161603}
  [\href{https://arxiv.org/abs/2001.05460}{{\ttfamily 2001.05460}}].

\bibitem{Basso:2022ruw}
B.~Basso, L.~J. Dixon, Y.-T. Liu and G.~Papathanasiou, \emph{{An Origin Story
  for Amplitudes}},
  \href{https://doi.org/10.1103/PhysRevLett.130.111602}{\emph{Phys. Rev. Lett.}
  {\bfseries 130} (2023) 111602}
  [\href{https://arxiv.org/abs/2211.12555}{{\ttfamily 2211.12555}}].

\bibitem{Basso:2014pla}
B.~Basso, S.~Caron-Huot and A.~Sever, \emph{{Adjoint BFKL at finite coupling: a
  short-cut from the collinear limit}},
  \href{https://doi.org/10.1007/JHEP01(2015)027}{\emph{JHEP} {\bfseries 01}
  (2015) 027} [\href{https://arxiv.org/abs/1407.3766}{{\ttfamily 1407.3766}}].

\bibitem{CosmicWebsite}
\url{http://www.slac.stanford.edu/~lance/Cosmic/}.

\bibitem{Gonch2}
A.~B. Goncharov, \emph{Galois symmetries of fundamental groupoids and
  noncommutative geometry},
  \href{https://doi.org/10.1215/S0012-7094-04-12822-2}{\emph{Duke Math. J.}
  {\bfseries 128} (2005) 209}
  [\href{https://arxiv.org/abs/math/0208144}{{\ttfamily math/0208144}}].

\bibitem{Brown:2011ik}
F.~Brown, \emph{{On the decomposition of motivic multiple zeta values}},
  {\emph{{Adv. Studies in Pure Math.}} {\bfseries 63} (2012) 31}
  [\href{https://arxiv.org/abs/1102.1310}{{\ttfamily 1102.1310}}].

\bibitem{Duhr:2012fh}
C.~Duhr, \emph{{Hopf algebras, coproducts and symbols: an application to Higgs
  boson amplitudes}},
  \href{https://doi.org/10.1007/JHEP08(2012)043}{\emph{JHEP} {\bfseries 1208}
  (2012) 043} [\href{https://arxiv.org/abs/1203.0454}{{\ttfamily 1203.0454}}].

\bibitem{Brown1102.1312}
F.~Brown, \emph{Mixed {T}ate motives over {$\mathbb{Z}$}},
  \href{https://doi.org/10.4007/annals.2012.175.2.10}{\emph{Ann. of Math. (2)}
  {\bfseries 175} (2012) 949}
  [\href{https://arxiv.org/abs/1102.1312}{{\ttfamily 1102.1312}}].

\bibitem{Brown:2015fyf}
F.~Brown, \emph{{Feynman amplitudes, coaction principle, and cosmic Galois
  group}}, \href{https://doi.org/10.4310/CNTP.2017.v11.n3.a1}{\emph{Commun.
  Num. Theor. Phys.} {\bfseries 11} (2017) 453}
  [\href{https://arxiv.org/abs/1512.06409}{{\ttfamily 1512.06409}}].

\bibitem{Beisert:2006ez}
N.~Beisert, B.~Eden and M.~Staudacher, \emph{{Transcendentality and Crossing}},
  \href{https://doi.org/10.1088/1742-5468/2007/01/P01021}{\emph{J. Stat. Mech.}
  {\bfseries 0701} (2007) P01021}
  [\href{https://arxiv.org/abs/hep-th/0610251}{{\ttfamily hep-th/0610251}}].

\bibitem{DelDuca:2016lad}
V.~Del~Duca, S.~Druc, J.~Drummond, C.~Duhr, F.~Dulat, R.~Marzucca et~al.,
  \emph{{Multi-Regge kinematics and the moduli space of Riemann spheres with
  marked points}}, \href{https://doi.org/10.1007/JHEP08(2016)152}{\emph{JHEP}
  {\bfseries 08} (2016) 152}
  [\href{https://arxiv.org/abs/1606.08807}{{\ttfamily 1606.08807}}].

\bibitem{vonManteuffel:2014ixa}
A.~von Manteuffel and R.~M. Schabinger, \emph{{A novel approach to integration
  by parts reduction}},
  \href{https://doi.org/10.1016/j.physletb.2015.03.029}{\emph{Phys. Lett. B}
  {\bfseries 744} (2015) 101}
  [\href{https://arxiv.org/abs/1406.4513}{{\ttfamily 1406.4513}}].

\bibitem{Peraro:2016wsq}
T.~Peraro, \emph{{Scattering amplitudes over finite fields and multivariate
  functional reconstruction}},
  \href{https://doi.org/10.1007/JHEP12(2016)030}{\emph{JHEP} {\bfseries 12}
  (2016) 030} [\href{https://arxiv.org/abs/1608.01902}{{\ttfamily
  1608.01902}}].

\bibitem{HyperlogProcedures}
O.~Schnetz. Computer program {\sc HyperlogProcedures},
  \url{https://www.math.fau.de/person/oliver-schnetz/}.

\bibitem{Dixon:2016epj}
L.~J. Dixon and I.~Esterlis, \emph{{All orders results for self-crossing Wilson
  loops mimicking double parton scattering}},
  \href{https://doi.org/10.1007/JHEP07(2016)116}{\emph{JHEP} {\bfseries 07}
  (2016) 116} [\href{https://arxiv.org/abs/1602.02107}{{\ttfamily
  1602.02107}}].

\bibitem{Georgiou:2009mp}
G.~Georgiou, \emph{{Null Wilson loops with a self-crossing and the Wilson
  loop/amplitude conjecture}},
  \href{https://doi.org/10.1088/1126-6708/2009/09/021}{\emph{JHEP} {\bfseries
  09} (2009) 021} [\href{https://arxiv.org/abs/0904.4675}{{\ttfamily
  0904.4675}}].

\bibitem{Dorn:2011gf}
H.~Dorn and S.~Wuttke, \emph{{Wilson loop remainder function for null polygons
  in the limit of self-crossing}},
  \href{https://doi.org/10.1007/JHEP05(2011)114}{\emph{JHEP} {\bfseries 05}
  (2011) 114} [\href{https://arxiv.org/abs/1104.2469}{{\ttfamily 1104.2469}}].

\bibitem{Dorn:2011ec}
H.~Dorn and S.~Wuttke, \emph{{Hexagon Remainder Function in the Limit of
  Self-Crossing up to three Loops}},
  \href{https://doi.org/10.1007/JHEP04(2012)023}{\emph{JHEP} {\bfseries 04}
  (2012) 023} [\href{https://arxiv.org/abs/1111.6815}{{\ttfamily 1111.6815}}].

\bibitem{CaronHuotprivate}
S.~Caron-Huot. private communication.

\bibitem{Basso:2014jfa}
B.~Basso, A.~Sever and P.~Vieira, \emph{{Collinear Limit of Scattering
  Amplitudes at Strong Coupling}},
  \href{https://doi.org/10.1103/PhysRevLett.113.261604}{\emph{Phys. Rev. Lett.}
  {\bfseries 113} (2014) 261604}
  [\href{https://arxiv.org/abs/1405.6350}{{\ttfamily 1405.6350}}].

\bibitem{Basso:2014nra}
B.~Basso, A.~Sever and P.~Vieira, \emph{{Space-time S-matrix and Flux-tube
  S-matrix IV. Gluons and Fusion}},
  \href{https://doi.org/10.1007/JHEP09(2014)149}{\emph{JHEP} {\bfseries 09}
  (2014) 149} [\href{https://arxiv.org/abs/1407.1736}{{\ttfamily 1407.1736}}].

\bibitem{Belitsky:2014sla}
A.~Belitsky, \emph{{Nonsinglet pentagons and NMHV amplitudes}},
  \href{https://doi.org/10.1016/j.nuclphysb.2015.05.002}{\emph{Nucl. Phys. B}
  {\bfseries 896} (2015) 493}
  [\href{https://arxiv.org/abs/1407.2853}{{\ttfamily 1407.2853}}].

\bibitem{Belitsky:2014lta}
A.~Belitsky, \emph{{Fermionic pentagons and NMHV hexagon}},
  \href{https://doi.org/10.1016/j.nuclphysb.2015.02.025}{\emph{Nucl. Phys. B}
  {\bfseries 894} (2015) 108}
  [\href{https://arxiv.org/abs/1410.2534}{{\ttfamily 1410.2534}}].

\bibitem{Basso:2014hfa}
B.~Basso, J.~Caetano, L.~Cordova, A.~Sever and P.~Vieira, \emph{{OPE for all
  Helicity Amplitudes}},
  \href{https://doi.org/10.1007/JHEP08(2015)018}{\emph{JHEP} {\bfseries 08}
  (2015) 018} [\href{https://arxiv.org/abs/1412.1132}{{\ttfamily 1412.1132}}].

\bibitem{Belitsky:2015efa}
A.~V. Belitsky, \emph{{On factorization of multiparticle pentagons}},
  \href{https://doi.org/10.1016/j.nuclphysb.2015.05.024}{\emph{Nucl. Phys.}
  {\bfseries B897} (2015) 346}
  [\href{https://arxiv.org/abs/1501.06860}{{\ttfamily 1501.06860}}].

\bibitem{Basso:2015rta}
B.~Basso, J.~Caetano, L.~Cordova, A.~Sever and P.~Vieira, \emph{{OPE for all
  Helicity Amplitudes II. Form Factors and Data Analysis}},
  \href{https://doi.org/10.1007/JHEP12(2015)088}{\emph{JHEP} {\bfseries 12}
  (2015) 088} [\href{https://arxiv.org/abs/1508.02987}{{\ttfamily
  1508.02987}}].

\bibitem{Basso:2015uxa}
B.~Basso, A.~Sever and P.~Vieira, \emph{{Hexagonal Wilson loops in planar ${
  \mathcal N }=4$ SYM theory at finite coupling}},
  \href{https://doi.org/10.1088/1751-8113/49/41/41LT01}{\emph{J. Phys. A}
  {\bfseries 49} (2016) 41LT01}
  [\href{https://arxiv.org/abs/1508.03045}{{\ttfamily 1508.03045}}].

\bibitem{Belitsky:2016vyq}
A.~Belitsky, \emph{{Matrix pentagons}},
  \href{https://doi.org/10.1016/j.nuclphysb.2017.08.011}{\emph{Nucl. Phys. B}
  {\bfseries 923} (2017) 588}
  [\href{https://arxiv.org/abs/1607.06555}{{\ttfamily 1607.06555}}].

\bibitem{Bartels:2008ce}
J.~Bartels, L.~Lipatov and A.~Sabio~Vera, \emph{{BFKL Pomeron, Reggeized gluons
  and Bern-Dixon-Smirnov amplitudes}},
  \href{https://doi.org/10.1103/PhysRevD.80.045002}{\emph{Phys.Rev.} {\bfseries
  D80} (2009) 045002} [\href{https://arxiv.org/abs/0802.2065}{{\ttfamily
  0802.2065}}].

\bibitem{Bartels:2009vkz}
J.~Bartels, L.~N. Lipatov and A.~Sabio~Vera, \emph{{N=4 supersymmetric Yang
  Mills scattering amplitudes at high energies: The Regge cut contribution}},
  \href{https://doi.org/10.1140/epjc/s10052-009-1218-5}{\emph{Eur. Phys. J. C}
  {\bfseries 65} (2010) 587} [\href{https://arxiv.org/abs/0807.0894}{{\ttfamily
  0807.0894}}].

\bibitem{Fadin:2011we}
V.~S. Fadin and L.~N. Lipatov, \emph{{BFKL equation for the adjoint
  representation of the gauge group in the next-to-leading approximation at N=4
  SUSY}}, \href{https://doi.org/10.1016/j.physletb.2011.11.048}{\emph{Phys.
  Lett. B} {\bfseries 706} (2012) 470}
  [\href{https://arxiv.org/abs/1111.0782}{{\ttfamily 1111.0782}}].

\bibitem{BrownSVHPLs}
F.~C. Brown, \emph{Single-valued multiple polylogarithms in one variable},
  \href{https://doi.org/http://dx.doi.org/10.1016/j.crma.2004.02.001}{\emph{C.
  R. Acad. Sci. Paris, Ser. I} {\bfseries 338} (2004) 527}.

\bibitem{Dixon:2012yy}
L.~J. Dixon, C.~Duhr and J.~Pennington, \emph{{Single-valued harmonic
  polylogarithms and the multi-Regge limit}},
  \href{https://doi.org/10.1007/JHEP10(2012)074}{\emph{JHEP} {\bfseries 1210}
  (2012) 074} [\href{https://arxiv.org/abs/1207.0186}{{\ttfamily 1207.0186}}].

\bibitem{DelDuca:2022skz}
V.~Del~Duca and L.~J. Dixon, \emph{{The SAGEX review on scattering amplitudes
  Chapter 15: The multi-Regge limit}},
  \href{https://doi.org/10.1088/1751-8121/ac845c}{\emph{J. Phys. A} {\bfseries
  55} (2022) 443016} [\href{https://arxiv.org/abs/2203.13026}{{\ttfamily
  2203.13026}}].

\bibitem{Bullimore:2011kg}
M.~Bullimore and D.~Skinner, \emph{{Descent Equations for Superamplitudes}},
  \href{https://arxiv.org/abs/1112.1056}{{\ttfamily 1112.1056}}.

\bibitem{CaronHuot:2011kk}
S.~Caron-Huot and S.~He, \emph{{Jumpstarting the All-Loop S-Matrix of Planar
  $\mathcal{N}=4$ Super Yang-Mills}},
  \href{https://doi.org/10.1007/JHEP07(2012)174}{\emph{JHEP} {\bfseries 1207}
  (2012) 174} [\href{https://arxiv.org/abs/1112.1060}{{\ttfamily 1112.1060}}].

\bibitem{2011arXiv1101.4497D}
M.~{Deneufch{\^a}tel}, G.~H.~E. {Duchamp}, V.~H.~N. {Minh} and A.~I. {Solomon},
  \emph{{Independence of hyperlogarithms over function fields via algebraic
  combinatorics}},  \href{https://arxiv.org/abs/1101.4497}{{\ttfamily
  1101.4497}}.

\bibitem{Duhr:2011zq}
C.~Duhr, H.~Gangl and J.~R. Rhodes, \emph{{From polygons and symbols to
  polylogarithmic functions}},
  \href{https://doi.org/10.1007/JHEP10(2012)075}{\emph{JHEP} {\bfseries 10}
  (2012) 075} [\href{https://arxiv.org/abs/1110.0458}{{\ttfamily 1110.0458}}].

\bibitem{Schnetz:2013hqa}
O.~Schnetz, \emph{{Graphical functions and single-valued multiple
  polylogarithms}},
  \href{https://doi.org/10.4310/CNTP.2014.v8.n4.a1}{\emph{Commun. Num. Theor.
  Phys.} {\bfseries 08} (2014) 589}
  [\href{https://arxiv.org/abs/1302.6445}{{\ttfamily 1302.6445}}].

\bibitem{Panzer:2016snt}
E.~Panzer and O.~Schnetz, \emph{{The Galois coaction on $\phi^4$ periods}},
  \href{https://doi.org/10.4310/CNTP.2017.v11.n3.a3}{\emph{Commun. Num. Theor.
  Phys.} {\bfseries 11} (2017) 657}
  [\href{https://arxiv.org/abs/1603.04289}{{\ttfamily 1603.04289}}].

\end{thebibliography}

\end{document}